\documentclass[usenatbib,useAMS]{mnras}
\usepackage{epsfig,psfig,graphicx,float,color}
\usepackage{subfigure}
\usepackage{mybibdefs}
\usepackage{multirow}
\usepackage{widetext}
\usepackage{url}
 \usepackage{color}
\usepackage{amsmath}

\def\apjl{ApJL}
\def\apjs{ApJS}

\def\aap{Astron. \& Astrophys.}

\newcommand\rr{\color{black}}
\newcommand\rrr{\color{black}}

\newcommand\bh{\color{black}}
\newcommand\bhN{\color{black}}

\newcommand\bhM{\color{black}}
\newcommand\bhP{\color{black}}
\newcommand\bhh{\color{black}}

\usepackage{graphicx}
\usepackage{amssymb}
\usepackage{epstopdf}

\title[Self-consistent redshift estimation using Cls]{Self-consistent redshift estimation using correlation functions without a spectroscopic reference sample}  
\author[Hoyle and Rau]{Ben  Hoyle$^{1,2}$ \& Markus Michael Rau$^{1,2,3}$\\\\
$^1$Universitaets-Sternwarte, Fakultaet fuer Physik, Ludwig-Maximilians Universitaet Muenchen, Scheinerstr. 1, D-81679 Muenchen, Germany\\
$^2$Max Planck Institute fuer Extraterrestrial Physics, Giessenbachstr. 1, D-85748 Garching, Germany\\
$^3$McWilliams Center for Cosmology, Department of Physics, Carnegie Mellon University, Pittsburgh, PA 15213, USA\\
\\
{\tt E-mail: benhoyle1212@gmail.com, markusmichael.rau@gmail.com}
 }
  
\begin{document}
\date{Accepted ----. Received ----; in original form ----.}
\pagerange{\pageref{firstpage}--\pageref{lastpage}} \pubyear{2010}
\maketitle
\label{firstpage}
\begin{abstract}
We present a new method to estimate redshift distributions and galaxy-dark matter bias parameters using correlation functions in a fully data driven and self-consistent manner. Unlike other machine learning, template, or correlation redshift methods, this approach does not require a reference sample with known redshifts. By measuring the projected cross- and auto- correlations of different galaxy sub-samples,  e.g., as chosen by simple cells in color-magnitude space, we are able to estimate the galaxy-dark matter bias model parameters, and the shape of the redshift distributions of each sub-sample. This method fully marginalises over a flexible parameterisation of the redshift distribution and galaxy-dark matter bias parameters of sub-samples of galaxies, and thus provides a general Bayesian framework to incorporate redshift uncertainty into the cosmological analysis in a data-driven, consistent, and reproducible manner. This result is improved by an order of magnitude by  including cross-correlations with the CMB and with galaxy-galaxy lensing. 

We showcase how this method could be applied to real galaxies. {\rr By using {\rrr idealised data vectors,} in which all galaxy-dark matter model parameters and redshift distributions are known, this method is demonstrated to recover unbiased estimates on important quantities, such as the offset $\Delta_z$ between the mean of the true and estimated redshift distribution and the 68\% and 95\% and 99.5\% widths of the redshift distribution to an accuracy required by current and future surveys}.

\end{abstract}

\section{introduction}
One of the {\bh greatest} challenges facing current and upcoming large area deep photometric surveys such as KIDS \citep[][]{2016arXiv160605338H}, DES \citep[][]{2005astro.ph.10346T}, Euclid \citep[][]{2011arXiv1110.3193L}, LSST \citep[][]{2009arXiv0912.0201L}, and current and future large area radio surveys, e.g., LOFAR \citep[][]{LOFAR} and the SKA \citep[][]{SKA}, is the ability to estimate and validate redshift estimates for galaxy samples. This validation procedure is currently very challenging because of the dearth of fully representative spectroscopic samples.

If representative spectra do exist one {\bh can} validate the redshift predictions which may be obtained using template codes \citep[][]{2000ApJ...536..571B,2006MNRAS.372..565F}, machine learning codes \citep[e.g., ][]{2003LNCS.2859..226T,2004PASP..116..345C,2007AN....328..852C,2008ASPC..394..521C,2010ApJ...715..823G,2013arXiv1312.1287B,tpz,2015MNRAS.449.2040H,2015MNRAS.452.3710R,2015arXiv150407255H}, or a combination of the two \citep[][]{2015arXiv150106759H,2015arXiv151008073S}. 

In particular these predictions should be validated on spectra which have not been used during the estimation or tuning of the redshift predictions, and which are representative  of the science sample of interest. This means the validation data must be unbiased in both observable (e.g. colour) space and in {\it redshift} space. These constraints increase the number of spectra required for the process of calibrating {\bh and validating} the redshift distributions of sub-samples of galaxies. There are ongoing efforts to attempt to obtain a sample of galaxies suitable for redshift validation \citep[e.g.][]{2017ApJ...841..111M}, but we note that even the most robust spectroscopic samples, e.g., those from the SDSS-III survey \citep[][]{2011AJ....142...72E}, also suffer from some amount of spectroscopic redshift mis-measurements \citep[][]{2012AJ....144..144B}.

If representative spectra do not exist, but the density of the combined science sample and spectroscopic reference sample is high enough, and spans approximately the {\it unknown} redshift range of the galaxy sub-sample of interest, one may use the cross correlation methods {\rr of \cite{2008ApJ...684...88N,2013MNRAS.431.3307S,2013arXiv1303.4722M}, which have been carefully compared and tested using simulations \citep{2017arXiv170900992G,2018MNRAS.474.3921S}, and further applied to a variety of datasets in \cite{2006ApJ...651...14S,2015MNRAS.447.3500R,2016arXiv160505501S,2016arXiv160607434K,2017arXiv171002517D,2018MNRAS.477.2196D,2017arXiv171207298C}. These approaches either assuming knowledge of galaxy-dark matter bias of both the spectroscopic reference and the galaxy sub-sample, or marginalise over the parameter[s] of a simple galaxy-dark matter bias model.}

If either of these conditions are not met for any of the galaxy sub samples, then these galaxies should not be used for cosmological analysis because uncertainties in the redshift distribution of the sample, and in the galaxy-dark matter bias model parameters, cannot be correctly propagated through a full cosmological pipeline, and lead to biases in the cosmological analysis.  In such cases one may remove, or cull, this data in a self-consistent manner, for example by using colour space culling techniques to determine which galaxies are representative of the assumed redshift distribution using e.g., correlation function comparison analysis (see Rau \& Hoyle in prep).

Even if representative spectra do exist, current analysis is not guaranteed to lead to unbiased cosmological parameters. For example if the inferred redshift distribution of the galaxy sub-sample is systematically wrong, {\bh e.g., due to systematics or statistical errors in the spectroscopic sample}, and the measured correlation function is compared to the expected correlation function in a cosmological context, this systematic may lead to biased cosmological parameters which may highlight unphysical tension between data sets, or preference for some cosmological models over others \citep[see e.g.][]{2006MNRAS.366..101H,2014MNRAS.444..129C,2015MNRAS.452.3710R,2016arXiv160700383R}.

In this work we propose a new method to estimate the redshift distribution for these galaxies sub-samples including those which would ordinarily have to be culled. We use the measured auto- and cross- projected correlation functions of various sub-samples of galaxies as a lever to  jointly constrain a flexible parametrisation of both redshift distributions and of the galaxy-dark matter bias that would produce such observed correlation functions. When embedded in a full cosmological analysis, this method naturally updates any knowledge one may have of the parameters describing the redshift distributions of galaxy samples, the galaxy-dark matter bias parameters, and the full set of cosmological and nuisance parameters.

This work is a natural extension of other works exploring how robust redshift estimates are, {\bhM for example by} selecting galaxy sub-samples which sit in separated redshift bins and cross correlating them \citep[][]{2010MNRAS.408.1168B,2012MNRAS.427..146H,2013MNRAS.431.1547B,2015arXiv151200600S,2016arXiv160800458B}. Once accounting for the cosmic magnification bias \citep[see][]{2005ApJ...633..589S}, the cross correlation signal between separated redshift bins should be zero, and if it is not, this extra power can be used to highlight the presence of anomalous data, which can be quantified through, e.g., the outlier fractions metric \citep[][]{2010MNRAS.405..359Z}. Recently \cite{2016arXiv160707910S} show that a correction to the potentially incorrect mean redshift value, of a sample of galaxies selected to sit within a redshift bin, can be obtained by cross correlating the galaxy sample with the Cosmic Microwave Background radiation. In \cite{2017MNRAS.466.3558M} the authors take the first steps towards the work presented here, by assuming that the mean and width of the redshift distributions for a few samples of galaxies with estimated redshifts can be left as a free parameter and estimated from comparisons of auto correlations, or cross correlations with spectroscopic data sets. They do however go one step further than this work, and also show how one can jointly recover both cosmological parameters and redshift distributions uncertainties.

The approach present here is different from those before, because we do not require redshift distributions of any of the samples a priori. We {\rrr demonstrate how one could} select galaxy sub-samples only in colour-magnitude space, or in the space of any observed property, which is both simpler than selecting in redshift space, and possible for all galaxy sub-samples. We furthermore characterise the full shape of the redshift distribution and simultaneously constrain galaxy-dark matter bias model parameters of {\it each} sub-sample in a self-consistent and data driven manner, and provide full error and covariances on all parameters. We furthermore show how the inclusion of galaxy-galaxy lensing, and galaxy- CMB correlation functions can lead to further improvements in the knowledge of the redshift distributions. This paper does not vary cosmological  parameters directly, but this would only lead to a small increase in the final dimensionality of the parameter space to be explored, and as motivated by \cite{2017MNRAS.466.3558M}, it has already been shown to lead to consistent results.

The format of the paper is as follows. We describe the {\rrr ideasised simulated data vectors, and how these could be obtained from real galaxy subsamples} in \S\ref{data}, and continue by describing the correlation function methodology and analysis in \S\ref{theory}. {\rr We show how the flexible function chosen to model the redshift distributions is consistent with real data in \S\ref{method}}. We present results in \S\ref{results}, discuss in \S\ref{discuss}, and summarise and conclude in \S\ref{conclusions}. {\bhP The analysis assumes a $\Lambda$CDM cosmology from \cite{PLANK2015} with $H_0,\sigma_8,\Omega_m,\Omega_b$=(67.8,{\rr 0.829,0.308},0.049).}


\section{Construction of data samples}
\label{data}
The main body of this paper uses idealised sets of simulations, in which {\rr mock} galaxy sub-samples are generated  {\rr by hand} with full knowledge of redshift distributions and galaxy-dark matter bias parameters, {\rr for details see \S\ref{ideal}}.  {\rr It is not possible to construct these perfect information samples} using either N-body simulations populated by galaxies, because is it unclear how the intrinsic galaxy-dark matter bias is interpreted by our model, {\rr or from the real data because} we may not be confident in the accuracy of the redshift distributions.  {\rr Given that we would expect this approach to eventually be applied to data, we use this section to showcase one possible techniques that would enable galaxy sub-samples to be constructed from observational data}.

\subsection{Example galaxy samples}
We {\rrr motivate the choices of simulated data vectors by } using 2 million real galaxies with photometry and spectroscopic redshifts from the Sloan Digital Sky Survey Data Release 12 \citep[hereafter SDSS,][]{2014ApJS..211...17A} {\rrr which are obtained} through the publicly available CasJobs server \citep[][]{10.1109/MCSE.2008.6}\footnote{skyserver.sdss3.org/CasJobs}. The approach developed in this paper works in the presence of hundreds of millions of sources{\rr, such that the measured correlation functions will not be dominated by shot noise, and therefore we only motivate how it could be implemented using this SDSS data}. For the importance of Poisson, and sub-Poissonian, shot noise corrections in a correlation function analysis see \cite{2017MNRAS.470.2566P} and references therein.

\subsection{Data Partitioning}
\label{data_clustering}
In this work we estimate redshift distributions {\rrr which could correspond} to  galaxy sub-samples that each reside in distinct partitions of the observable data space. One plausible method to generate these partitions could be to first approximately estimate the redshifts of galaxies using a photometric redshift code, and then to select data which sit in different tomgraphic bins \citep[e.g.,][]{2017arXiv170801532H,2017MNRAS.465.1454H}. Another method could be to partition the data based on those regions where one trusts the redshift estimates the most \citep[e.g.,][]{2017MNRAS.466.3558M}. {\rrr Here we show how we could partition a data sample, such as the SDSS,} by constructing simple cells in the colour-magnitude space of the full sample. For this purpose we use the scikit-learn \citep[][]{scikit-learn} implementations of the Decision Tree algorithm \citep[see][for a recent implementation and use case]{2017MNRAS.468..769G}. We note that any other manifold learning or space partitioning algorithm would be equally suitable \citep[e.g.][]{2015ApJ...813...53M,2016arXiv160302763M}. {\rr We note that, in general, there there not a single {\it best} method to partition the data space, given the different observable spaces of observational surveys and differing science goals.} {\rrr These sub-samples chosen by color are to motivate the choices of simulated data vectors used later in the paper.}

\subsubsection{Decision Tree partitioning}
\label{sdss_dt}
{\rr We show case the use of a single decision tree as a (unsupervised learning) data partitioning method by ensuring that the input and target features are the same.} The resulting decision tree is very simple to interpret, consisting of trivial divisions in the input space which are represented by a set of {\tt if} statements. The depth of the decision tree describes how many data points reside in each final partition at each end of the decision tree, which is called a `leaf node'. {\rr We note that each node splitting the data has a corresponding ID, and therefore the IDs of the final leaf nodes do not increment consecutively.}

{\rr The selected SDSS spectroscopic data sample has 2 million galaxies, however we would expect the approach described in this paper to be applicable to {\rrr much larger} current and future datasets, such as DES, LSST, or Euclid. {\rrr We proceed with this relatively small sample by sub dividing the data into cells in order to obtain some realistic spectroscopic redshift distributions, used later to motivate the construction of the idealised simulated data vectors}. We choose the following dereddened magnitudes $u,g,r,i,z$ and the following constructed colours $r-i, g-u, g-r$ as the feature space when growing the decision tree. We generically refer to the boundaries of each leaf node as a  ``colour cell".

\subsubsection{Properties of the colour cells}
In Fig. \ref{KMClus_col} we show a few projections through colour-magnitude space of the data associated to some of the colour cells for the SDSS galaxies. Each panel shows a different projection through the input space, and we have highlighted a few colour cells using different coloured points.
\begin{figure*}
\includegraphics[scale=0.45,clip=true,trim=20 20 20 20]{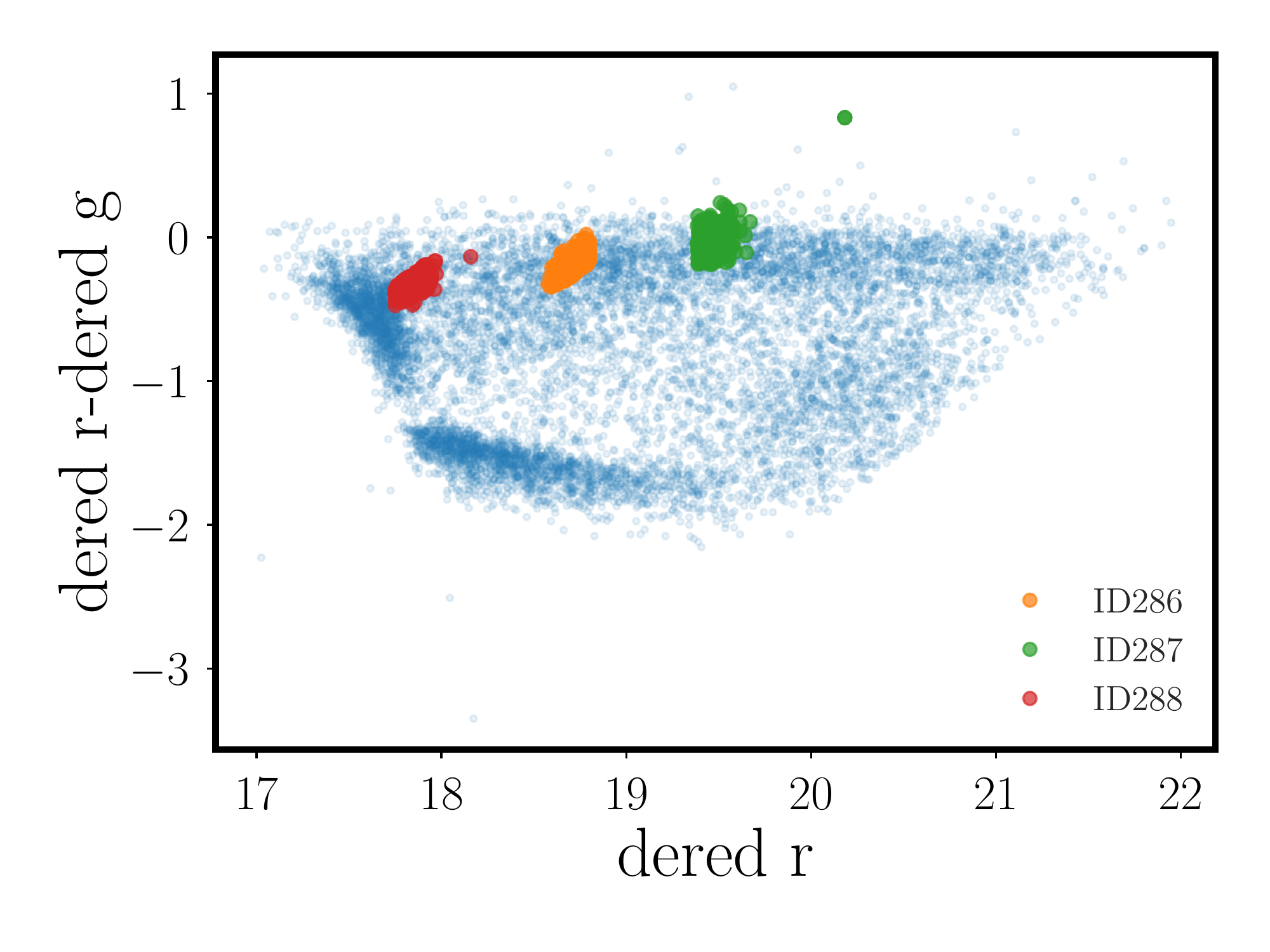}
\includegraphics[scale=0.45,clip=true,trim=20 20 20 20]{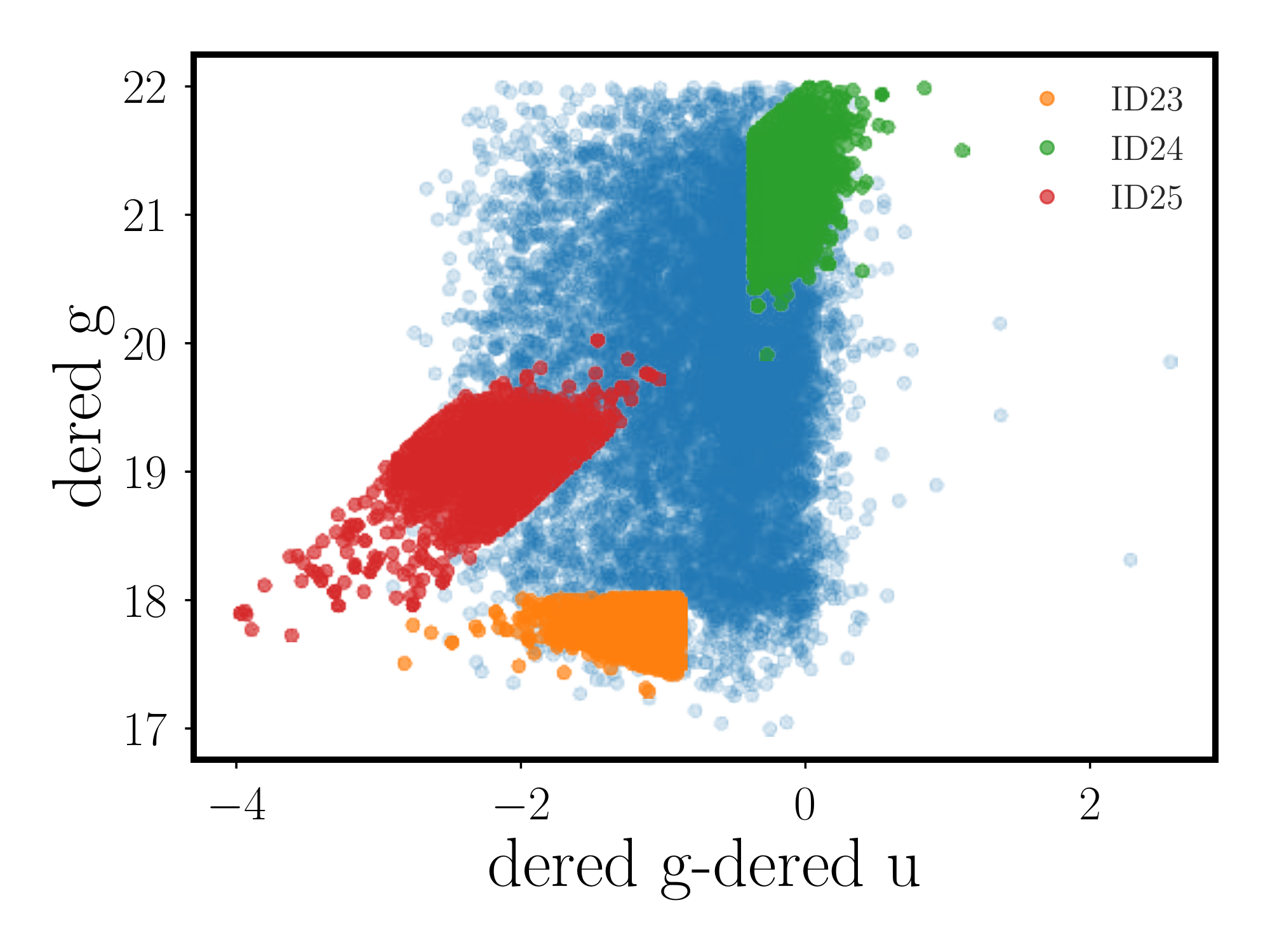}
\caption{ Projections through colour-magnitude space of the photometric data. Each panel shows a different projection of randomly selected colours and magnitudes through the {\rr SDSS data. The color cells are generated using a single Decision Tree in an unsupervised learning analysis. Each panel distinguishes a few randomly selected `colour cells' from the full distribution using different coloured points.}}
\label{KMClus_col}
\end{figure*}

The right hand panel of Fig. \ref{KMClus_col} hints at the well known cloud of red galaxies, and the separated, in $g-r$ colour, bluer main sample galaxies. This projection uses the apparent $r$ band magnitude, instead of the often used absolute magnitude, for which one requires redshifts, which smears these clouds. The colour cells naturally identify different regions of the colour space, which could correspond to different galaxy types, with different properties.

{\rr In Fig.\ref{num_KMClus} we show the true redshift changes between colour cells by presenting the mean and standard deviation of the galaxy redshifts of the SDSS data residing in each colour cell, as defined by the Decision Tree algorithm as a function of colour cell ID. The colour cells have been sorted in order of increasing redshift}.
\begin{figure}
\includegraphics[scale=0.435,clip=true,trim=20 25 20 20]{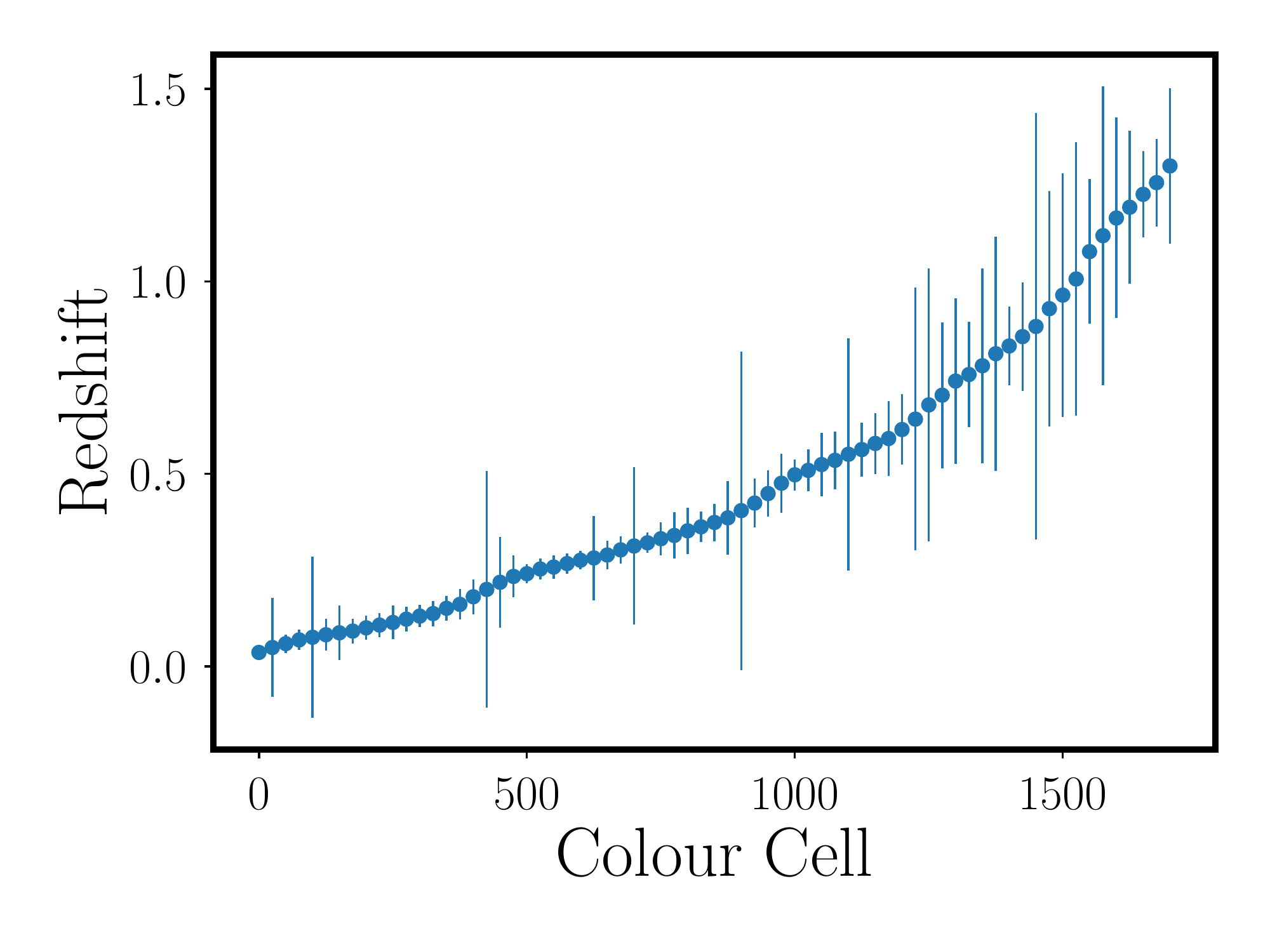}
\caption{The mean and standard deviation of SDSS galaxy redshifts of data residing in each colour cell. The colour cells have been sorted in order of increasing redshift}
\label{num_KMClus}
\end{figure}

In Fig. \ref{num_KMClus} it can be seen that different colour cells have galaxies with different mean redshifts, and redshift distributions, as shown by the width of the 1-sigma error bar. This relationship between colour and redshift is exactly the property which allows normal photometric redshift codes to assign a redshift to a galaxy based on its photometric properties. We model the full redshift distribution in each colour cell in more detail in \S\ref{param_dndz}. 

{\rr 
\subsection{The idealised galaxy sub-samples}
\label{ideal}
To test the method presented in this paper we required exact knowledge of all redshift distributions and galaxy-dark matter parameter values for each of the galaxy sub-samples. To obtain this {\rrr idealised simulated} data vector we construct redshift distributions either by hand for some analyses, or by using those dndz's obtained from the SDSS colour-cells (see \S\ref{sdss_dt}) in other analyses. Finally we randomly {\rr assign} values for the adopted three parameter galaxy-dark matter bias model for each sub-sample. We refer to each galaxy sub-sample generically as a colour-cell, even though these sub-samples do not actually refer to real galaxies.
}

\section{The projected correlation function, $C_l$}
\label{theory}
In this work we compare sets of {\rr idealised} correlation functions with predicted correlation functions in which the full redshift distributions and galaxy-dark matter bias model parameters are free parameters to be explored until good fits are obtained. We use a standard $\chi^2$ analysis to perform this comparison, described in detail at the end of this section. Through minimising and exploring around the minimum value of $\chi^2$, the redshift distributions and galaxy-dark matter bias model parameters {\bhM can be jointly estimated and interdependencies quantified, leading to a self-consistent analysis}.  In this paper we use idealised simulations{\bh, both with and without noise,} in {\bh which knowledge of the redshift distributions and galaxy-dark matter bias model parameters are known exactly. This enables a quantitative examination of this new self-consistent method.}

Here we briefly introduce the modelling of the projected correlation function, and describe how it {\rrr could be measured from a real data sample}. We note that making the measurements of projected correlation functions on data does not require any knowledge of redshift distributions or of the galaxy-dark matter bias parameters.} 

\subsection{Modelling the projected correlation function}
The large scale distribution of matter can be predicted from first principles by following the evolution of density perturbations from the early Universe to late times. 

We first define the density field of matter fluctuations as
\begin{equation}
\delta_m = \frac{\rho_{m}}{\bar{\rho}_{m}} - 1\;,
\end{equation}
where $\rho_{m}$ is the spatially varying matter density in the Universe
with a mean of ${\bar \rho_{m}}$. The matter over density $\delta_{m}$ is related to the measured galaxy overdensity $\delta_{g}$ through the bias model \citep{1993ApJ...413..447F},
\begin{equation}
\delta_g = b(k,z)\delta_{m}
\end{equation}
which may vary as a function of scale $k$, and redshift $z$.  In this work we adopt the following three parameter scale and redshift dependent galaxy-dark matter bias model for {\it each} colour cell given by,
\begin{equation}
   b(k, z)  = k^{b_k}\big( b_0 + b_1 \times (1 - 1/(1 + z) \big)\,.
\label{eq:bias}
\end{equation}
Where $b_k, b_0, b_1$ are model parameters to be estimated jointly with the redshift distribution for each colour cell. In particular, if $b_k=0, b_1=0$ we obtain a scale independent bias $b=b_0$. We note that any well motivated galaxy-dark matter bias model could be adopted, and we would suggest any flexible model, for example that of \cite{2015MNRAS.448.1389C}. We note that our galaxy-dark matter bias model {\bh is similar to } \cite{1538-4357-461-2-L65}, but with an additional scale dependent pre-factor.

{\rr We note that for either small values of $b_1$, or for redshift distributions with a small width, we can effectively rewrite Equ. \ref{eq:bias} into a two parameter galaxy-dark matter bias model.
\begin{eqnarray*}
   b(k, z)  &=& k^{b_k} *  b^* \\
   b^* &\approx& b_0 + b_1 \times (1 - 1/(1 + <z>)\,,
\label{eq:bias1}
\end{eqnarray*}
where $<z>$ is the mean value of the redshift distribution.}

Moving to Fourier space, we define the galaxy power spectra between any pair of colour cells $x$ and $y$ as:
\begin{equation}
(2\pi)^3 \delta_{\rm D}(k-k^\prime) P_{xy} (k) \equiv \langle \delta_{x}(\mathbf{k}) \, \delta^{\star}_{y}(\mathbf{k^\prime}) \rangle\;,
\end{equation}
where $\mathbf{k}$ denotes a wave vector of amplitude $k$ and the angled brackets indicates {\bh that} the average is performed over all Fourier modes within a spherical shell, and $\delta_{\rm D}$ is
the usual Dirac delta function. The galaxy power spectrum between colour cells $x$ and $y$ can be directly related to the matter power spectrum $P(k)$ though the bias model,
\begin{equation}
\label{eq:shot}
P_{xy}(k) \simeq b_{x}(k,z)b_{y}(k,z) \, P(k)
\end{equation}
where we have assumed the shot noise contribution is small \citep[see][for a fuller treatment]{2017MNRAS.470.2566P}.

This work does not assume knowledge of the redshifts distributions in the colour cells therefore we are forced to measure the angular power spectrum $C^{xy}_l$ which is a projection of $P_{xy}(k)$ onto the sky, given by

\begin{equation}
\label{eq:class_cl}
C^{xy}_l = 4 \pi \int \frac{\rm d k}{k} P_{\rm ini}(k) \Delta_l^x(k)\Delta_l^y(k) \: ,
\end{equation}
where $P_\mathrm{ini}$ denotes the (dimensionless) primordial power spectrum, and the transfer
function for the matter component $\Delta^x_l(k)$
is given by
\begin{equation}
\Delta^x_l(k) = \int {\rm d} z \: b_{x}(k,z) \frac{ {\rm d} n_x}{{\rm d} z} j_l(k \chi(z)) {\mathcal{D}}(k,z) \: .
\end{equation}
Here ${\mathcal{D}}(k,z)$ is the growth rate of the
fluctuations, $\chi(z)$ is the comoving distance,  and ${\rm d} n/{\rm d} z$ is the galaxy redshift distributions for each colour cell $x$, and $j_l$ is a Bessel function. In this work we model ${\rm d} n/{\rm d} z$ by a normalised sum of five Gaussian distributions called a Gaussian mixture model, {\rr described in more detail in \S\ref{param_dndz}}, and use the bias model $b_x$ given by Equ. \ref{eq:bias}.

The error  on the correlation function $\sigma \big(C_l \big)$, for a tracer population with number density $N$ {\bhP per steradian}, is given by \citep[see, e.g.,][]{2004MNRAS.349..603E}
\begin{equation}
\label{eq:cl_err}
\sigma \big(C_l \big) = \sqrt{\frac{2}{(2l +1)f_{sky}}} {\bh \Big( C_l  + \frac{1}{N}\Big)}\: ,
\end{equation}
where $f_{sky}$ is the fractional sky coverage, which we set to have a value 0.25, similar to that of the SDSS and the final footprint of DES. If we assume $N$ to be large, then the corresponding term can be neglected from Equ. \ref{eq:cl_err}. In reality one should carefully account for the effect of the survey mask on the term $1/N$, which introduces a scale dependence that can be modelled \citep[see e.g.,][for details]{2017MNRAS.470.2566P}.

In practice, the expected projected correlation functions are obtained by passing `trial' redshift distributions and galaxy-dark matter bias model parameters into the Cosmic Linear Anisotropy Solving System \citep[{\tt CLASS}, v2.5][]{2011arXiv1104.2932L}\footnote{\url{http://class-code.net/}}. 
{\tt CLASS} solves sets of hierarchy Boltzmann differential equations following the density perturbations in the dark matter, baryons, photons and other relevant particle species. 

{\tt CLASS}  is not constrained to predicting only the density-density perturbations, as presented above, but also all combinations of galaxy-galaxy lensing \citep[see, e.g.][]{2017arXiv170801537P}, and correlations with the Cosmic Microwave Background radiation (CMB), for example as measured by the WMAP and Plank Collaborations \citep[][]{2013ApJS..208...20B,2015arXiv150201582P} {\bhP see e.g., \cite{2016MNRAS.456.3213G} for a recent measurement.} For full details about the predicted correlation functions see \cite{2011arXiv1104.2932L}. We reiterate that all of these projected correlation functions do not require knowledge of redshift distributions, or of galaxy-dark matter model parameters, when being measured in the data, however modelling them is sensitive to these {\rr parameters as shown in} Equ. \ref{eq:class_cl}.

Furthermore, in this work we neglect contributions to the correlation function and the covariance matrices which are caused by galaxy lensing shape noise. We assume that the number density of data would be large enough to make this simplification. In practice this term should be included as appropriate, but this is well within the bounds of current galaxy lensing surveys, such as KIDS \citep[][]{2017MNRAS.465.1454H}, and DES \citep[][]{2017arXiv170801537P}. We also assume that noise from CMB lensing can be well modelled and included in the analysis from e.g., the PLANK experiment, see e.g. \cite{2016MNRAS.456.3213G}. We note that the primary goal of this paper is to provide a proof of concept of this new redshift estimation technique, and not to accurately model a state of the art survey. These extensions to the method are left to future work.

{\rr 
\subsection{The idealised data vectors}
To construct the data vector corresponding to the {\rrr simulated idealised} correlation functions, we insert known ${\rm d} n/{\rm d} z$'s and galaxy-dark matter bias parameter values, as described in \S\ref{ideal} into {\tt CLASS} to produce correlation functions. {\rrr We choose to generate up to 12 separate redshift distributions, which would correspond to 12 color cells in a real galaxy sample. We therefore also refer to each of these idealised redshift distributions as a color cell.} These correlation functions and their associated errorbars c.f. Equ. \ref{eq:cl_err}, produce the idealised {\rrr simulated} data vector $C_l^D$. We chose to assign galaxy-dark matter bias model parameter values between the ranges $0.4<b_0<1.6$,  $0<b_1<0.15$ , $-0.075<b_k<0.075$, drawn at random for each of the color-cells. We use up to 12 colour cells in this analysis, corresponding to 12 different ${\rm d} n/{\rm d} z$'s and different galaxy-dark matter bias model parameters. 
}

\subsection{Comparing measured and {\rr modelled} correlation functions}
We use the standard $\chi^2$ statistic to evaluate how well the model projected correlation functions $C_l^M$'s, as estimated from {\tt CLASS} for a given set of redshift distributions parameters and galaxy-dark matter bias model parameters, fit the measured $C_l^D$'s, from the data. The $\chi^2$ function is written generally as

\begin{equation}
\chi^2 = \sum_{i=1}^N{\bh \sum  (C_l^D - C_l^M)\, \Sigma_i^{-1}(l,l') \, (C_{l'}^D - C_{l'}^M)^T} \, ,
\end{equation}
where the outer sum runs over all $N$ measured auto- and cross- correlation functions $i=1..N$ and the {\bh inner} sum runs over all $l$ values in each measured correlation function. The covariance matrices $\Sigma$, would normally correspond to the combined covariance of the model and the data between all $l$ values for each correlation function $i$.

In this work we bin the correlation function in bins of width {\bh $\Delta(l)=15$} which allows the approximation that the covariance matrix may be expressed by only the diagonal terms with values denoted by $\sigma_l^2$. In practice one would not make this approximation because the geometry of the mask can introduce off diagonal components into the covariance matrix. We also assume that the uncertainty on the model can be neglected. This allows the $\chi^2$ to be expressed by

\begin{equation}
\chi^2 = \sum_{i=1}^N \sum_l \Big( \frac{(C_l^D - C_l^M)^2}{\sigma_l^2} \Big) \, .
\end{equation}
This is the quantity which will be used to judge the quality of the fit of the modelled with the measured projected correlation functions.

In most of the analyses presented here, the data vector $C_l^D$, is generated using the true redshift distributions, and galaxy-dark matter bias parameter values. The model vector $C_l^M$ is calculated for the trial redshift distributions and trial galaxy-dark matter bias parameter values as our sampler explores the high dimensional parameter space. We additionally do not add Gaussian random noise to the data vector, except in \S\ref{galgallenserr}. This allows us to monitor the minimum value of $\chi^2$ which will exactly equal 0 if the correct redshift distributions and galaxy-dark matter bias model parameters are found.

\section{Methods}
\label{method}
In this section we demonstrate that the model chosen to parametrise the redshift distribution ${\rm d} n/{\rm d} z$, is general enough to describe the redshift distributions we encounter in the SDSS datasets. We describe and critique the galaxy-dark matter bias model adopted and show how the predicted correlation function changes in the face of changing redshift distributions and galaxy-dark matter bias parameters. We end this section with a description of the algorithm to explore the high dimensional parameter space.

\subsection{Parameterising the redshift distribution using a Gaussian Mixture Model}
\label{param_dndz}

\begin{figure}
\includegraphics[scale=0.45,clip=true,trim=20 0 10 10]{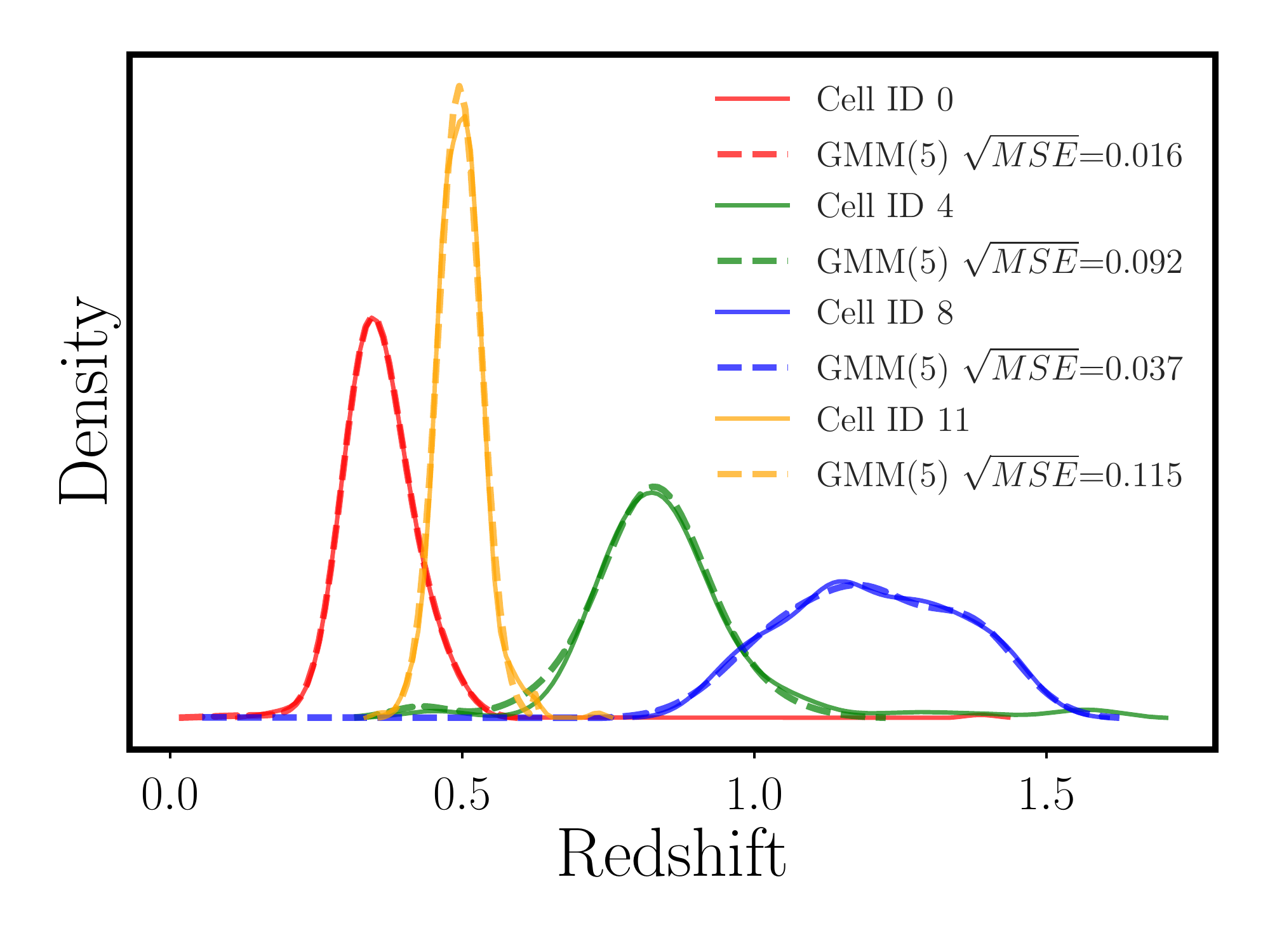}
\caption{The solid lines {\rr show the redshift distributions of SDSS data associated to a few colour cells, as shown in the legend. The dashed lines of the same colour show a reasonable} fitting five Gaussian mixture model to the redshift distributions.}
\label{sdss_KMClus_col}
\end{figure}

We choose to model the redshift distribution as a Gaussian Mixture Model consisting of 5 Gaussians, hereafter denoted GMM(5). We have empirically chosen to use 5 Gaussians because this provides enough flexibility to {\rr fit all of the observed redshift distributions from data sitting in each of the colour cells of the SDSS. We find that a lower number of Gaussians does not provide an adequate fit to the redshift distribution such that the mean squared error between the true redshift distribution and the best fit redshift distribution is large.} Using more than 5 Gaussians increases the size of the parameter space to be explored, but does {\rrr not} improve the mean squared error noticeable. {\rr Of course a GMM(5) is not flexible enough to fit all distributions. We leave an fuller exploration of this problem to future work.}
{\rr 
For illustrative purposes only, we use the {\tt scipy.stats.optimise} \citep[][]{scipy} package to obtain a reasonable GMM(5) fit to each of the redshift distributions in each colour cell of the SDSS data. In Fig. \ref{sdss_KMClus_col} we show the redshift distributions of data belonging to four colour cells using the coloured solid lines, and the best fit GMM(5) approximation to the redshift distribution using the dashed lines of the same colour. We note that the lines often sit directly on top of each other. In the legend of each panel we show the ID for the colour cell, corresponding to the redshift distributions shown in the results panel of the same number, in the results section \S\ref{gg-corrl-sdss}, and the value of the square root of the mean squared error (MSE) between the true and approximated distributions. We choose to show the square root of the MSE because this value is very small.
}

We see that the GMM(5) parametrisation provides a {\rr reasonable} fit to the redshift distributions in each cell.  We note that during the actual analysis, the exploration of the best fitting parameter space will be more complete than that of the {\tt minimise} package which may fall into local minima.

\subsection{The effect of galaxy-dark matter bias model on $C_l$'s}
For illustration purposes we show how the modelled galaxy-galaxy projected correlation function changes with varying the redshift distributions and with galaxy-dark matter bias model parameter values. We note that in general these results are sensitive to the {\bhP choice} of galaxy sub-sample, the scales of interest, and the choice of correlation function probed.

\begin{figure*}
\includegraphics[scale=0.315,clip=true,trim=20 87 15 25]{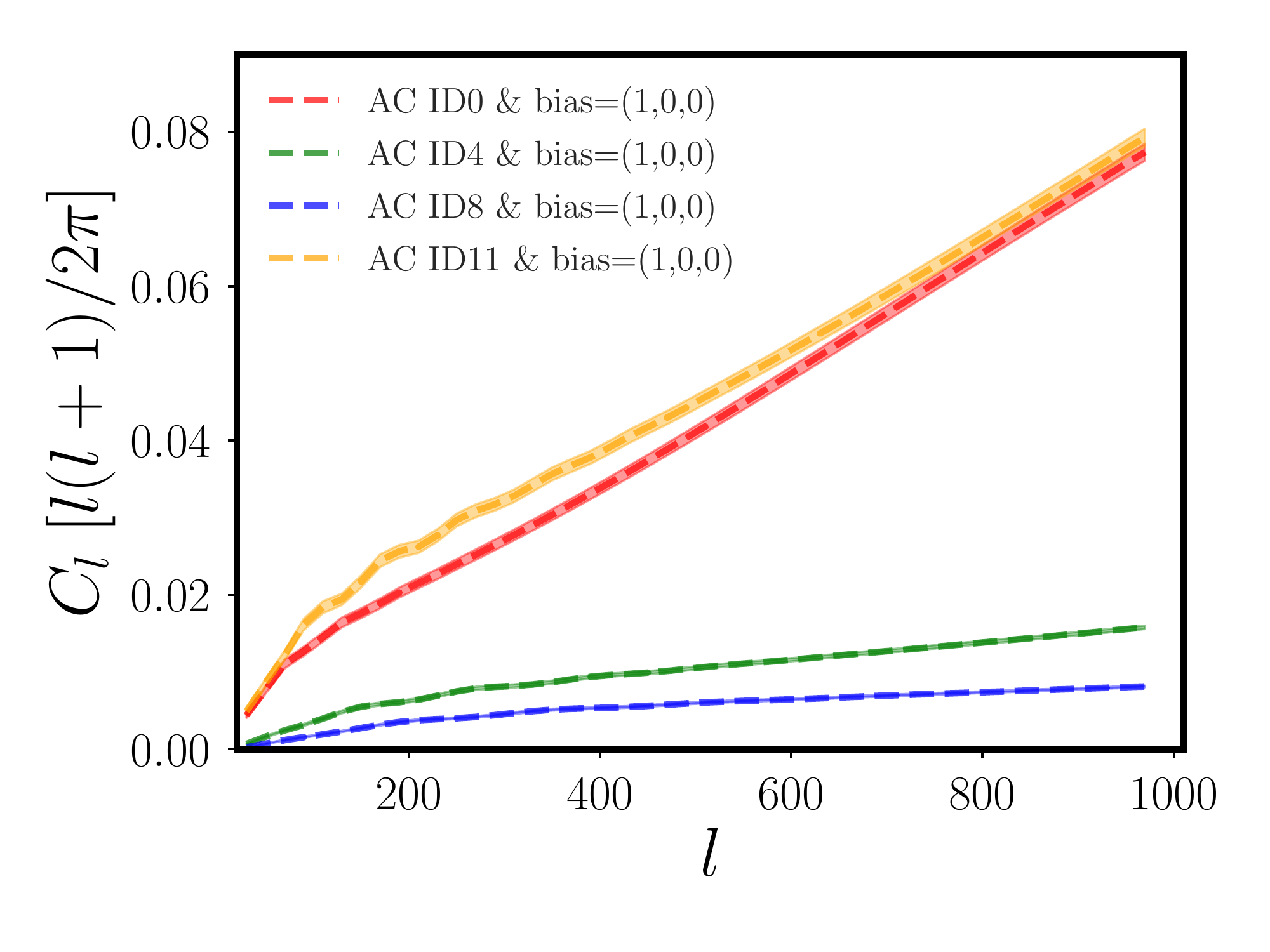}
\includegraphics[scale=0.315,clip=true,trim=55 87 15 25]{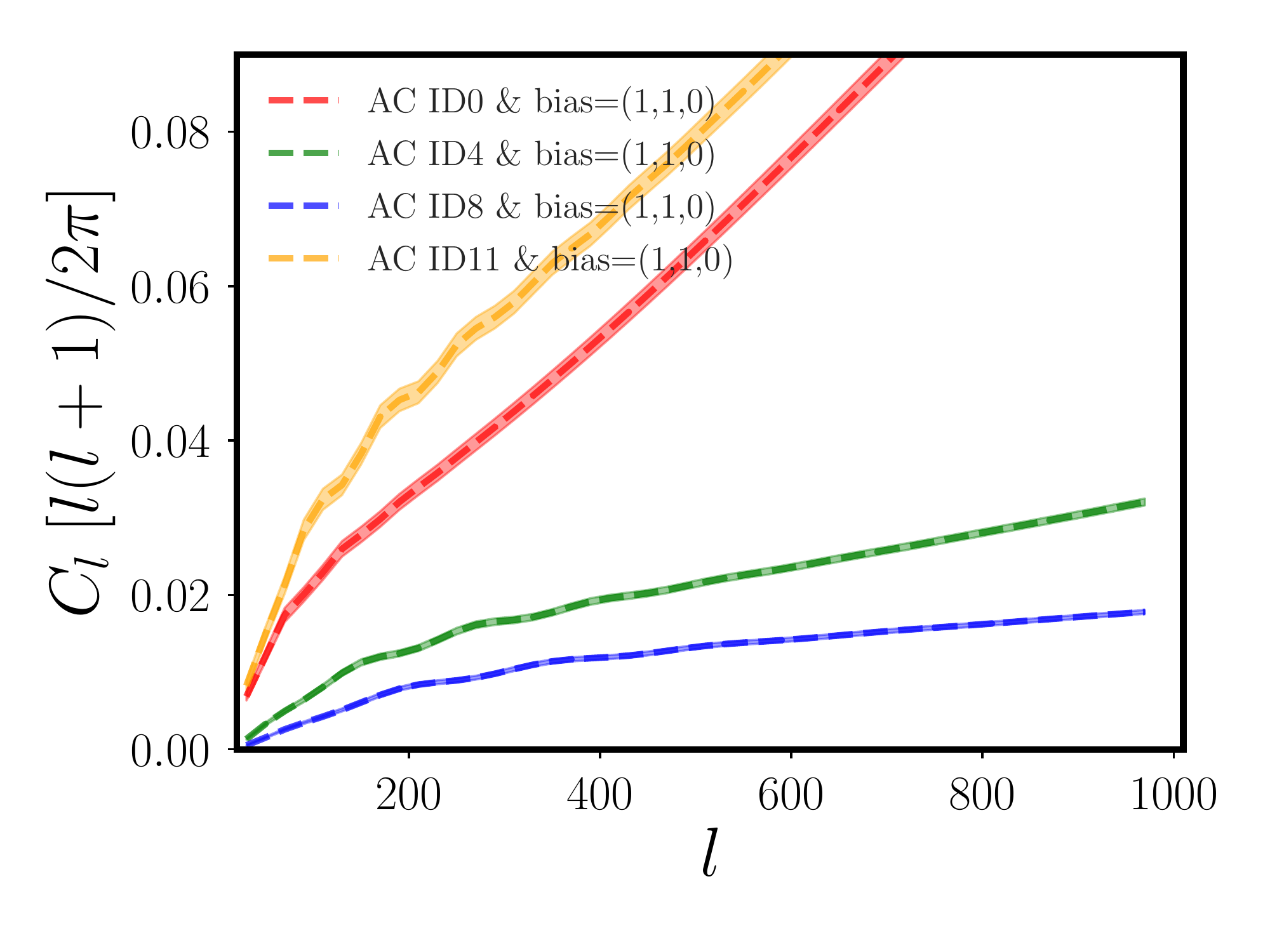}
\includegraphics[scale=0.315,clip=true,trim=55 87 25 25]{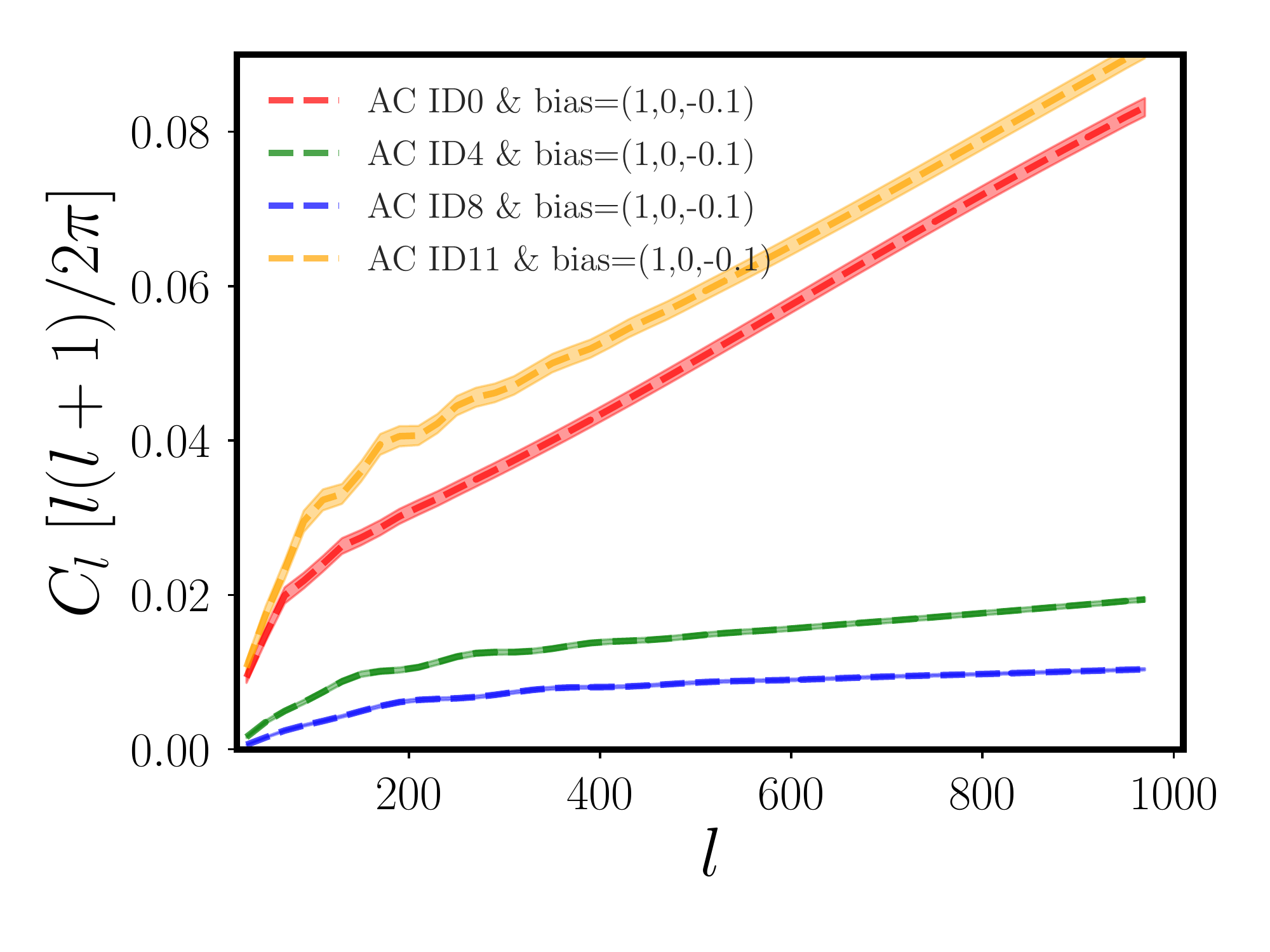}\\
\includegraphics[scale=0.32,clip=true,trim=20 20 20 25]{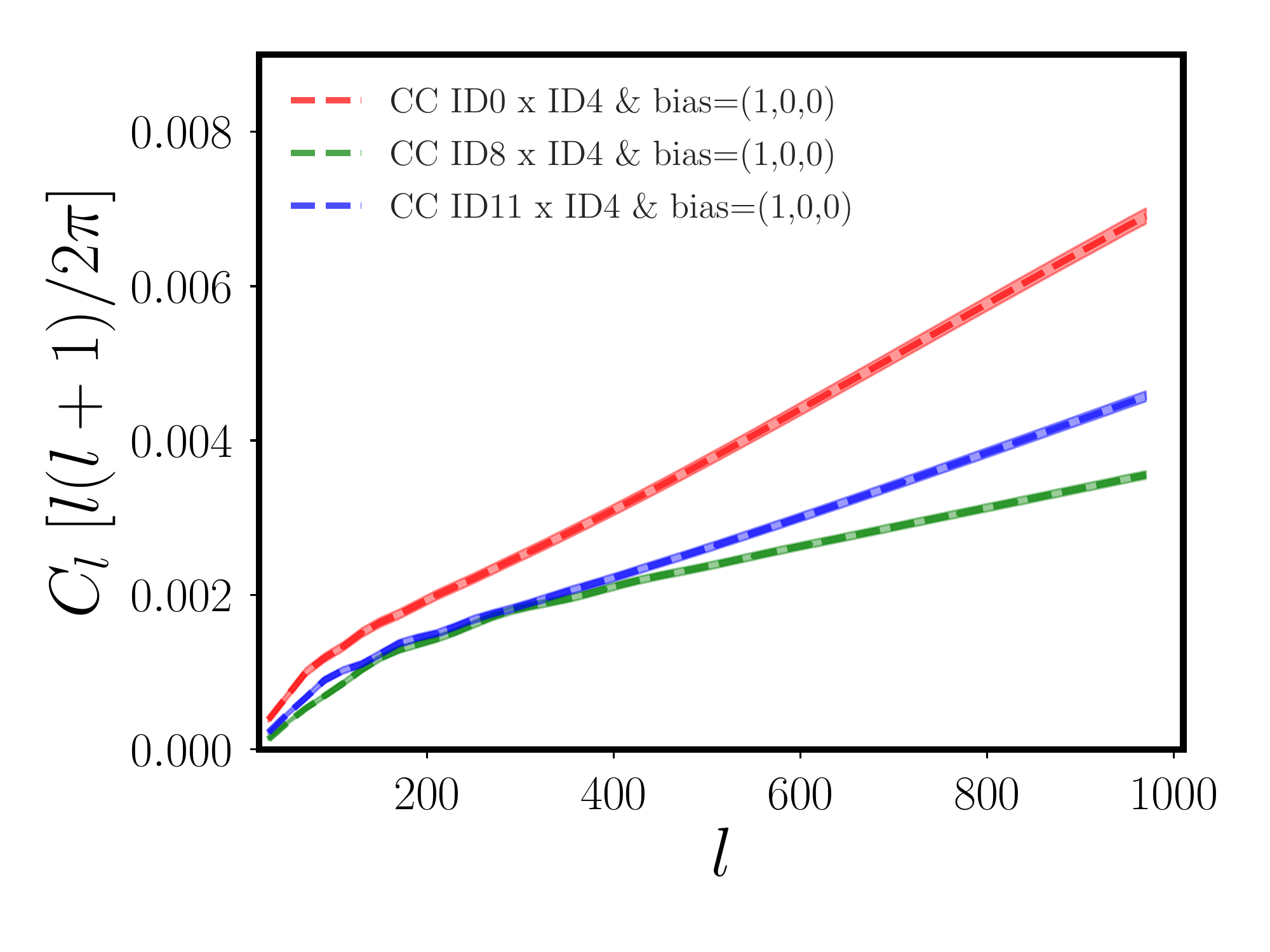}
\includegraphics[scale=0.32,clip=true,trim=55 20 20 25]{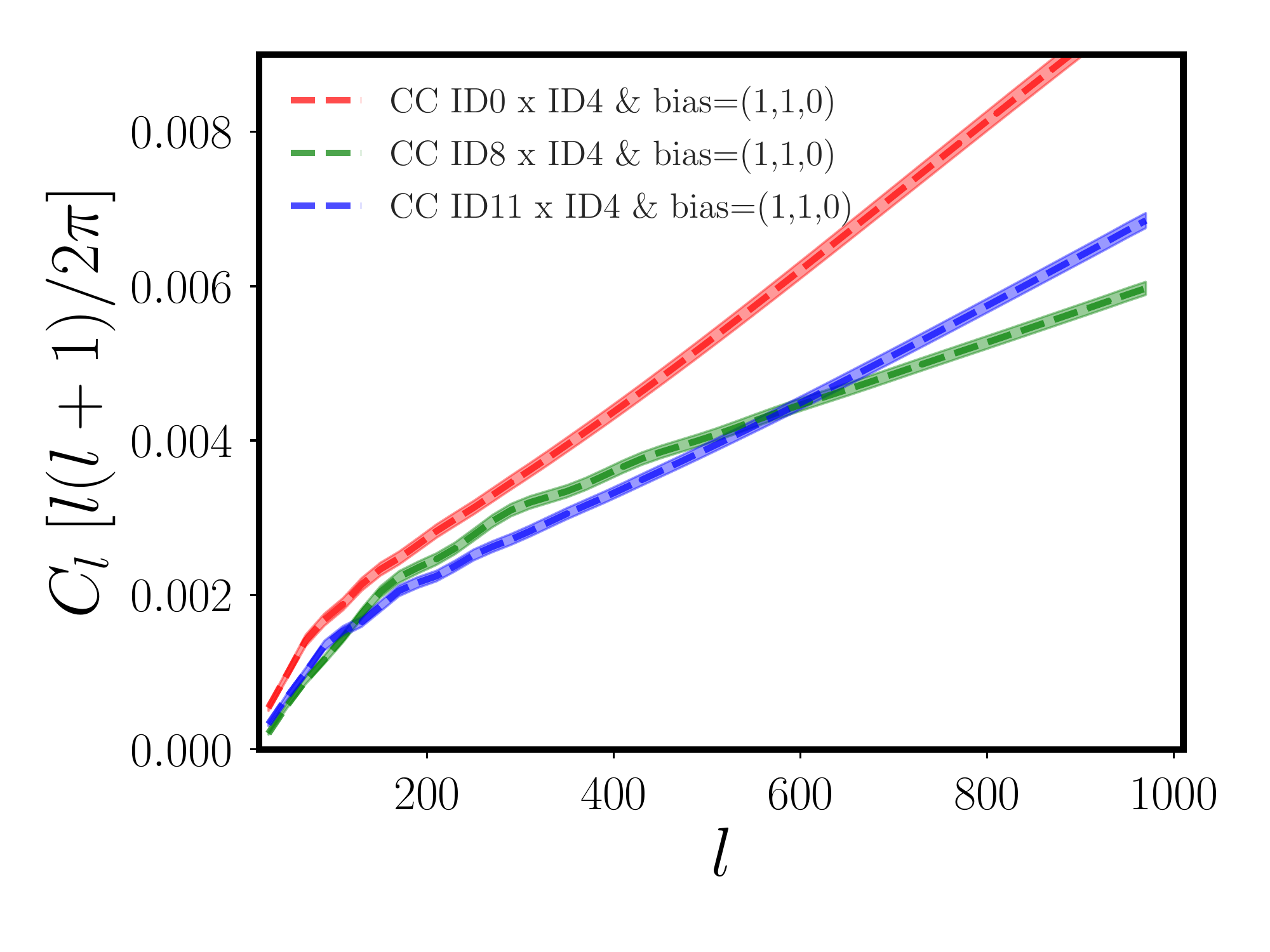}
\includegraphics[scale=0.32,clip=true,trim=55 20 20 25]{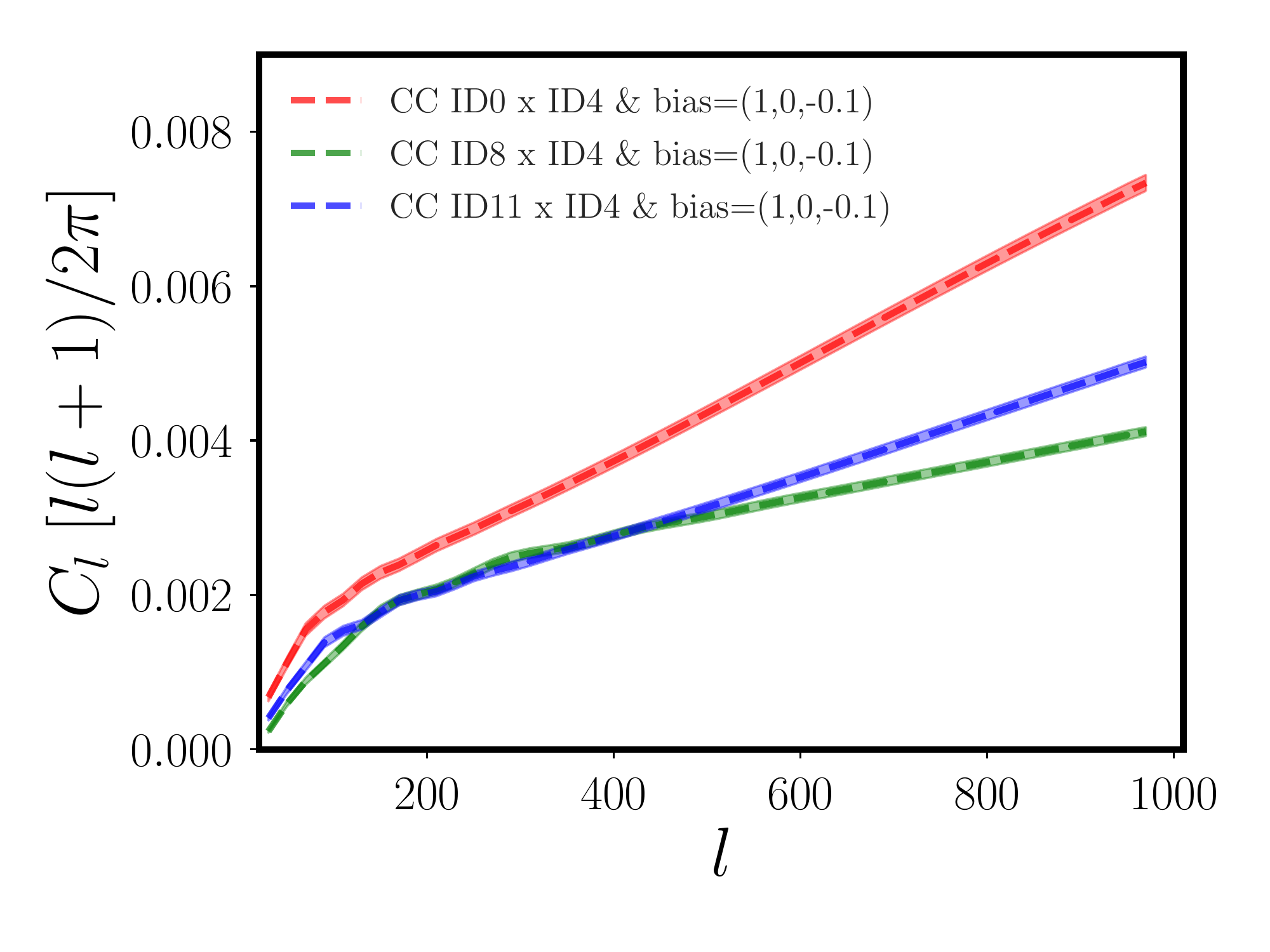}
\caption{{\rr A pedagogical insight into complex effect of the galaxy-dark matter bias model. We show the predicted projected correlation functions $C_l$ as a function of scale $l$ for the SDSS redshift distributions identified by the cell IDs 0, 4, 8, 11, which are presented in Figs. \ref{sdss_KMClus_col} and \ref{dndnz_sdss_ncl}. The top (bottom) panels shows the predicted projected auto-correlation AC, (cross-correlation CC,) function. Each row shows the effect of varying the galaxy-dark matter bias model parameters in the order ($b_0, b_1, b_k$) from right to left, while fixing the redshift distributions. The y-axis range of the lower panel is an order of magnitude less than the upper panel.}}
\label{dndnz_cls2}
\end{figure*}

In the top panels of Fig. \ref{dndnz_cls2} we show the projected correlation functions $C_l$ as a function of scale $l$ for simulated galaxies residing in the four different colour cells with ID's 0,4,8,11. Each panel shows the effect of assuming the correct redshift distribution ${\rm d} n/{\rm d} z$, but varying the galaxy-dark matter bias model parameters in the order ($b_0, b_1, b_k$), see legend for specific values. The top panels show the predicted auto- correlation functions (denoted by AC), and the lower panels show the predicted cross- correlations (CC), between selected colour cells.

{\rr
Examining the first panel of Fig. \ref{dndnz_cls2}, we can clearly see how sensitive the modelled cross- correlation signal is to differences in redshift distributions, for fixed galaxy-dark matter bias model parameters. The differences in the lines are those predicted from different redshift distributions. Moving from the right to left of Fig. \ref{dndnz_cls2} we see the complex effect that the galaxy-dark matter bias model has on the model correlation functions, and how this changes for each of different redshift distributions. The lower panels of Fig. \ref{dndnz_cls2} show that the cross-correlations also contain valuable information which can be leveraged when estimating redshift distribution and galaxy-dark matter bias model parameters, however the amount of predictive power varies {\rr through} the redshift overlap of the sub-samples, and is often smaller than the auto correlation by many factors}.


\subsection{Sampling high dimensional parameter space}
\label{highd_space}
For each colour cell we parametrise the redshift distribution using a five Gaussian mixture model, which requires 15 parameters to be estimated\footnote{There are actually 14 free parameters in a GMM(5) because of the normalisation constraint.}. We further use a three parameter bias model. Therefore each colour cell is described by 18 parameters and if there are 12 colour cells then we have to explore a 12*18 $\approx 216$ dimensional parameter space. With the future addition of interesting cosmological parameters, this could expand by a further $\approx$10. Trying to explore such a large dimensional parameter space often leads to very low acceptance rates. 

We investigated a variety of parameter exploration methods, including a normal MCMC walker, the {\tt emcee}\footnote{\url{http://dfm.io/emcee/current/}} \citep{emcee} routine, and the Hybrid technique \citep[][]{2007A&A...464.1167E}. We find that Gibbs sampling, and Gibbs block-sampling lead to very reasonable acceptance rates.

\subsubsection{Gibbs sampling and Gibbs block-sampling}
The Gibbs Sampling algorithm is very similar to a normal MCMC walk through parameter space, in which a random step is taken in each parameter dimension, and the likelihood calculated at the new location. The step is accepted if the ratios of new to old likelihoods is greater than a randomly generated number between 0 and 1, otherwise the step is rejected and a new step from the previous position in parameter space is evaluated. The final list of accepted parameter positions approximates the likelihood surface of interest. Gibbs sampling differs from the MCMC algorithm by holding all parameters fixed expect one, or a natural grouping of parameters in the Gibbs block-sampling algorithm, during each sub-step. The next parameter in the list is then varied and all others held fixed.  This means that many one dimensional, or a few dimensional, steps are explored, which while being computationally slower, often leads to much higher acceptance rates. We direct the reader to \cite{Gibbs} for more details.

\subsubsection{Other techniques to reduce the volume of parameter space}
To further reduce the complexity of the parameter volume to be explored, we apply the following constraints: Each Gaussian in the five Gaussian mixture model, must be centred in redshift at a value larger than the previous Gaussian, and the first Gaussian must have a centre $z>-0.5$. We note this is a very conservative prior, because redshifts smaller than 0 are unphysical in a cosmological setting, but this choice enables a larger range of redshift distributions to be modelled. During the walks, we randomly decide whether to perform a block Gibbs sampling step, or the more computationally expensive full Gibbs sampling. For each colour cell we arrange all the galaxy-dark matter bias model parameters into one block, and all the redshift distribution parameters into a second block. During block Gibbs sampling, all the parameters in a block are varied simultaneously which requires two calls to the likelihood function for each colour cell. In normal Gibbs sampling, each of the 18 parameters is varied one at a time, with the others held fixed. This requires 18 calls to the likelihood function for each colour cell for each Gibbs sample.

The most important approach to reduce the computational cost of this analysis is to break the initial problem of exploring the joint parameter space for N colour cells, into one of exploring the parameter space for one colour cell at a time. We allow the chains to run until the measured $\chi^2$ stops improving. Then we randomly select three out of the N colour cells and explore the combined parameter space ten times, and repeat this procedure until the minimum $\chi^2$ value stops reducing. We note that this leads to large jumps in the total $\chi^2$ value because redshift distributions, and bias model parameters may be pulled in incorrect directions during these local searches. We repeat this entire procedure once more by randomly selecting five of the N distributions. This approach and a summary of the results are described in more detail in \S\ref{evol_chi2}. We note that this technique is itself similar to Gibbs sampling, but also helps to pull distributions out of local minimum.

In effect the above procedure moves the walkers from a random initial point in the high dimensional parameter space, towards a more reasonable starting position for the subsequent Gibbs sampling.  We reiterate, that if any information is known a priori about the redshift distribution or galaxy-dark matter bias model parameters of the galaxies in any of the colour cells, this can be used as an initial starting point in the parameter space, which will reduce the time taken to find a global minimum, and explore around it.

We then continue to perform Gibbs and Gibbs-block sampling, attempting to identify a global minimum, and to explore around the minimum to estimate parameter uncertainties and covariances.

We adopt prior ranges quoted in Table \ref{prior_ranges},  which are chosen to enable all of the redshift distributions to be well modelled for each of the different colour cells generated from the SDSS data. If one is very confident in priors on any of the known parameters, these could be included in the analaysis. Here we assume only flat priors across the range of model parameters. We choose the galaxy-dark matter bias model parameter ranges, such that they allow reasonable values, such as $b=$0.9-1.1 for e.g., main sample galaxies at low redshift $z<0.4$, $b=$1.5 for e.g., LRGs at higher redshift $z\approx 0.6$ and $b=$2.5 for e.g., Quasars at high redshifts $z>2$. We reiterate that the prior ranges, and adopted functions to model the redshift and galaxy-dark matter bias can be updated to allow any functional form, as long as they can be  incorporated into cosmological pipelines such as {\tt CLASS}, or {\tt CAMB}.

\begin{table}
\centering
\renewcommand{\footnoterule}{}
\begin{tabular}{l c c c }
\hline
parameter & min & max  & sampling width \\
 \hline  \hline 
GMM mean & -0.5 & 7 & 0.01  \\
GMM sigma & 0.005 & 2 & 0.01 \\
GMM amplitude  & 0 & 6 & 0.06 \\
galaxy-dm bias $b_0$ & 0 & 4 & 0.20 \\
galaxy-dm bias $b_1$  & -2 & 2 & 0.02  \\
galaxy-dm bias $b_k$  & -1 & 1 & 0.01 \\
 \hline 
\end{tabular}
\caption{The prior parameter ranges for each colour cell, and the value of the sampling width, describing the width of a Gaussian that a random step is drawn from during the parameter space exploration.  The Gaussian parameters describe the five Gaussians in the Gaussian Mixture Model (GMM) which models the redshift distributions, and the bias parameters correspond to the galaxy-dark matter bias model. These ranges can be easily extended given sufficient motivation.}
\label{prior_ranges}
\end{table}

\section{Analysis and Results}
\label{results}
We present an overview of how to interpret the main result figures in \S\ref{overview} followed by {\bhN the details of each analysis performed using either different combinations of probes, or differences in true redshift distributions.} We conclude with an overview of the results in \S\ref{summ_res}, and with a description of the evolution of the $\chi^2$ function during one of the correlation function analyses in \S\ref{evol_chi2}.

\subsection{Overview of the results panels}
\label{overview}
Here we present an overview of the results panels shown in Figs. \ref{dndnz_ncl}, \ref{dndnz_sdss_ncl}, \ref{dndnz_ncllclscl}, and \ref{dndnz_sim_ncl_noise} which have many features in common. We discuss each figure in turn in a subsequent sub-section.

Each figure shows twelve different panels, which correspond to each of the twelve colour cells in this analysis. Each panel has a number label in the bottom left hand corner. Each panel shows the completely random initial guess for the redshift distribution using the grey starred line. This redshift distribution evolves as we explore the high dimensional parameter space, as shown by the thin grey lines. A final draw from the posterior is shown by the  thick red continuous line. In each panel the thick grey dashed line shows the true redshift distribution, which we are trying to estimate using the data driven correlation method presented here. In the legend of each panel we {\bhP quote} the mean of the true redshift distribution and the redshift range of the true redshift distribution spanning the 68\%, and 95\%, and 99.5\% percentiles around the median, denoted here as $\sigma_{68, 95, 99.5}$. The legend of each figure also shows how wrong the initial value of $\Delta_z=<z_{true}> - <z_{est}>$ is, in each colour cell. The legend additionally shows the final value of $\Delta_z$ between a final draw from the posterior for the redshift distributions and the truth. We note that many works refer to $\Delta_z$  as `bias' but we do not use this term here for clarity, especially when discussion the unrelated galaxy-dark matter bias. The legend also shows the percentage error, defined as $|\sigma_{x}^{true}-\sigma_x^{predict}|/\sigma_x^{true}*100\%$, for the values of $\sigma_{68,95,99.5}$ as measured on a final draw from the posterior of the redshift distribution. 

Finally each colour-cell panel has a sub panel showing either the posterior for each of the three galaxy-dark matter bias model parameters $b_0, b_1, b_k$, or the scale dependent parameter $b_k$ and the composite parameter {\rr $b^*$ (see Equ. \ref{eq:bias1})}.  We reiterate that the initial guesses for these parameter values are drawn randomly from within the prior ranges. For clarity we label {\bhP the galaxy-dark matter bias model parameters in only the first sub-panel of each figure.} The light grey vertical lines correspond to the true parameter values, and the coloured lines correspond to the width of a Gaussian fit to the posterior distribution.

The density of the thin grey lines in the redshift distribution highlight local minima corresponding to where the redshift distribution remained stationary for some time in the walk through the redshift distribution parameter space. The final sets of thin grey lines around the thick red continuous line also highlight allowed ranges of draws from the posterior, which are consistent with the measured correlation functions.

\subsection{Galaxy-Galaxy correlation functions}
\label{gg-corr}
In this sub-section we present the constraints on the redshift distributions and galaxy-dark matter bias parameter values obtained using only the cross- and auto- correlation {\rrr simulated data vectors, that would correspond to} the angular positions of data residing in each colour cell. In this sub-section we do not use the correlation functions available from galaxy-galaxy lensing, or CMB lensing around the galaxy positions. We present the results of these idealised {\rrr data vectors which, by construction, } have known redshift distributions, and with known galaxy-dark matter bias parameter values.

\begin{figure*}
\includegraphics[scale=0.3125,clip=true,trim=15 85 15 25]{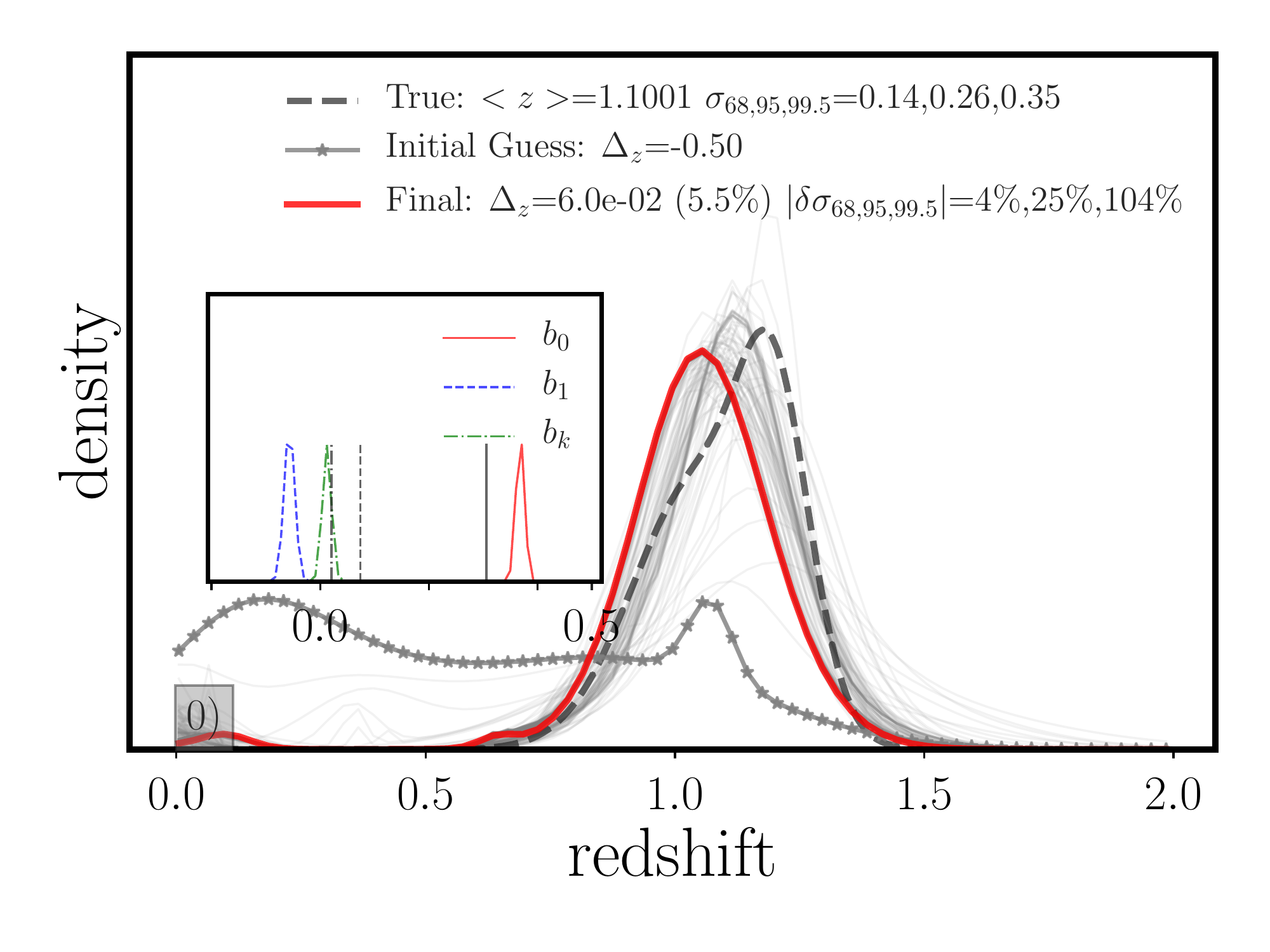}
\includegraphics[scale=0.3125,clip=true,trim=55 85 15 25]{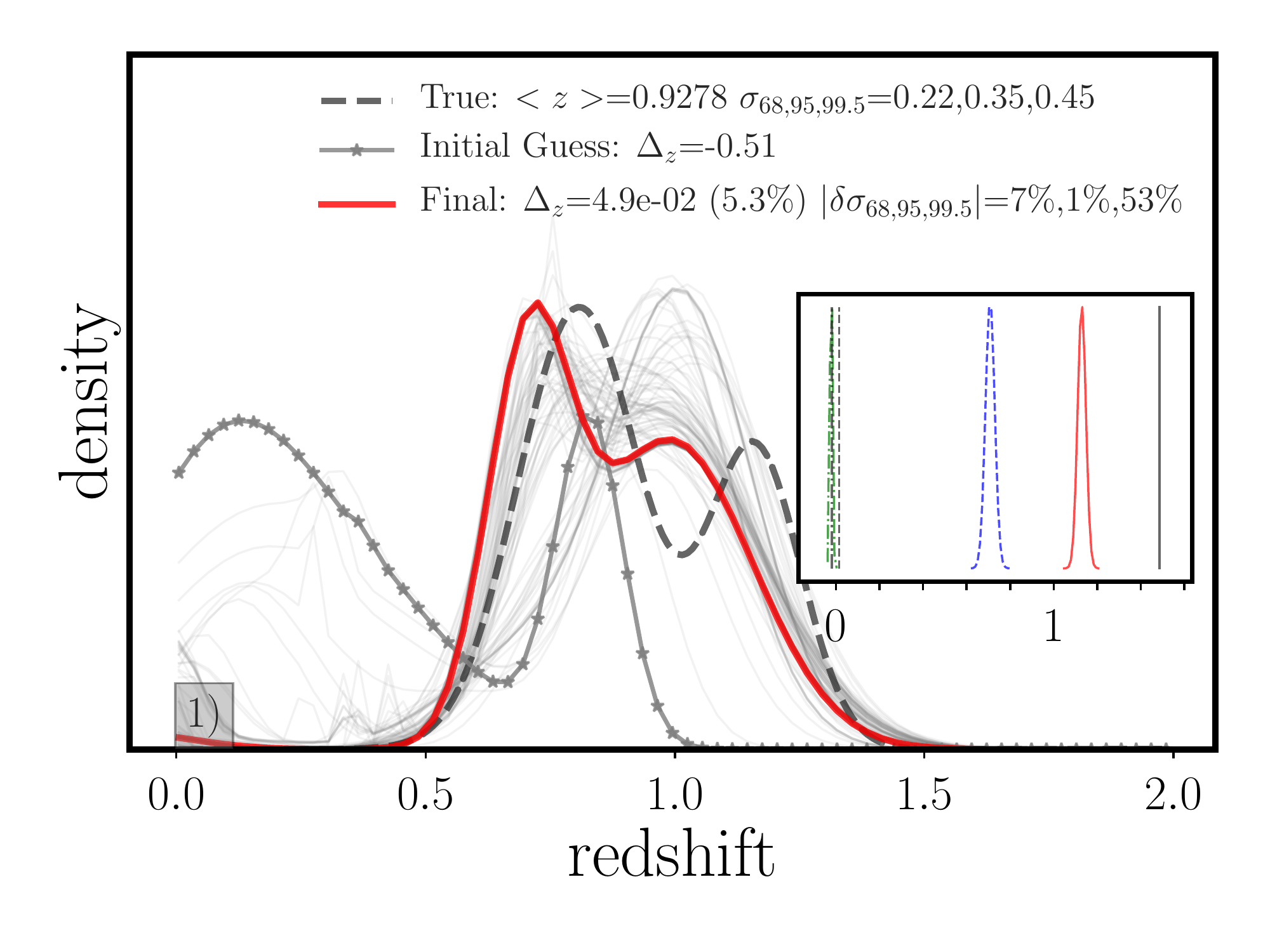}
\includegraphics[scale=0.3125,clip=true,trim=55 85 15 25]{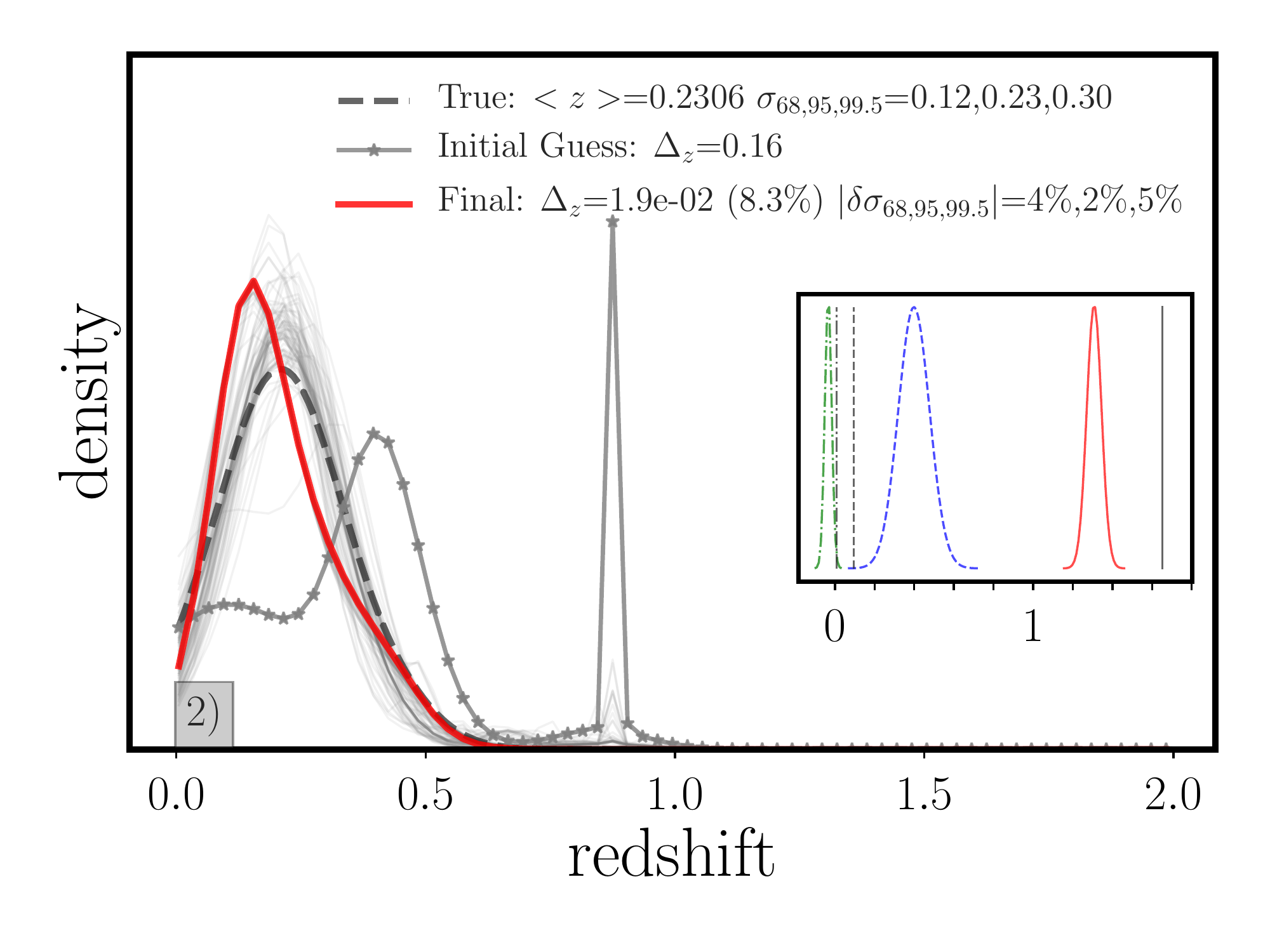}\\
\includegraphics[scale=0.3125,clip=true,trim=15 85 15 25]{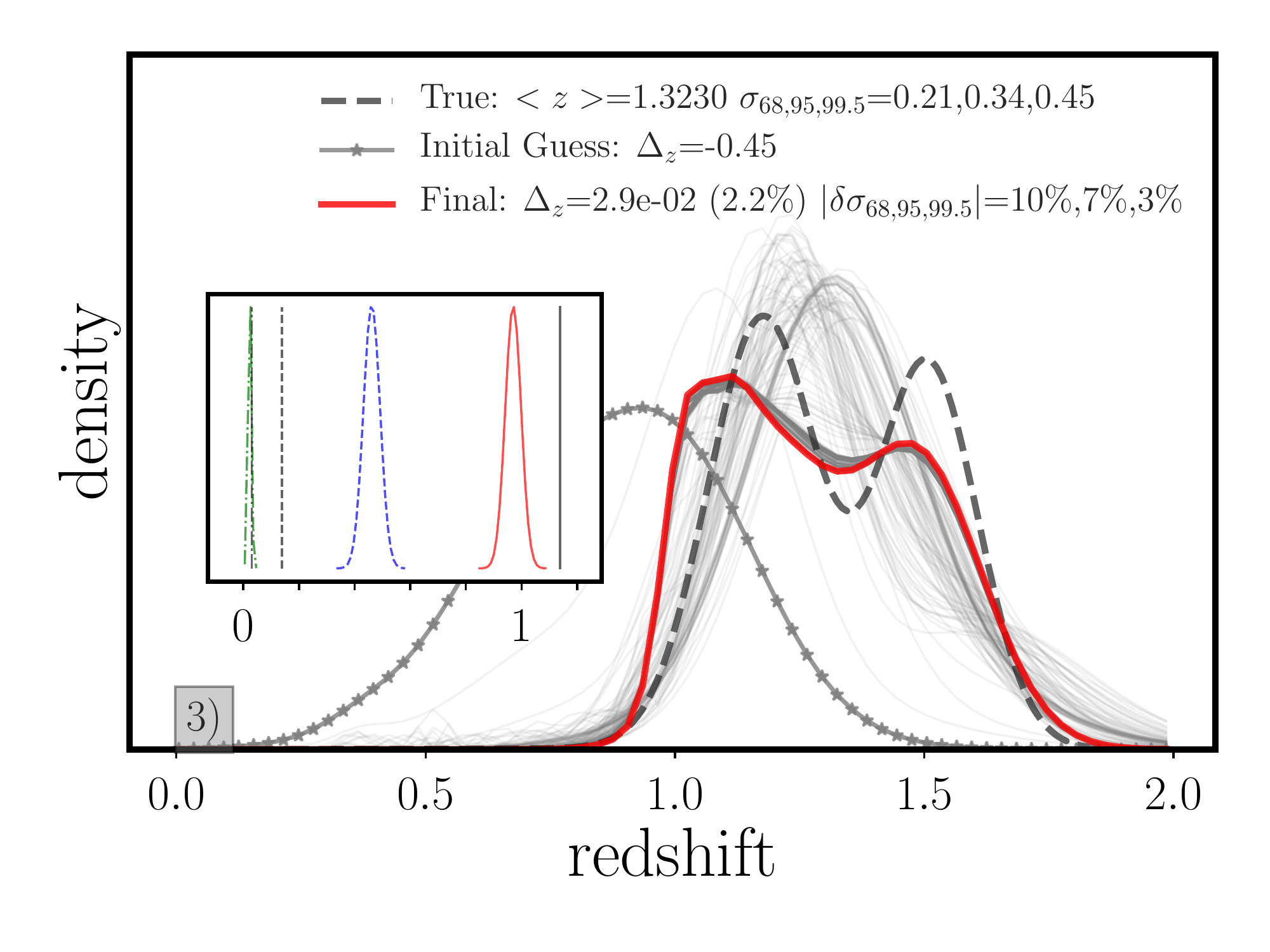}
\includegraphics[scale=0.3125,clip=true,trim=55 85 15 25]{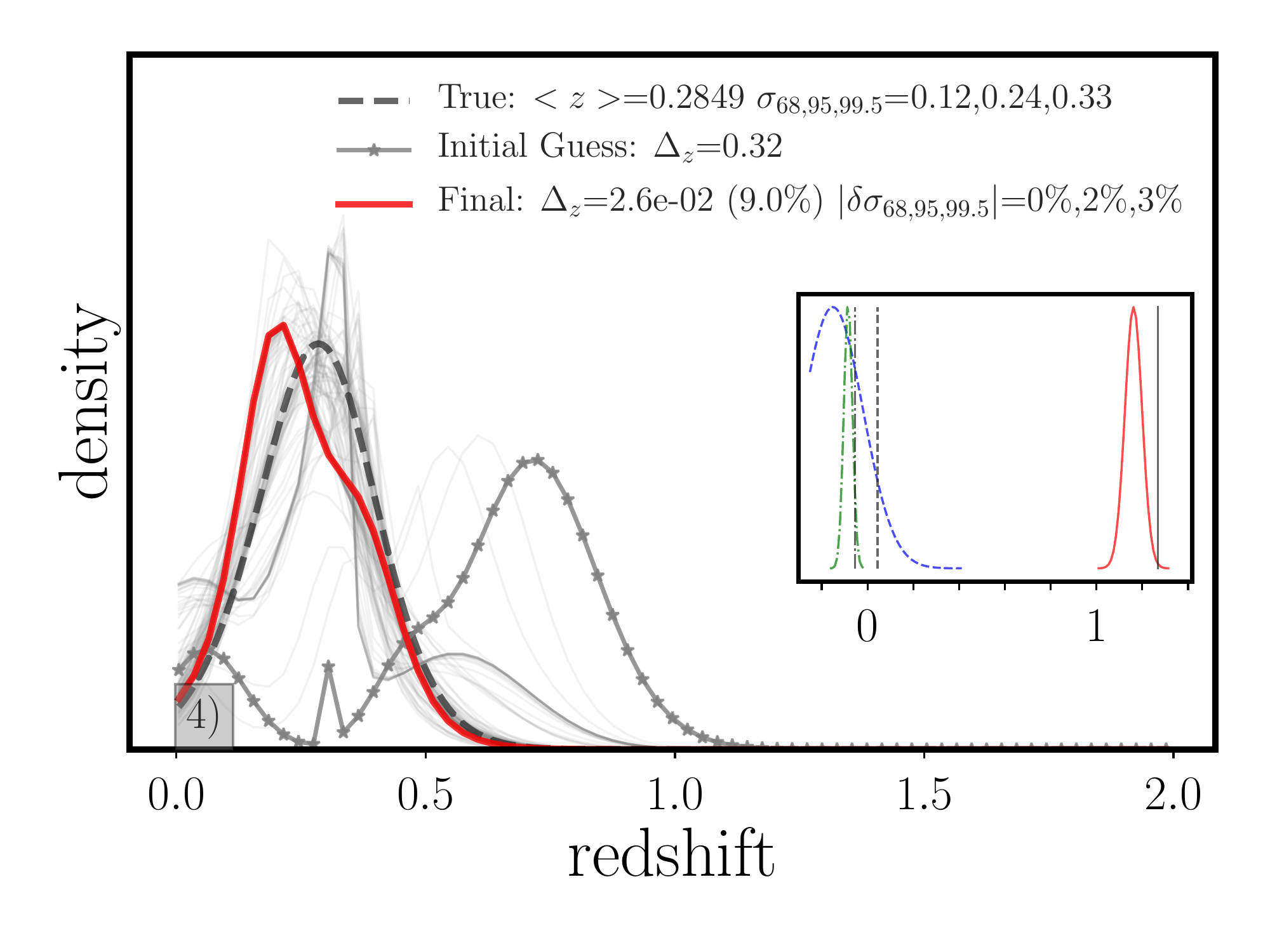}
\includegraphics[scale=0.3125,clip=true,trim=55 85 15 25]{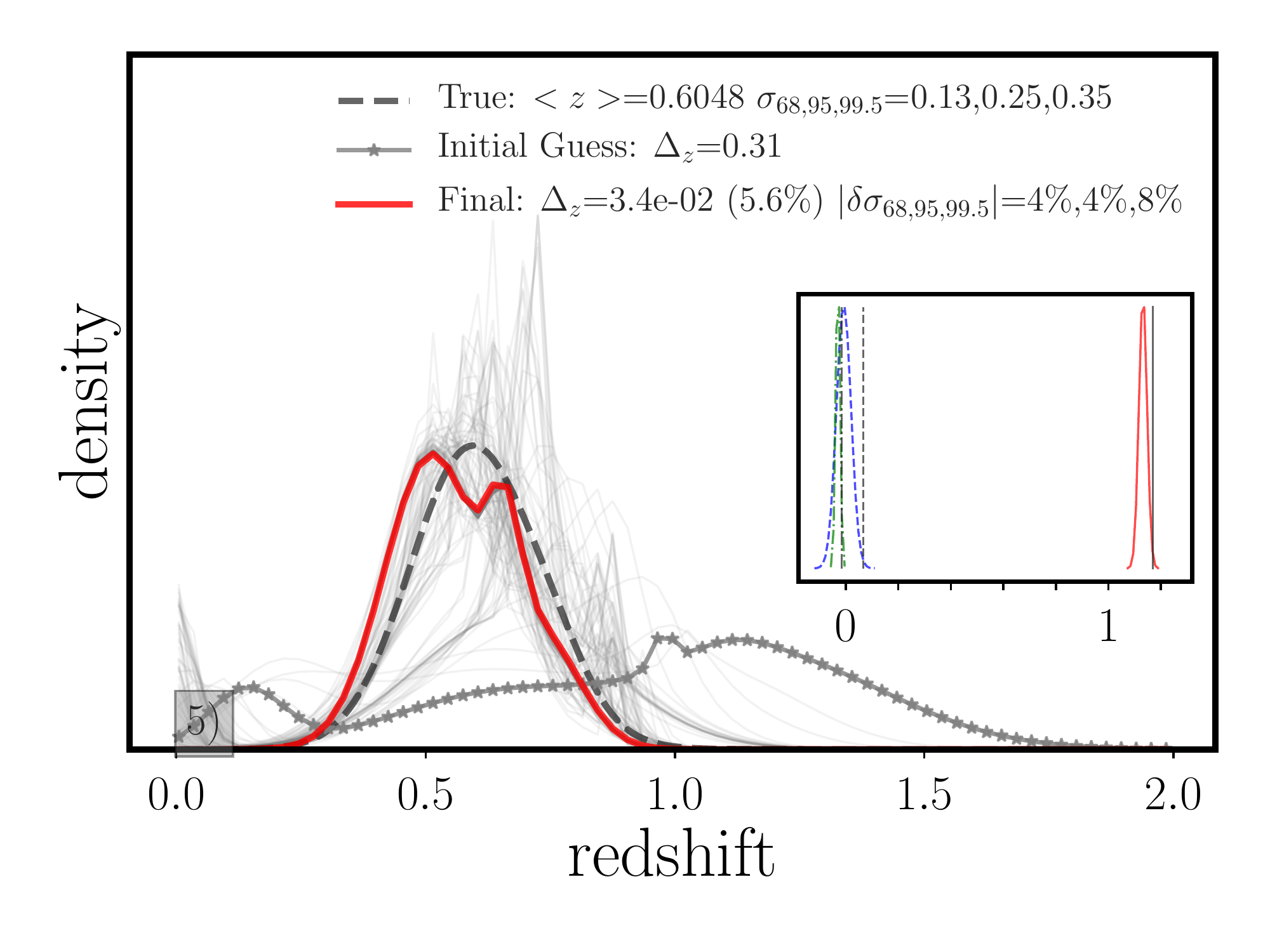}\\
\includegraphics[scale=0.3125,clip=true,trim=15 85 15 25]{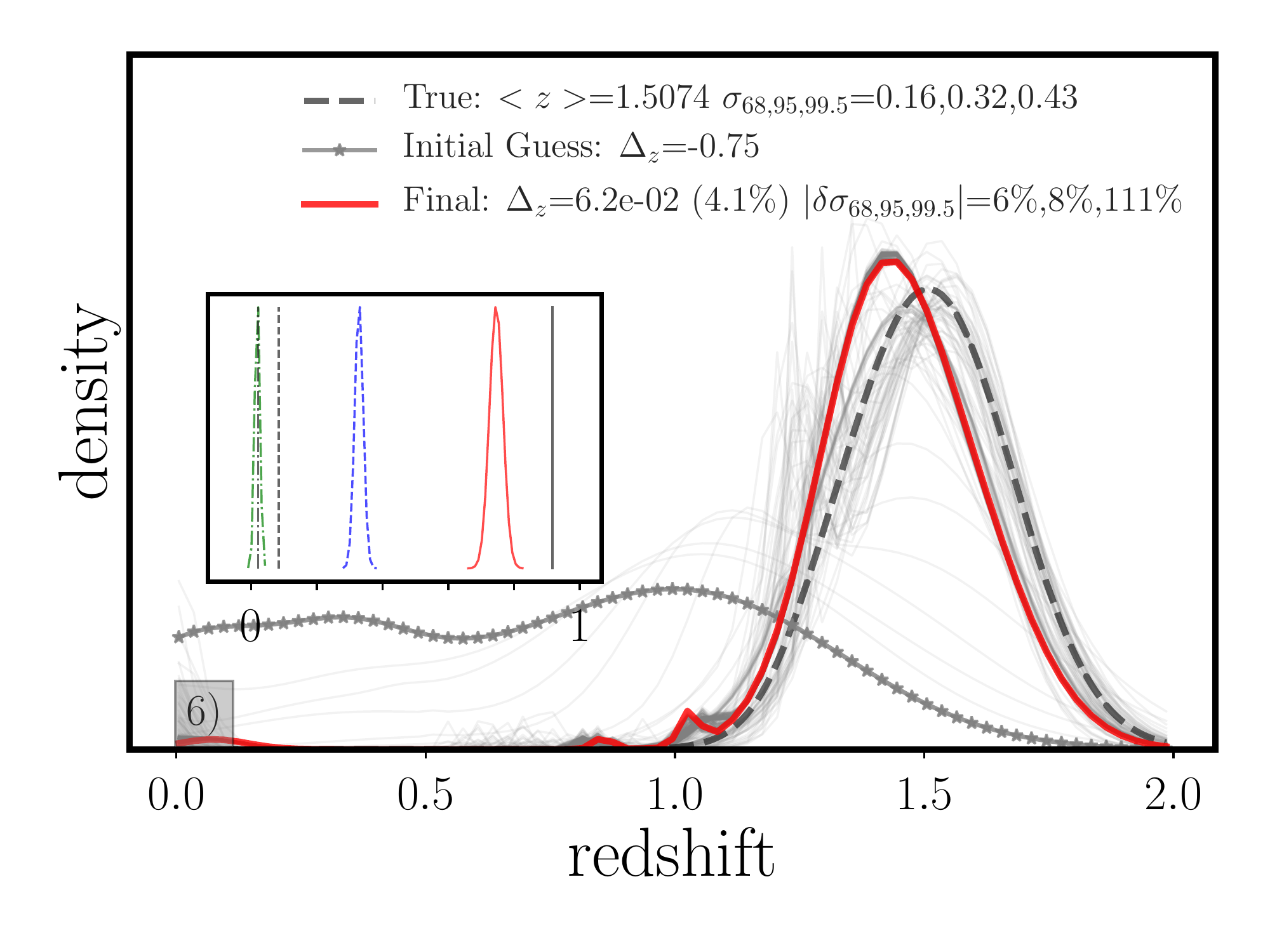}
\includegraphics[scale=0.3125,clip=true,trim=55 85 15 25]{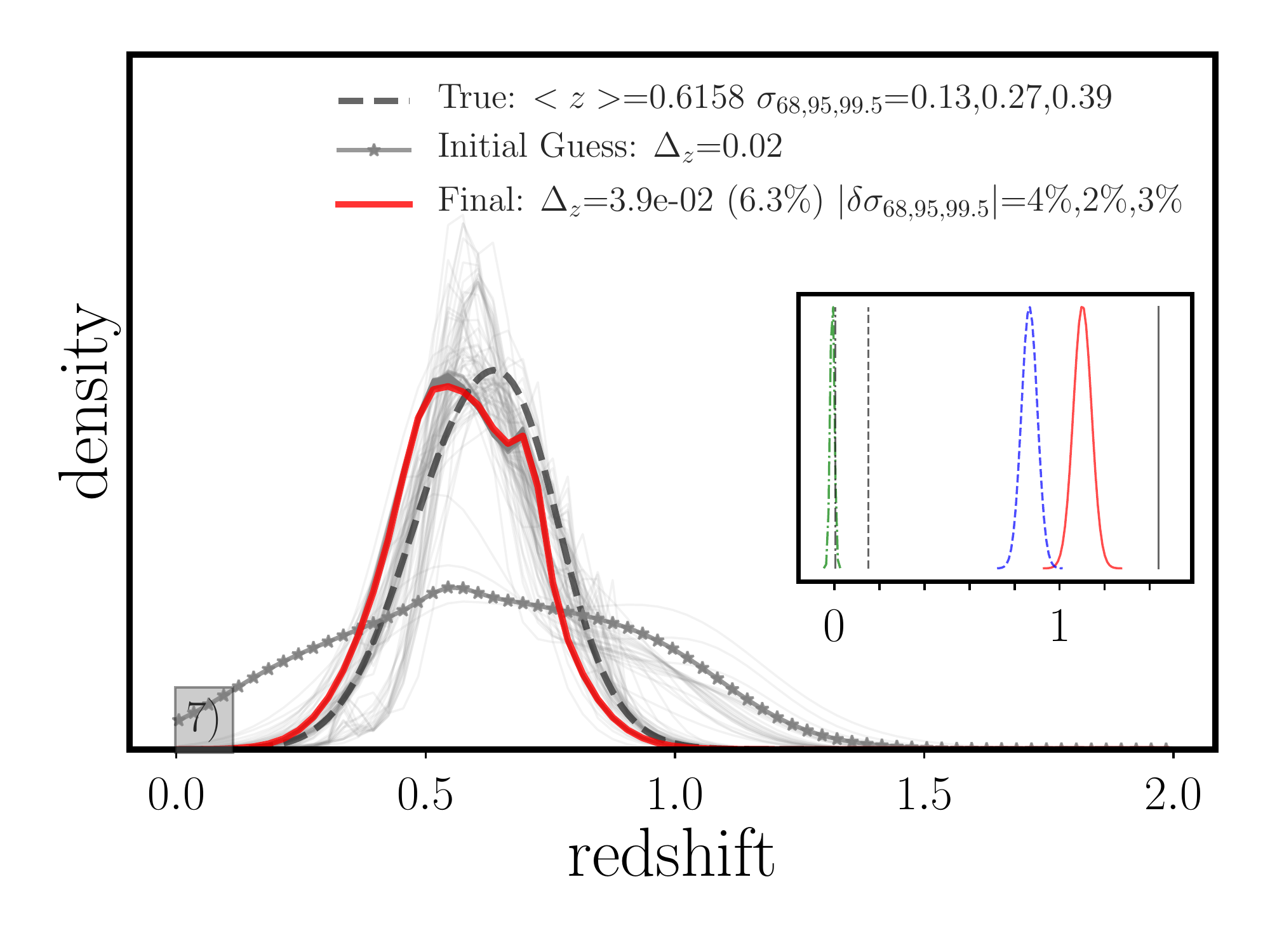}
\includegraphics[scale=0.3125,clip=true,trim=55 85 15 25]{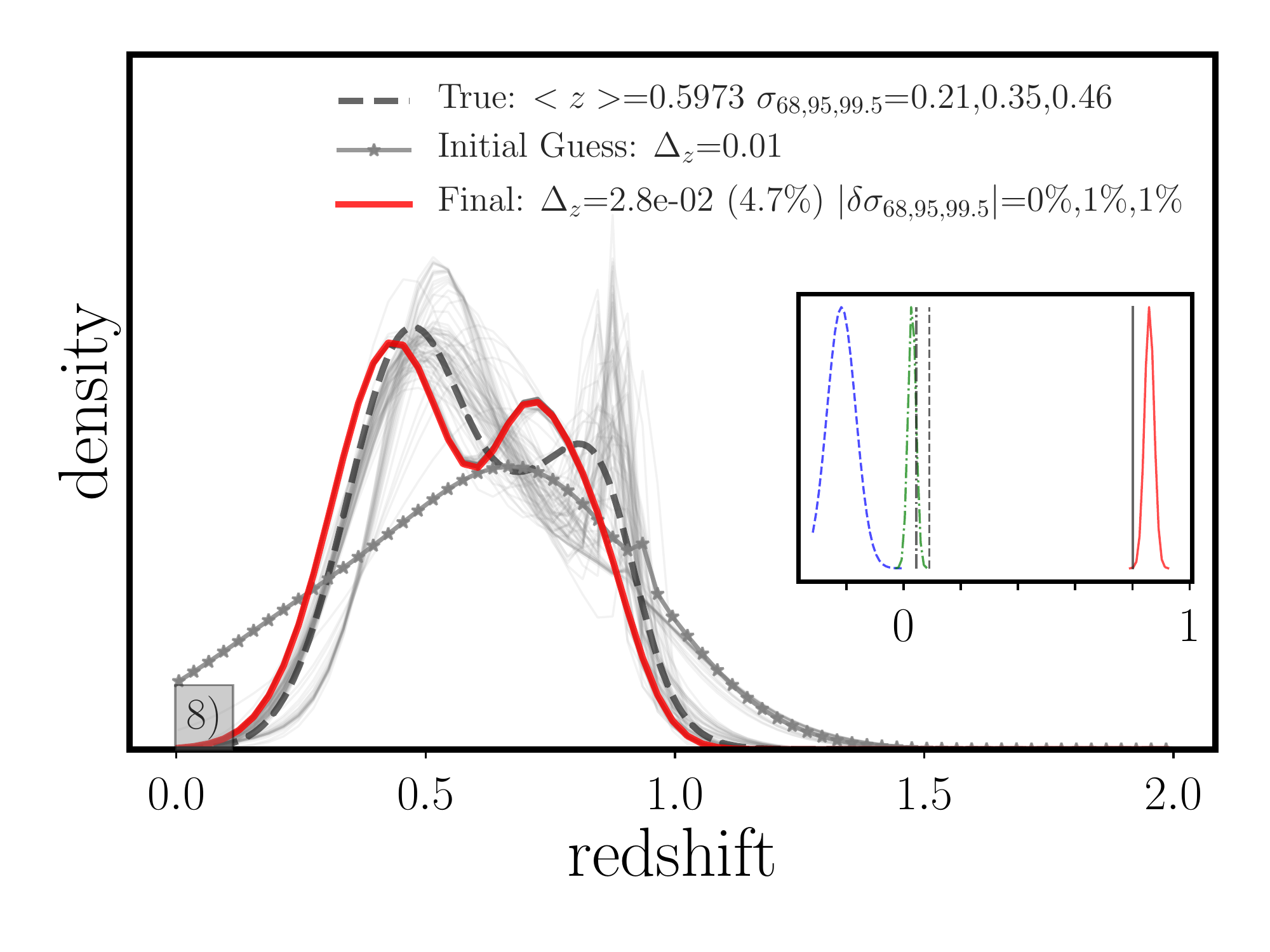}\\
\includegraphics[scale=0.3125,clip=true,trim=15 0 15 25]{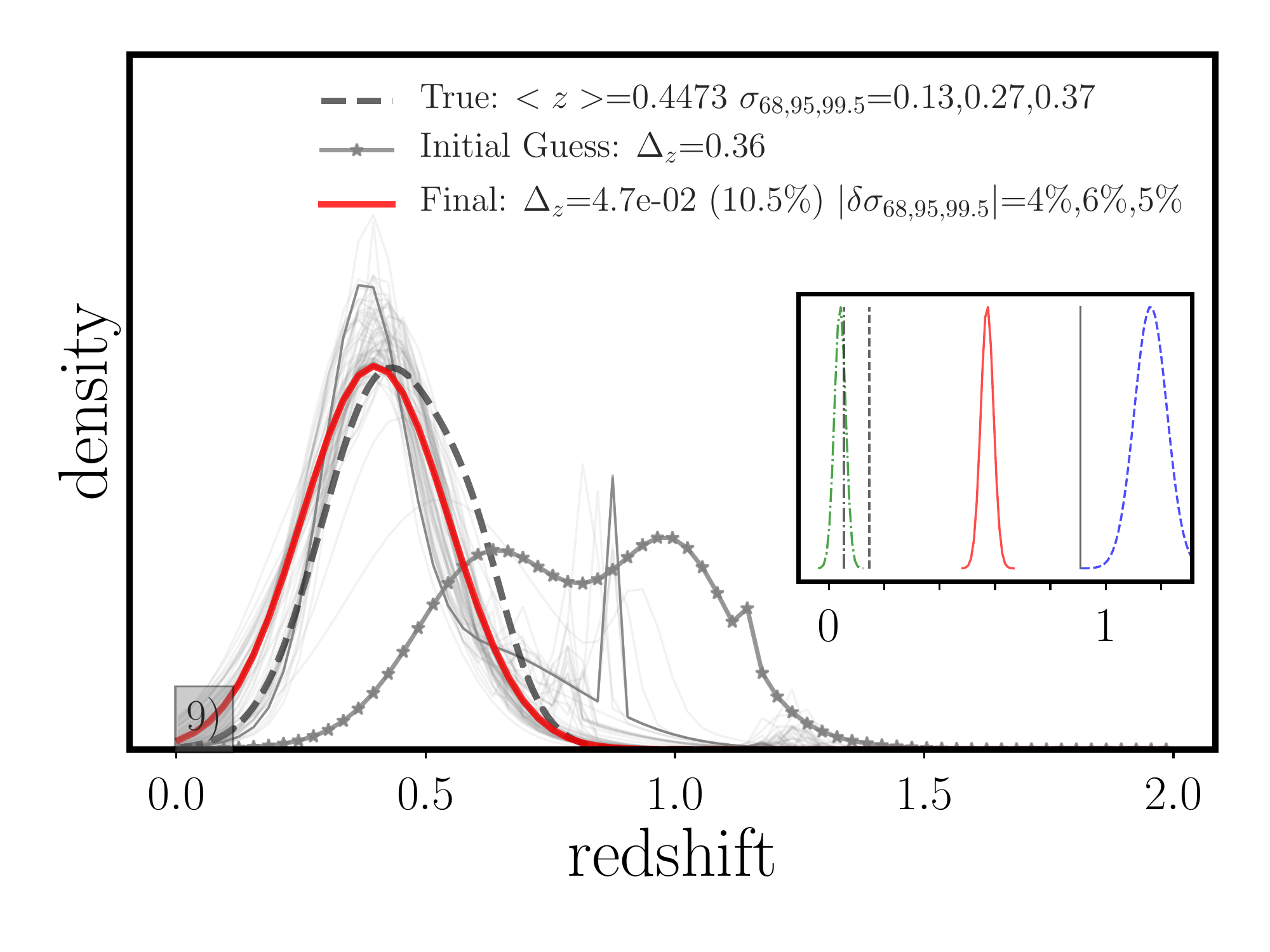}
\includegraphics[scale=0.3125,clip=true,trim=55 0 15 25]{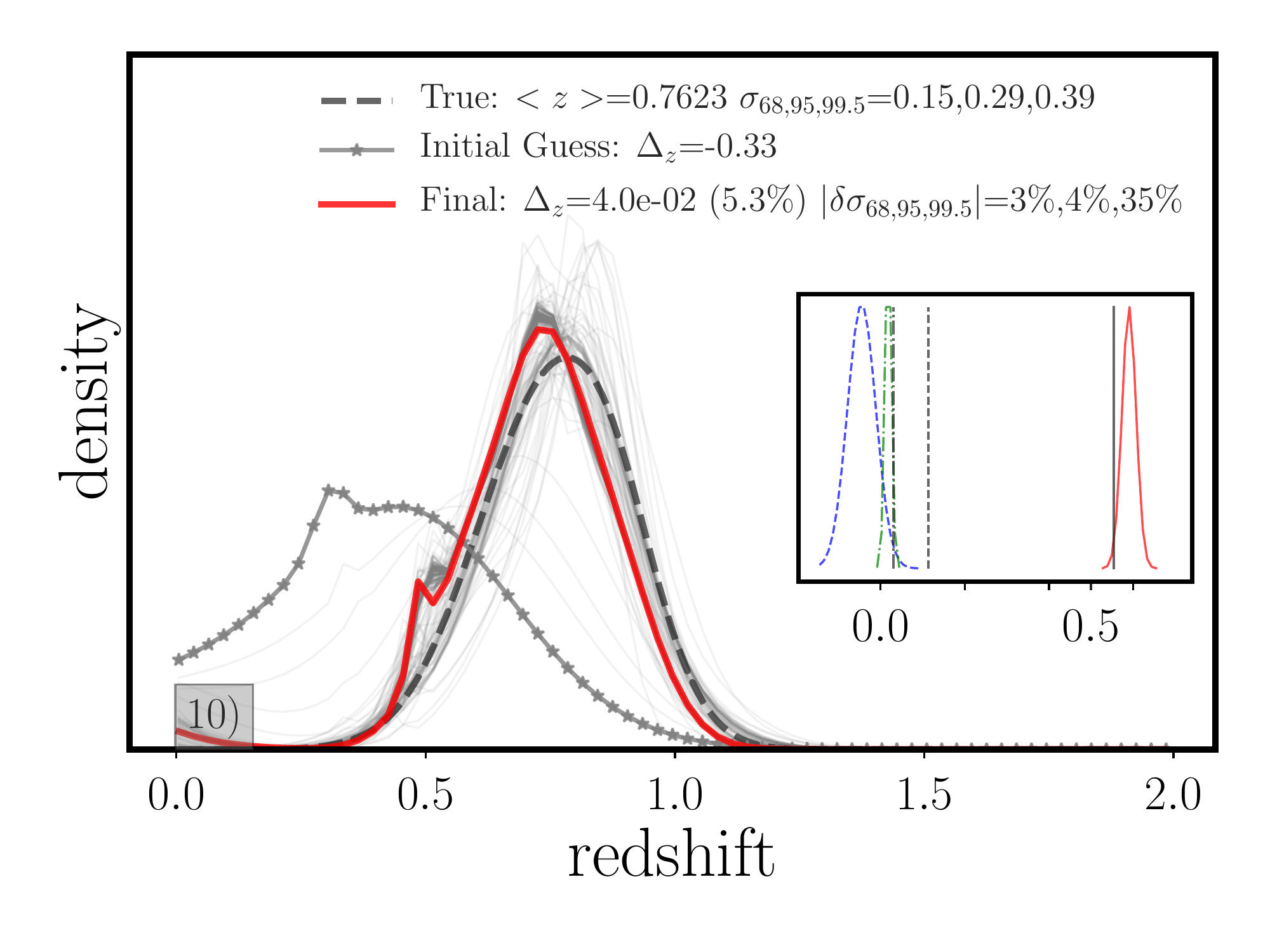}
\includegraphics[scale=0.3125,clip=true,trim=55 0 15 25]{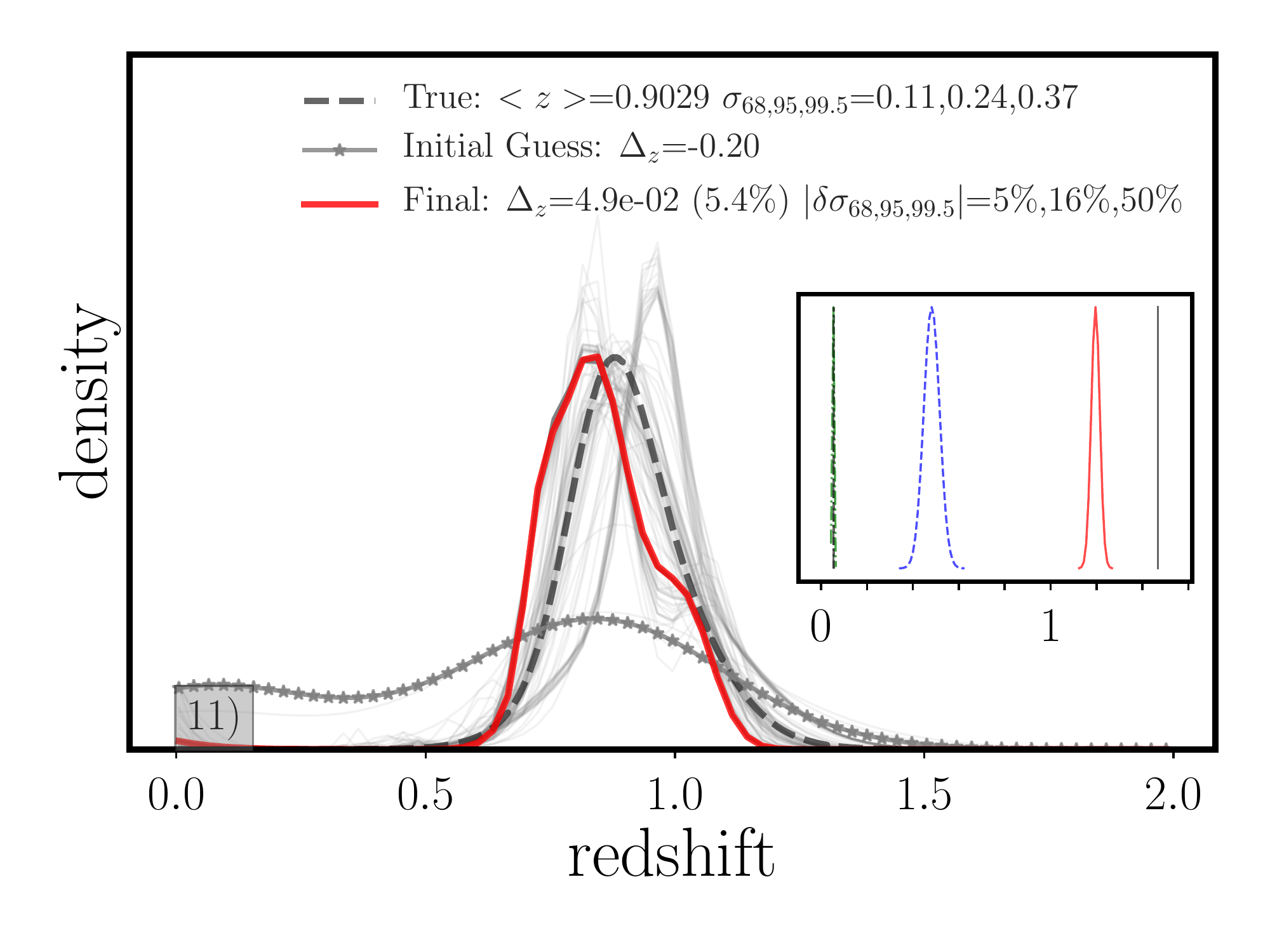}
\caption{The redshift distributions, and galaxy-dark matter bias parameter values as estimated through the use of the data driven correlation function method, using only the {\rrr simulated data vectors that would correspond to the} angular positions of galaxies in colour cells. Each of the twelve different panels correspond to one of the twelve colour cells in this analysis. Each panel shows the completely random initial guess for the redshift distribution using the grey starred line.  This redshift distribution evolves as we explore the high dimensional parameter space, as shown by the thin grey lines. A final draw from the posterior is shown by the red thick continuous line. In each panel the thick grey dashed line shows the true redshift distribution which we are trying to estimate using the correlation method presented here. In the legend of each panel we show the mean of the true redshift distribution and the redshift range of the true redshift distribution spanning the 68\%, and 95\%, and 99.5\% percentiles around the median, denoted here as $\sigma_{68, 95, 99.5}$. The legend of each figure also show how wrong the initial value of $\Delta_z=<z_{true}> - <z_{est}>$ is in each colour cell. The legend additionally shows the final value of $\Delta_z$ between a final draw from the redshift posterior distributions and the truth.  The legend also shows the percentage error, defined as $(\sigma_{x}^{true}-\sigma_x^{predict})/\sigma_x^{true}*100\%$, for the values of $\sigma_{68,95,99.5}$ as measure on a final draw from the posterior distribution. Finally each colour-cell panel has a sub panel showing a Gaussian fit to the posterior estimates for each of the three galaxy-dark matter bias model parameters $b_0, b_1, b_k$. For clarity we only label $b_0, b_1, b_k$ in panel 0), but in all panels they are highlighted using a red solid, blue dashed and green dash-dotted lines respectively. The light grey vertical lines in each sub-panel correspond to the true parameter values.  The analysis in these panels corresponds to that described in \S\ref{gg-corr}.
}
\label{dndnz_ncl}
\end{figure*}

Fig. \ref{dndnz_ncl} shows how unimportant the initial ``random-guess'' redshift distribution, as shown by the starred grey line, is to the ability of the procedure to estimate the general shape of the true redshift distribution, shown by the thick grey dashed line. A final draw from the posterior distribution, shown by the thick red solid line, {\bhN is} in good qualitative agreement with the true redshift distribution. We find that those distributions which are smooth and have one well defined redshift peak, are well fit by this analysis, for example these are the panels labelled $2,4,9,6,7,10,11$. For those redshift distributions which show more than one well defined peak, the method is still able to estimate the {\rr summary} metrics of the distribution, such as the mean of the distribution and the quantiles to a similar accuracy as the colour cells with a single peak, however the individual draw from the posterior appears to be a lot more discrepant. If we concentrate on the light grey thin lines, which correspond to the algorithm searching through the redshift distribution parameter space, we find that range of explored solutions and draws from the posterior often fully encompasses the truth redshift distribution. {\rr We note that the chains exploring the high dimensional parameter space are still converging to the best fit values. Given more calls to the likelihood function we would expect this envelope to fully cover the truth redshift distributions, as motivated in \S\ref{space_conv}}

The smaller sub-panels of each panel in Fig. \ref{dndnz_ncl}, show posterior distributions of the galaxy-dark matter bias model parameters. We find that we are very sensitive to the scale dependent bias parameter value, shown by the dash-dotted line, recovering the correct value at below $\sim$1$\sigma$, but much less sensitive to the redshift dependent and scale independent galaxy-dark matter bias parameters $b_0$ and $b_1$. 

Indeed we find that the choice of truth values for $b_1$, {\rrr which are used to generate the idealised data vectors}, and are between the values 0.01 and 0.15, mean that the full galaxy-dark matter bias model $b(k,z)$ is very insensitive to this range of parameter values.  {\rr On average, and for the current state of this analysis, the parameters $b_0$ and $b_1$ are offset from the truth values by many (e.g. 10) $\sigma$, and also have a very large scatter (again, of order 5) in units of $\sigma$. By examining the full Gibbs chains we find that these parameters are highly correlated and have a correlation coefficient value of -0.8.  If we concentrate further on the $b_1$ parameter in relation to Equ. \ref{eq:bias} we find that for a redshift distribution which spans a range in redshift of 0.5, which are typical of those in this analyses, the contribution from this parameter will only change $b(k,z)$ by 16\% of $b_1$ between the minimum and maximum redshift. Given that we chose relatively small values of $b_1$, bounded by $0.01<b_1<0.15$, we see that this addition to $b(k,z)$ will change by approximately  $0.002-0.03$.  This indicates that the adopted galaxy-dark matter bias model is a relatively poor decision, and quite unconstraining for our choices of $b_1$ parameter values. We note that any other reasonable galaxy-dark matter bias model could be adopted, and we would suggest one that is expected to be well constrained by the data. However this {\rrr discrepancy is also} due to the continuing convergence of the chains exploring the parameter space. We note that this parameters can be constrained by this analysis, {\rrr once we find the true minimum of the likelihood surface}, which we show in \S\ref{global_min}.

In light of this, we constructed a new galaxy-dark matter bias parameter which is a combination of $b_0$ and $b_1$ scaled by the average value of the redshift distribution $\bar{z}$, and given by $b^*$, see Equ. \ref{eq:bias1}. This new parameter is also calculated for each colour cell. This parameter is shown in the final column of Table \ref{final_results}, and will be adopted in the following analysis. We note that the composite parameter shows much more consistency with the truth values. }

\subsection{Gal-gal \& lensing correlation functions I}
\label{gg-lcorr}
\begin{figure*}
\includegraphics[scale=0.3125,clip=true,trim=15 85 15 25]{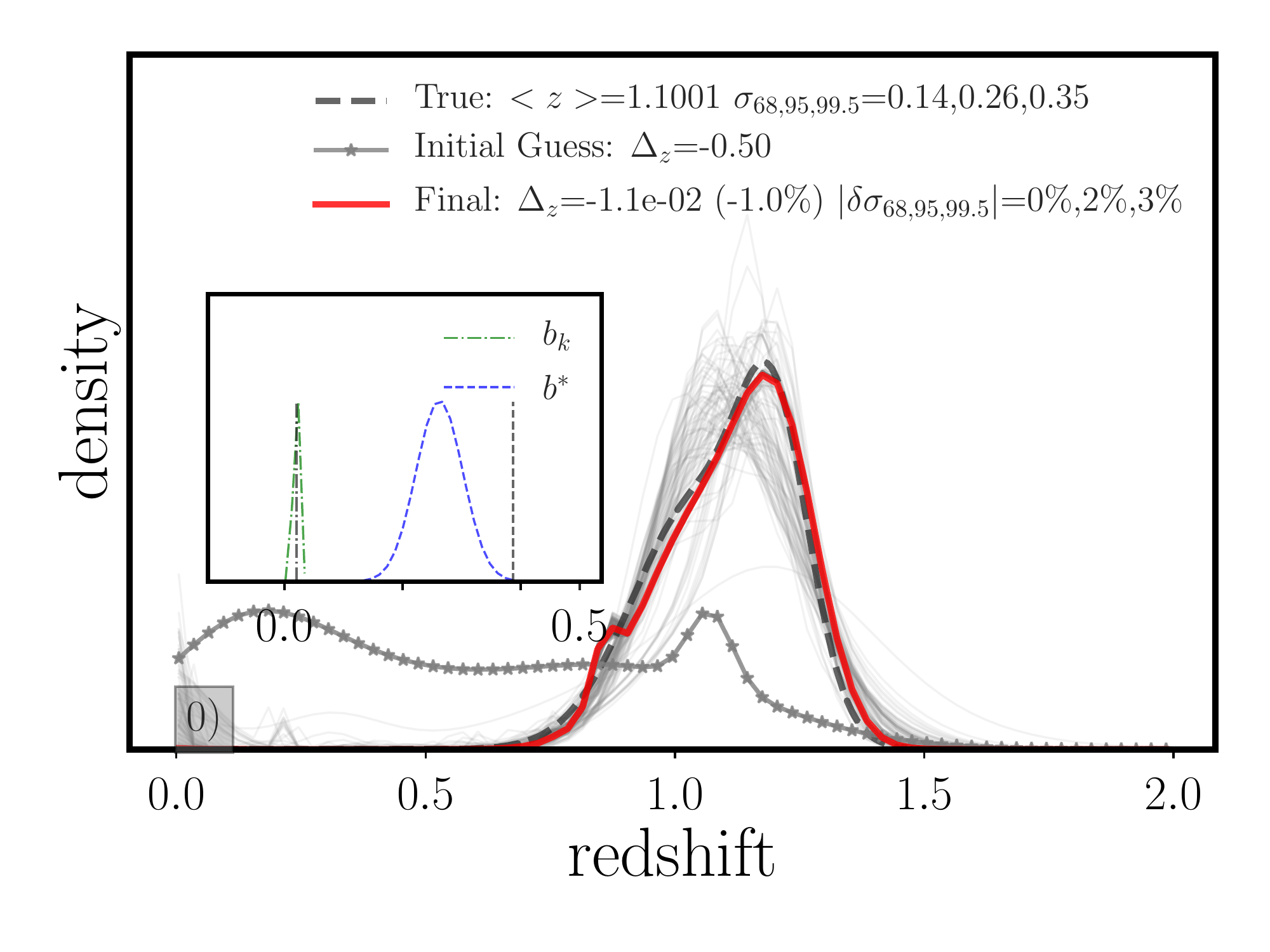}
\includegraphics[scale=0.3125,clip=true,trim=55 85 15 25]{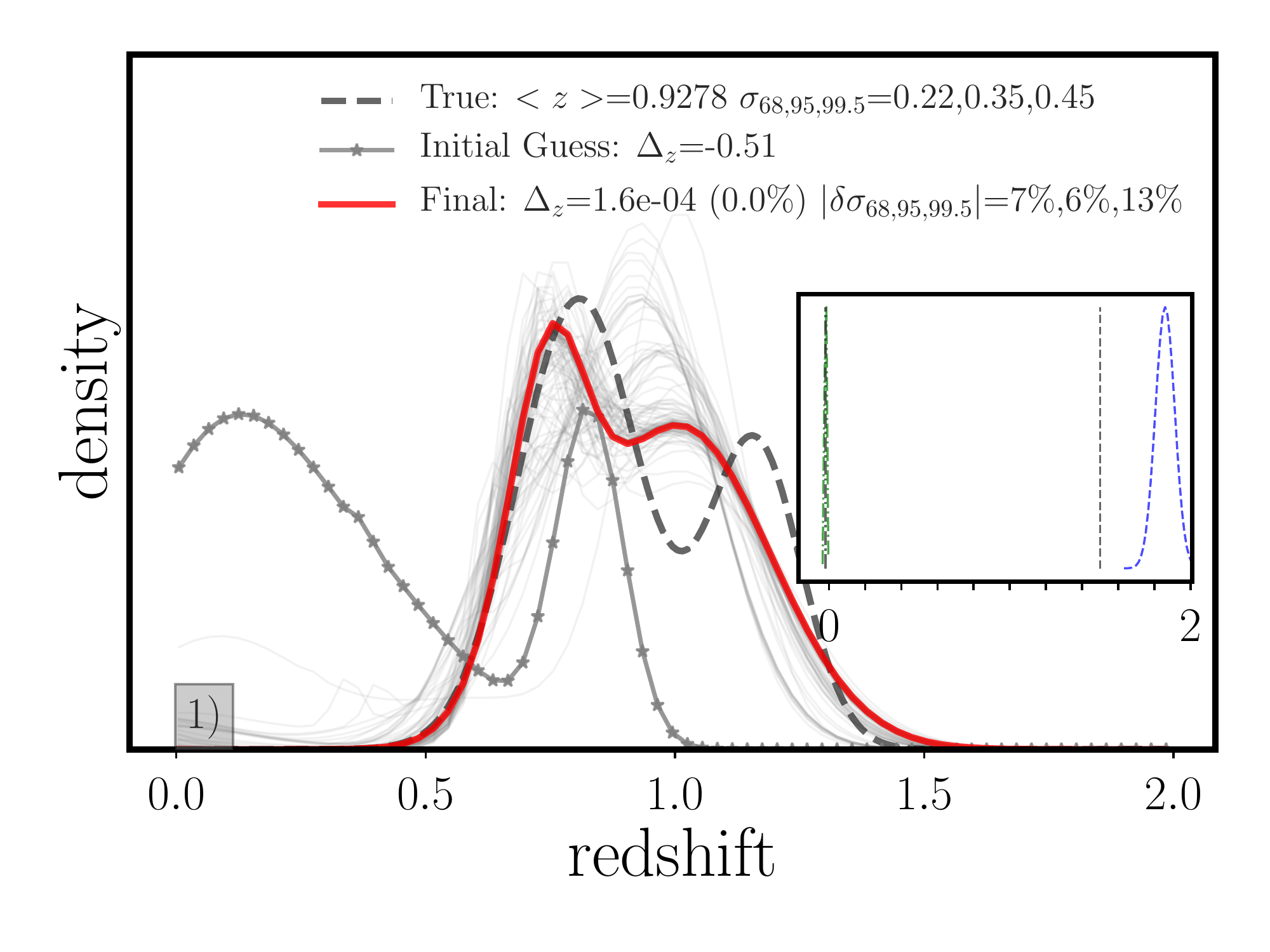}
\includegraphics[scale=0.3125,clip=true,trim=55 85 15 25]{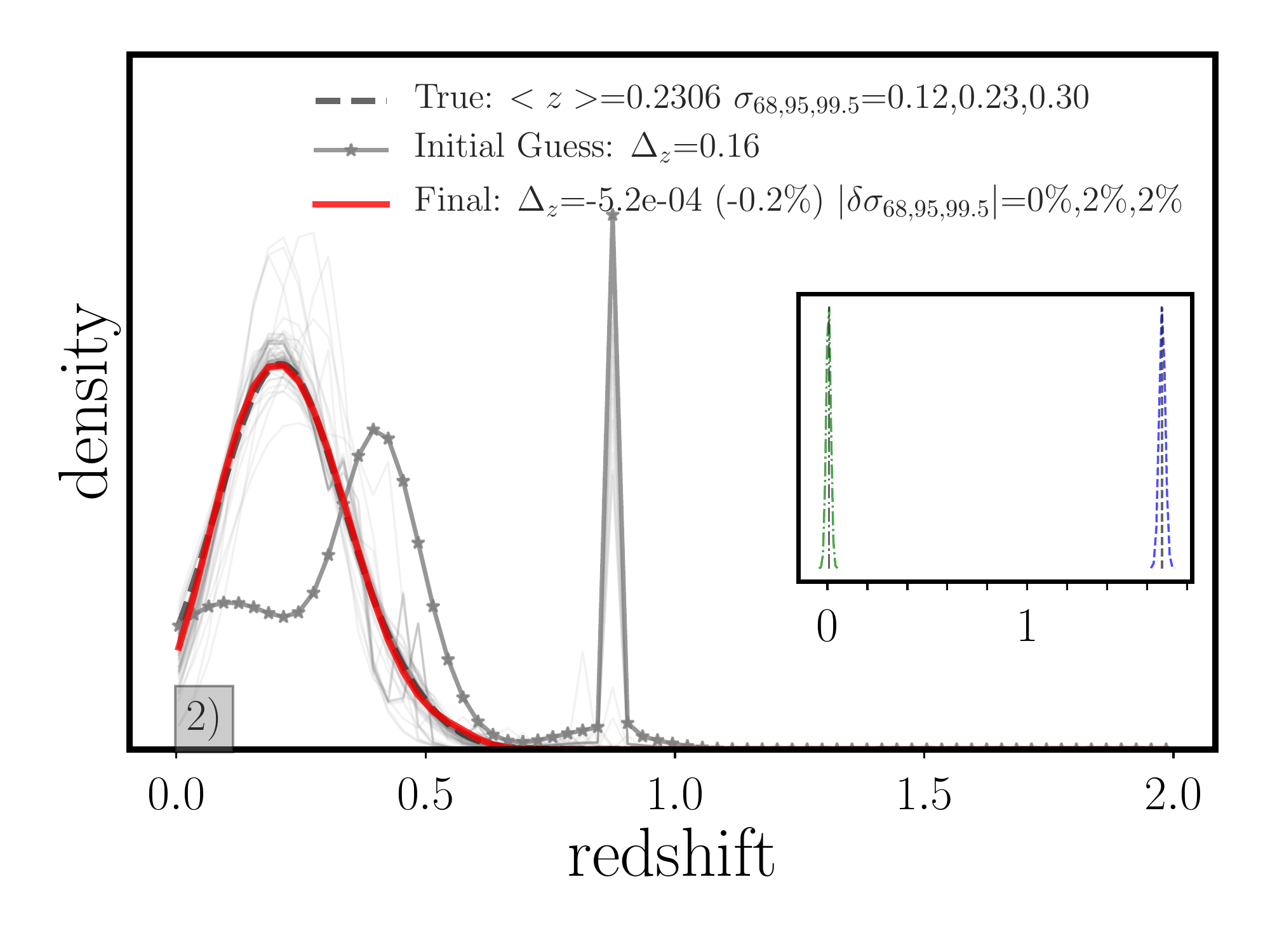}\\
\includegraphics[scale=0.3125,clip=true,trim=15 85 15 25]{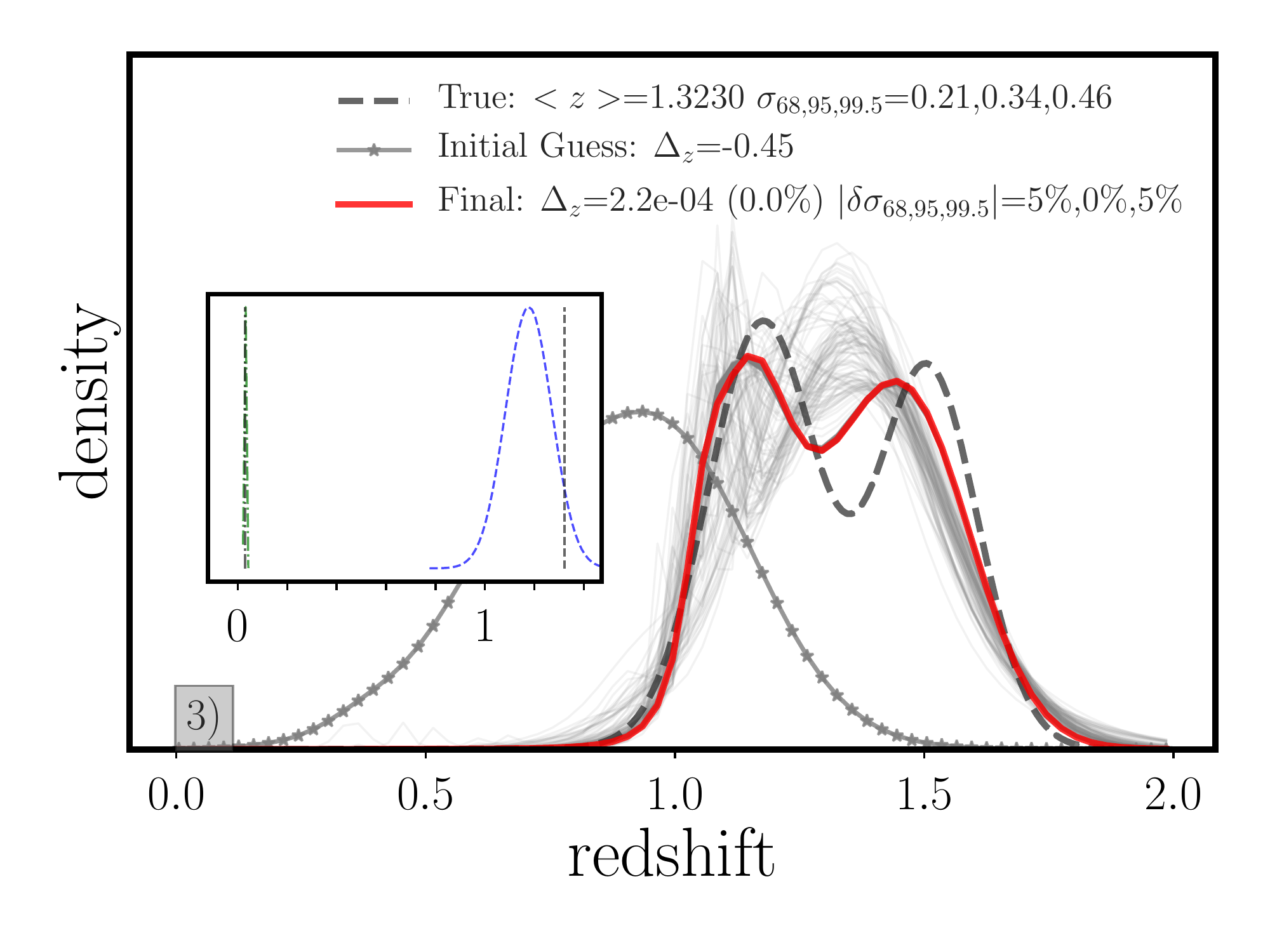}
\includegraphics[scale=0.3125,clip=true,trim=55 85 15 25]{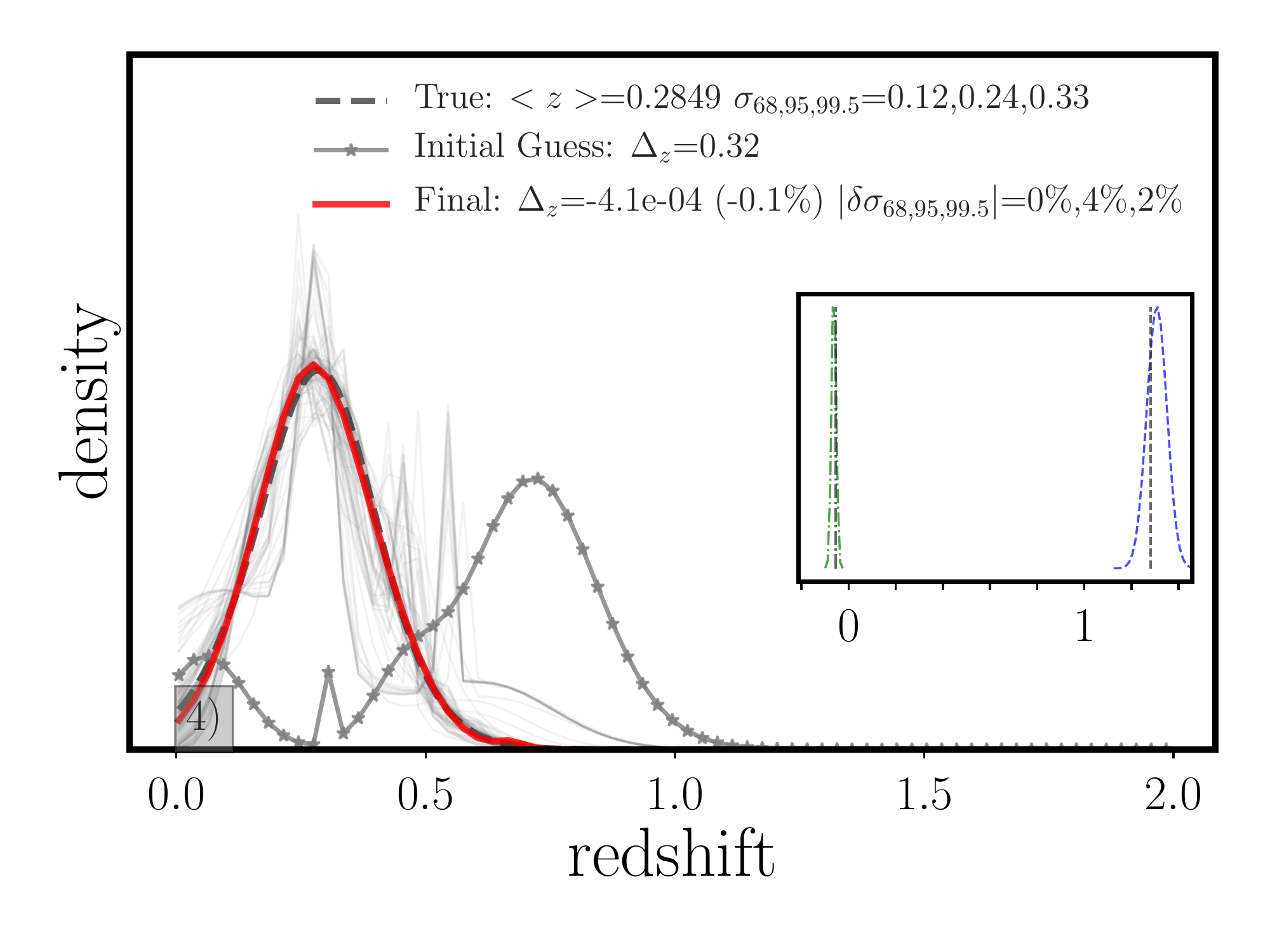}
\includegraphics[scale=0.3125,clip=true,trim=55 85 15 25]{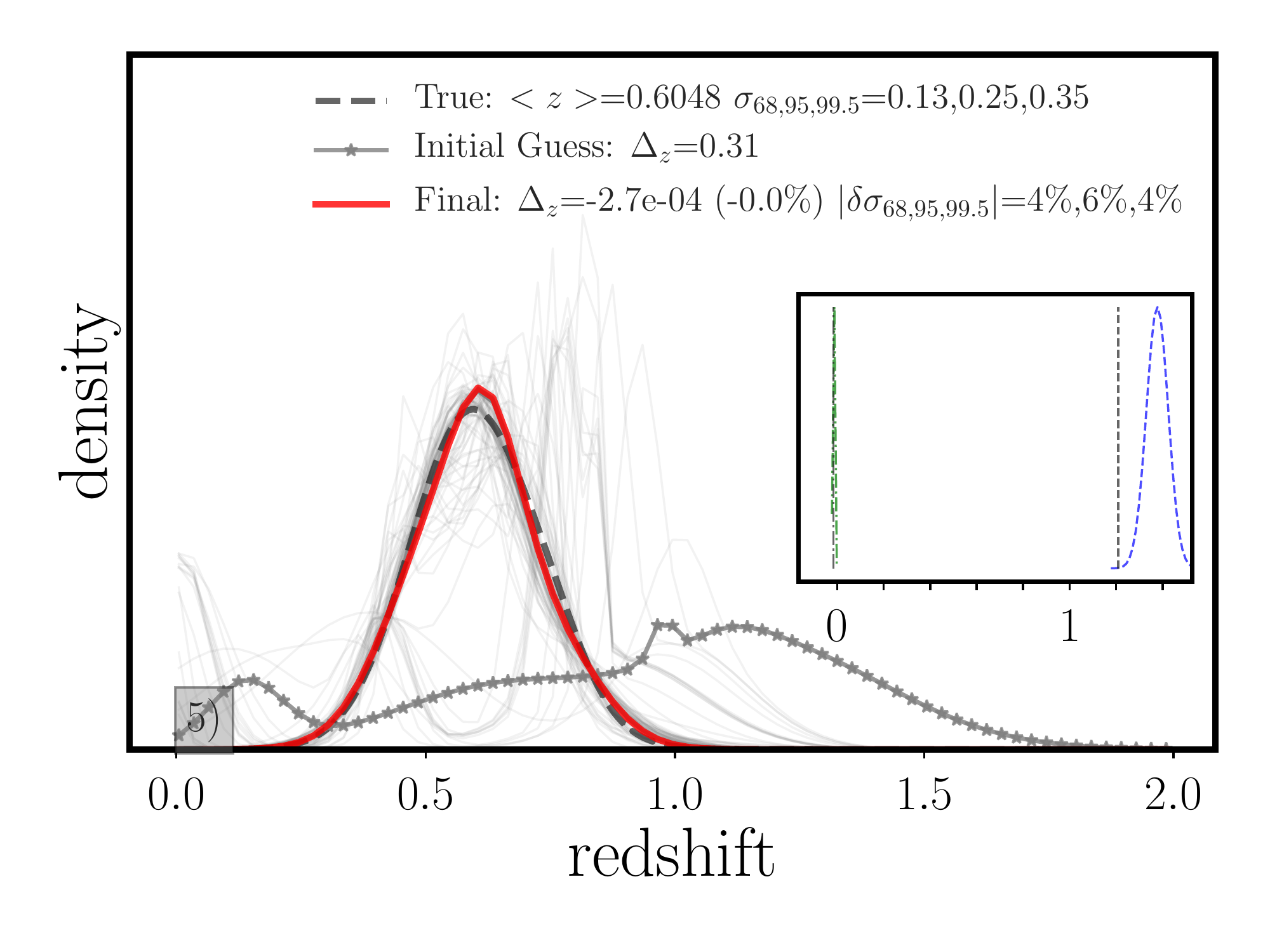}\\
\includegraphics[scale=0.3125,clip=true,trim=15 85 15 25]{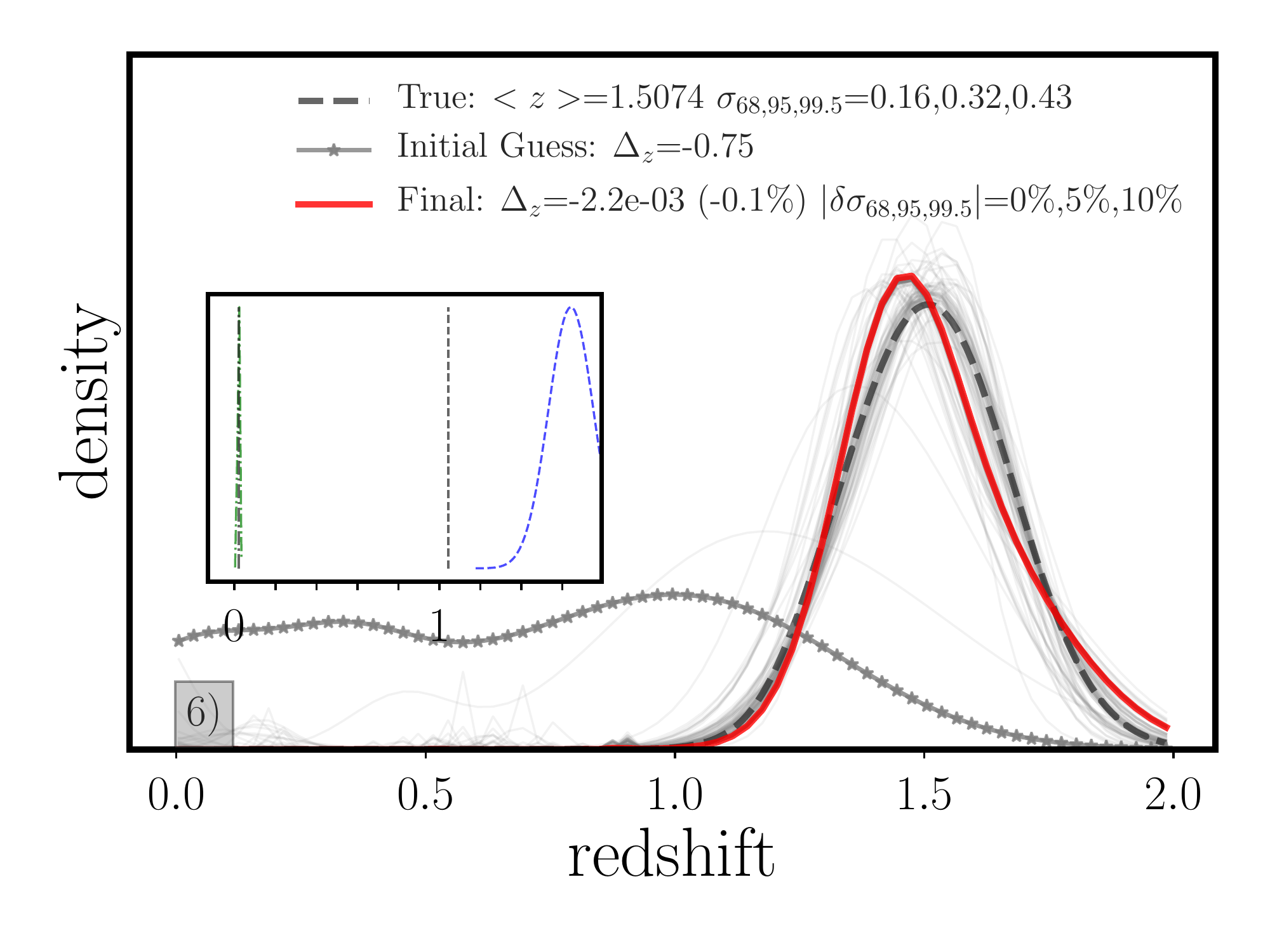}
\includegraphics[scale=0.3125,clip=true,trim=55 85 15 25]{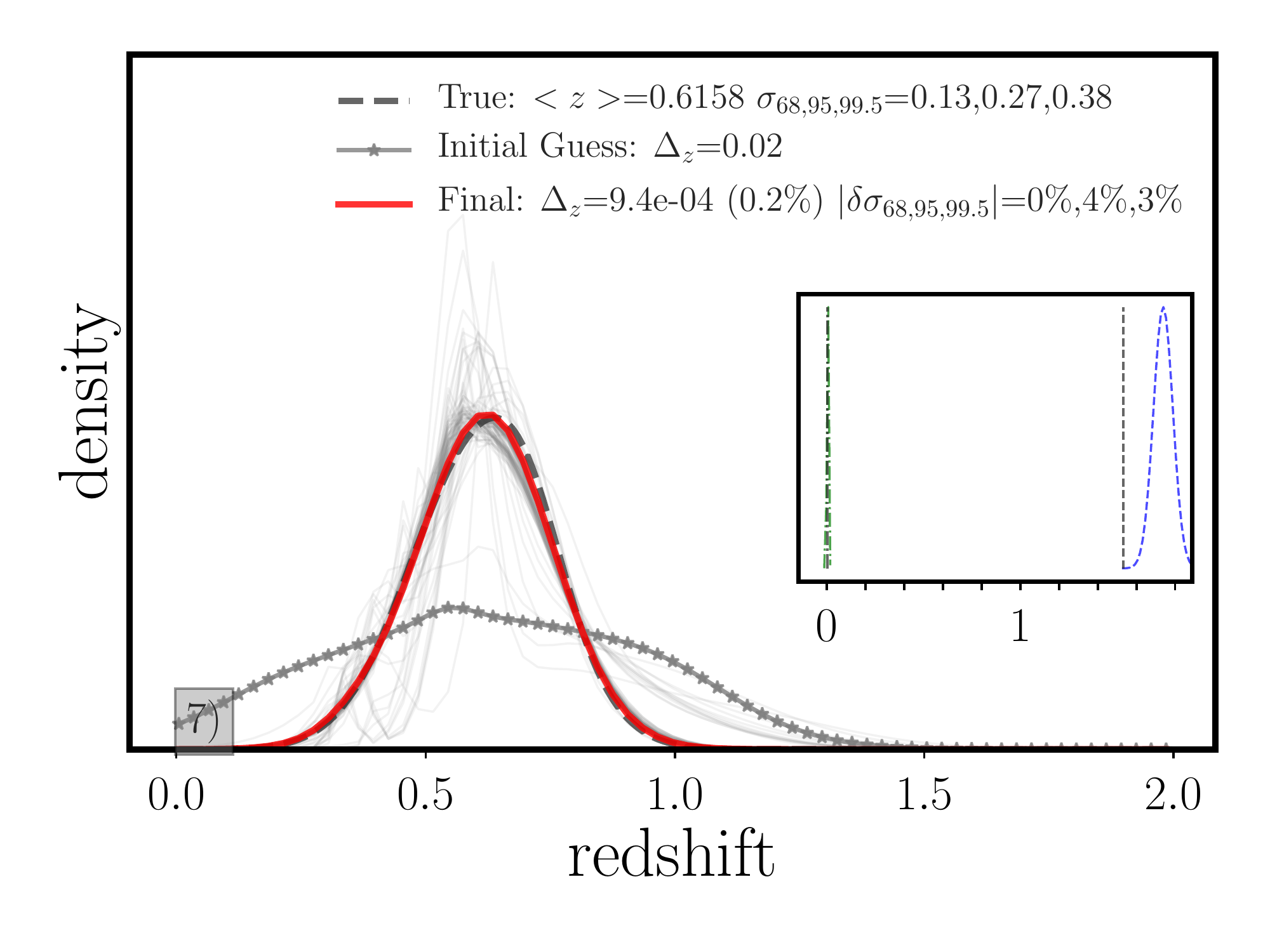}
\includegraphics[scale=0.3125,clip=true,trim=55 85 15 25]{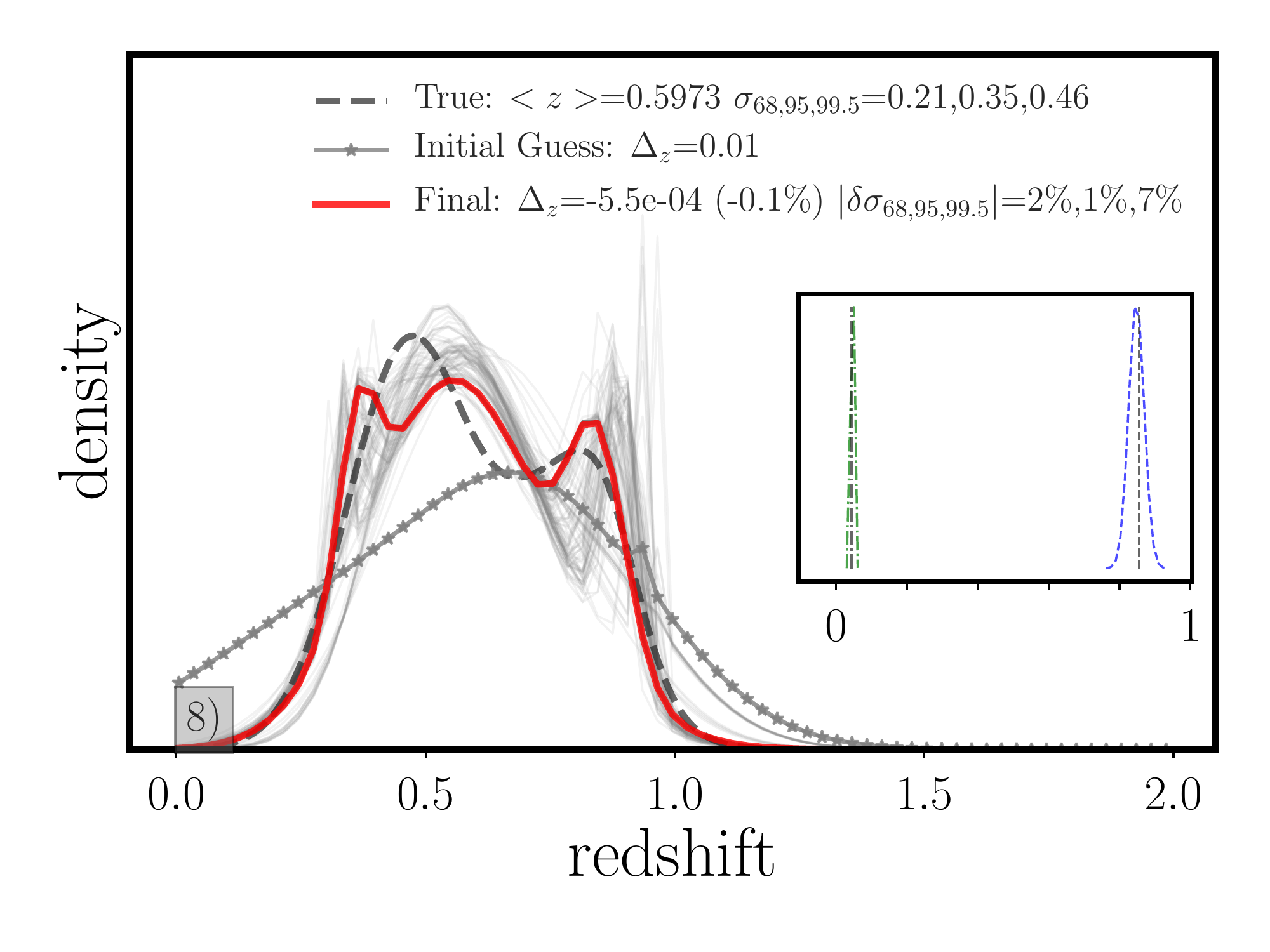}\\
\includegraphics[scale=0.3125,clip=true,trim=15 0 15 25]{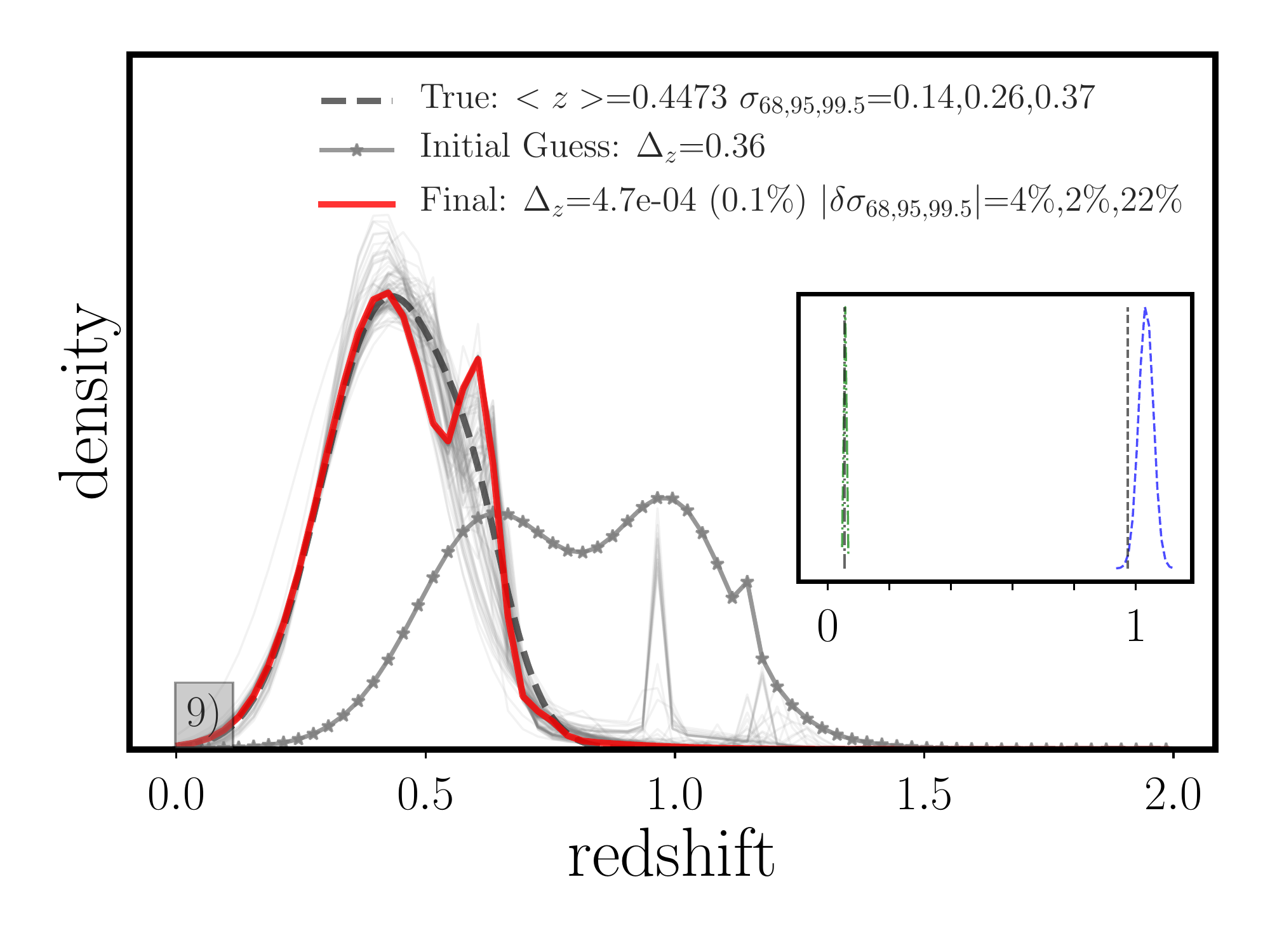}
\includegraphics[scale=0.3125,clip=true,trim=55 0 15 25]{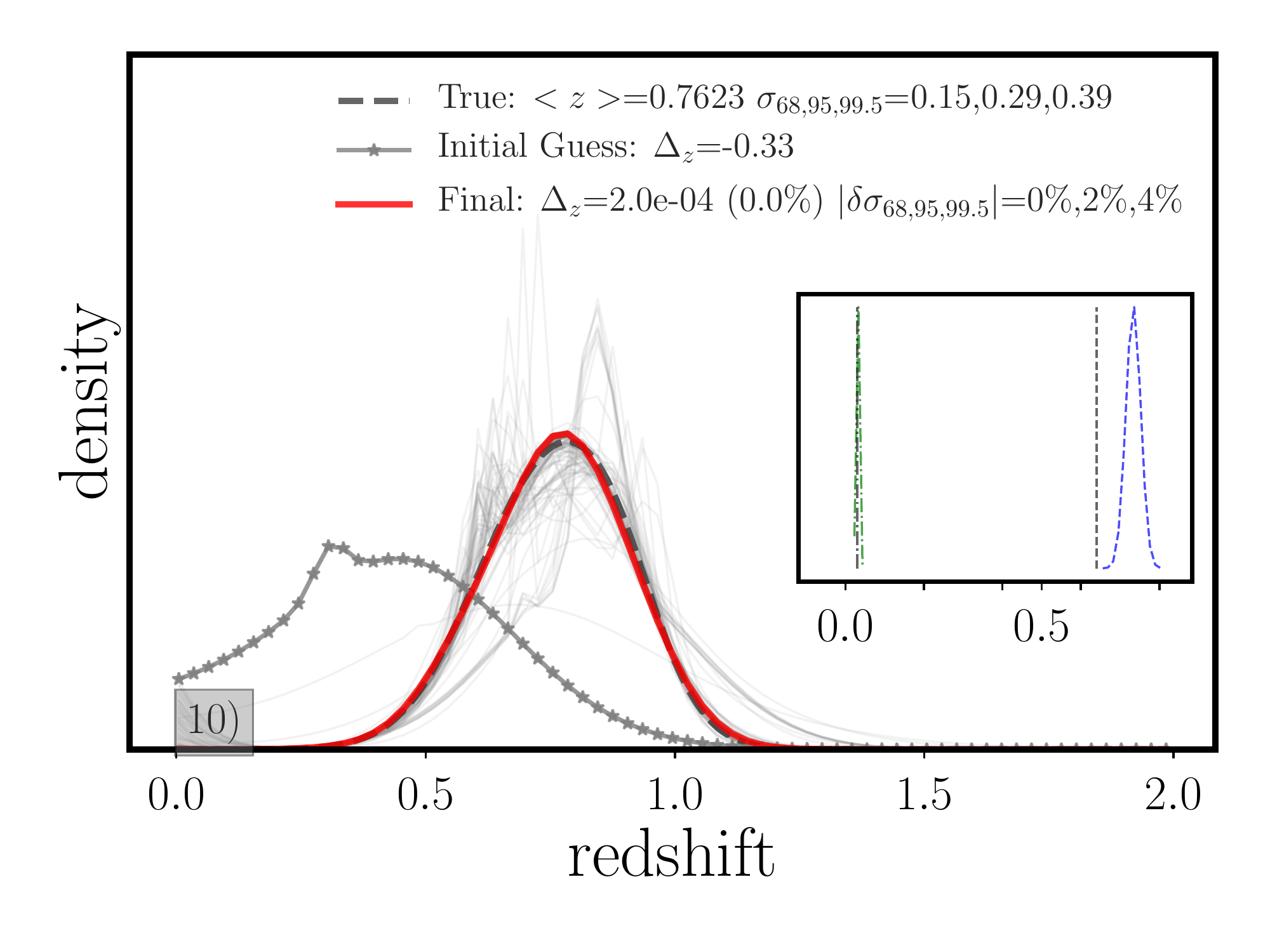}
\includegraphics[scale=0.3125,clip=true,trim=55 0 15 25]{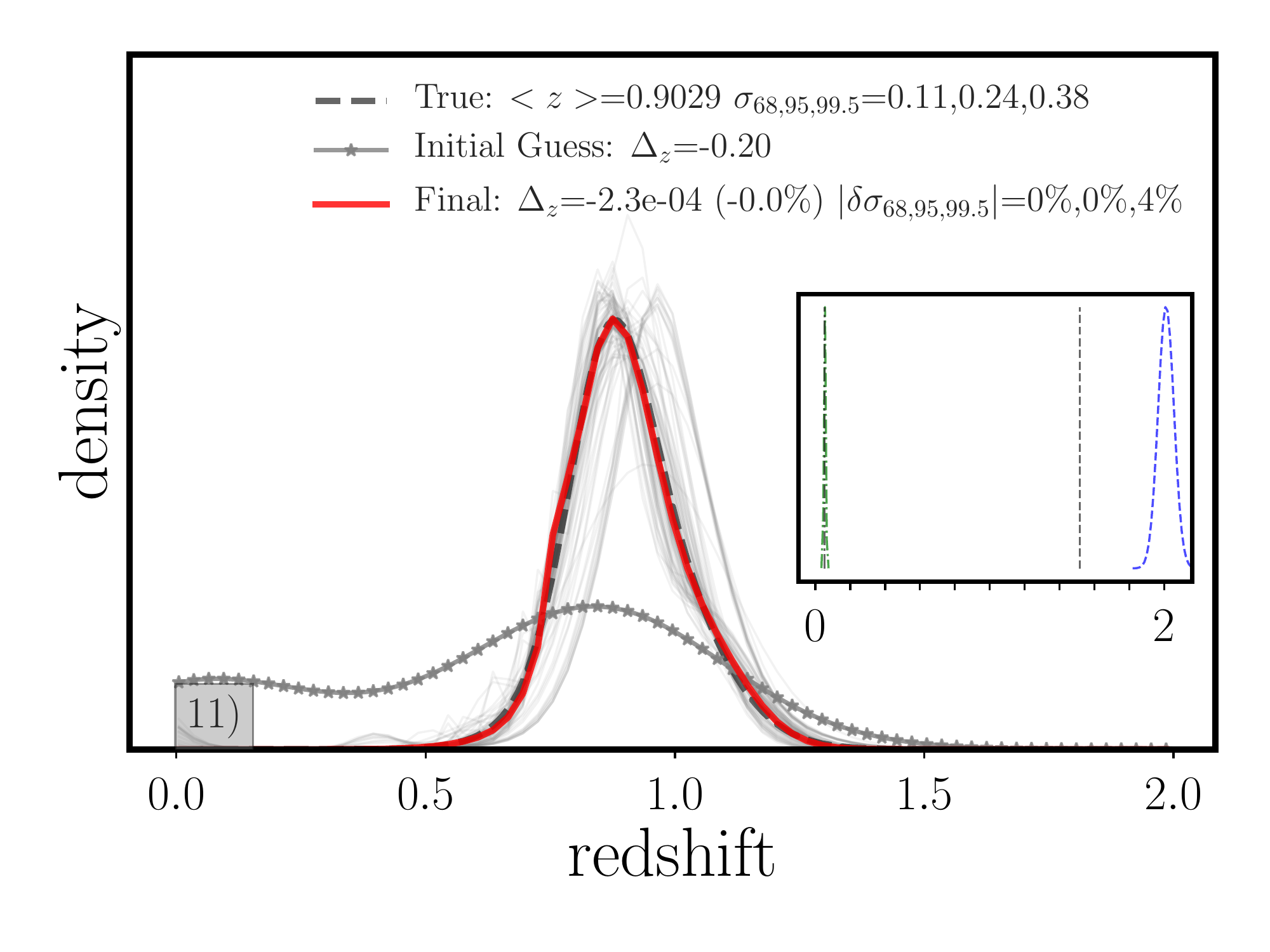}
\caption{The redshift distributions, and galaxy-dark matter bias parameter {\rr values estimated using {\rrr simulated data vectors corresponding to} the positional correlation functions of the galaxies in each colour cell,} the galaxy-galaxy lensing between colour cells, and the correlation of galaxy positions and CMB lensing. This figure can be understood as per Fig. \ref{dndnz_ncl}. The sub panels show the scale dependent galaxy-dark matter bias parameter $b_k$ and a composite parameter $b^*$, in which $\bar{z}$ is the mean redshift of the colour-cell. {\bhP The analysis in these panels corresponds to that described in \S\ref{gg-lcorr}}. }
\label{dndnz_ncllclscl}
\end{figure*}

In this section we use the full range of correlation functions between the colour cells; including galaxy-galaxy auto- and cross- correlations, galaxy-galaxy lensing correlation functions, and galaxy-CMB lensing correlation functions. For ease of comparison to the previous section we use the same true redshift distributions, and galaxy-dark matter bias model parameters as \S\ref{gg-corr} for each colour cell. We also use the same initial guesses for the parameter values before commencing the exploration of parameter space. We present the results of this analysis in Fig. \ref{dndnz_ncllclscl}, which should be read as per Fig. \ref{dndnz_ncl}, except for the galaxy-dark matter bias panels, which show the scale dependent bias parameter $b_k$ and the composite parameter $b^*$.

Examining the redshift distributions in Fig. \ref{dndnz_ncllclscl} we find that one final draw from the posterior is often a very convincing match to the true redshift value. For example panel numbers 0,2,4,5,7,10,11 show excellent agreement, which is also mimicked by the accuracy of the estimation of the mean and widths of the redshift distributions. Concentrating on the panels with a bi-modal true redshift distribution, e.g., 1, 3, and 8, one sees that a draw from the posterior appears to find difficultly accurately reproducing the full shape of these distributions to high accuracy.  We expect this discrepancy to become reduced or to disappear as the codes continue to explore the high dimensional parameter space {\rr and the chains converge around the best fit parameter values}. We do note that the summary statistics relating to these distributions are still in excellent agreement with those measured on the true redshift distributions.

Concentrating again on the smaller sub-panels in each panel of Fig. \ref{dndnz_ncl}, we find that the scale dependent galaxy-dark matter bias model parameter $b_k$, is very well constrained by this analysis, showing an offset from the truth values consistent with 0, and a scatter around the truth values at level of $0.4\sigma$. The composite parameter $b^*$ is also quite well constrained at the level $(2.2\pm2.9)\sigma$. The value $\pm2.9\sigma$ suggests that the uncertainties on the composite parameter are slightly small. We note that these {\rr chains are still exploring the high dimensional parameter space and converging}, and we would expect these value to become closer to 1, see the discussion in \S\ref{global_min} for more details. The individual parameters $b_0$ and $b_1$ are less well constrained and exhibit a large offset and scatter.  Summary statistics as measured on all 12 colour cells, for the original galaxy-dark matter bias parameter values and the composite values are shown in Table \ref{final_results}.

\subsection{Gal-gal \& lensing correlation functions II}
\label{gg-corrl-sdss}
\begin{figure*}
\includegraphics[scale=0.3125,clip=true,trim=15 85 25 25]{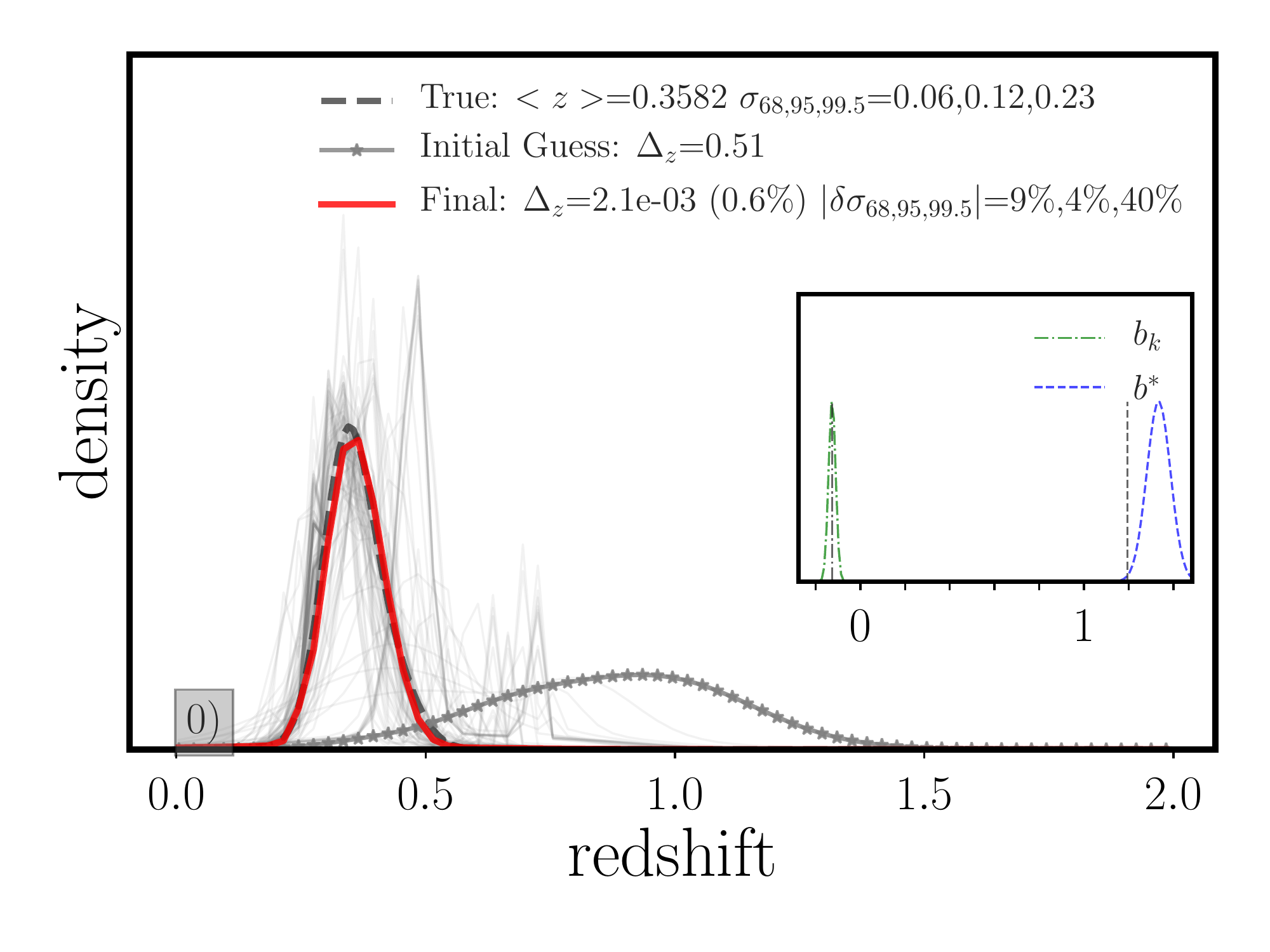}
\includegraphics[scale=0.3125,clip=true,trim=55 85 25 25]{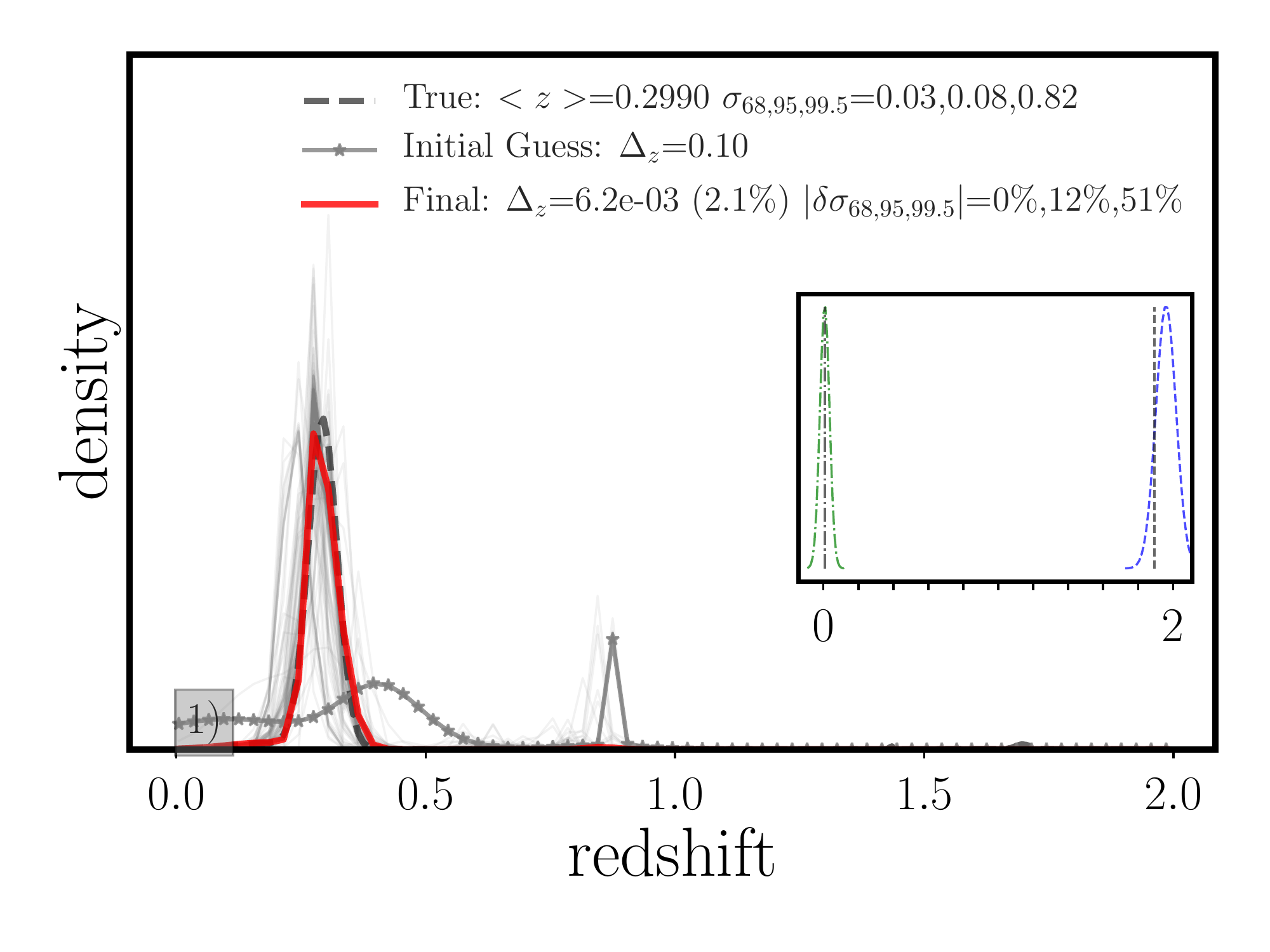}
\includegraphics[scale=0.3125,clip=true,trim=55 85 25 25]{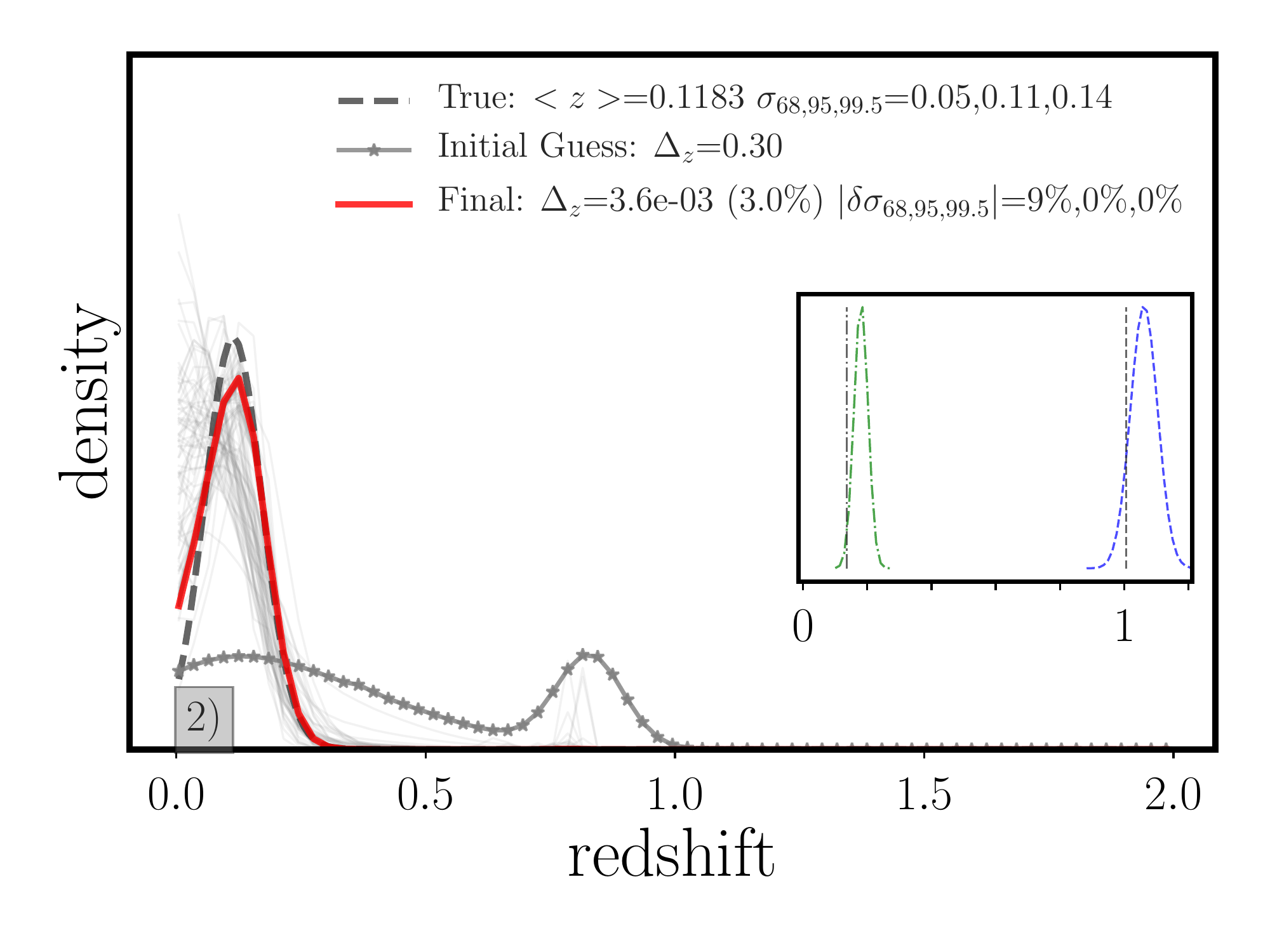}\\
\includegraphics[scale=0.3125,clip=true,trim=15 85 25 25]{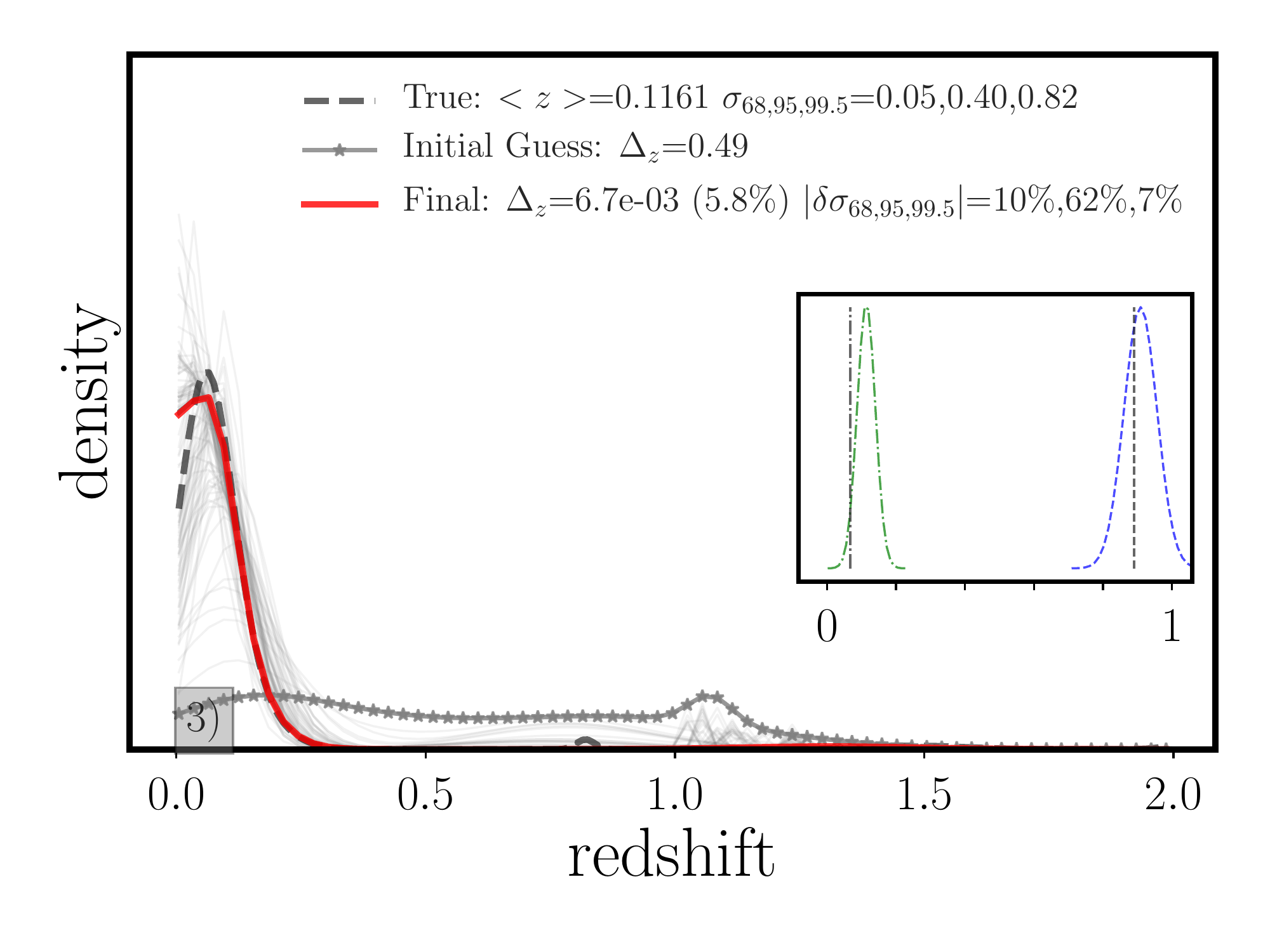}
\includegraphics[scale=0.3125,clip=true,trim=55 85 25 25]{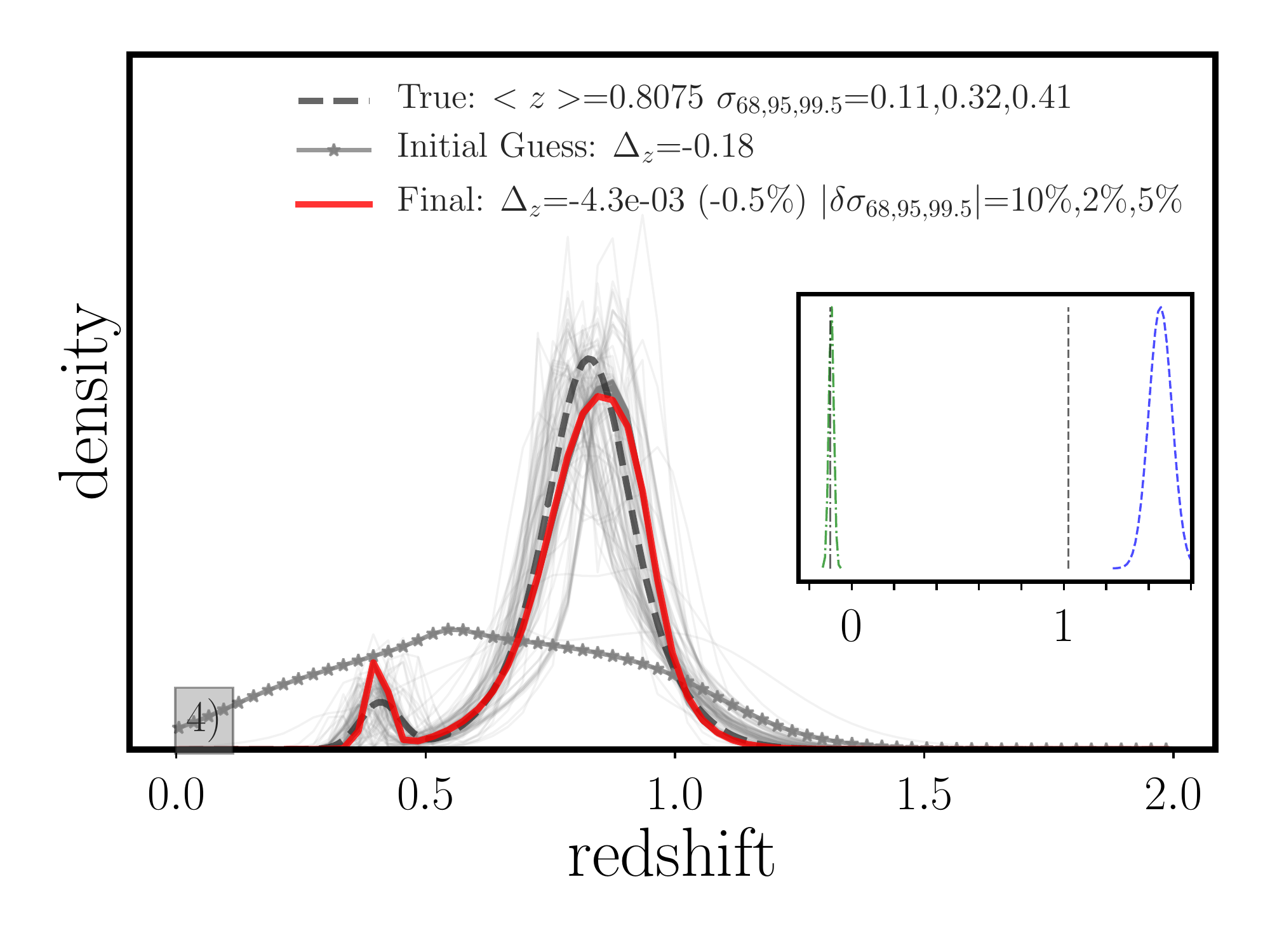}
\includegraphics[scale=0.3125,clip=true,trim=55 85 25 25]{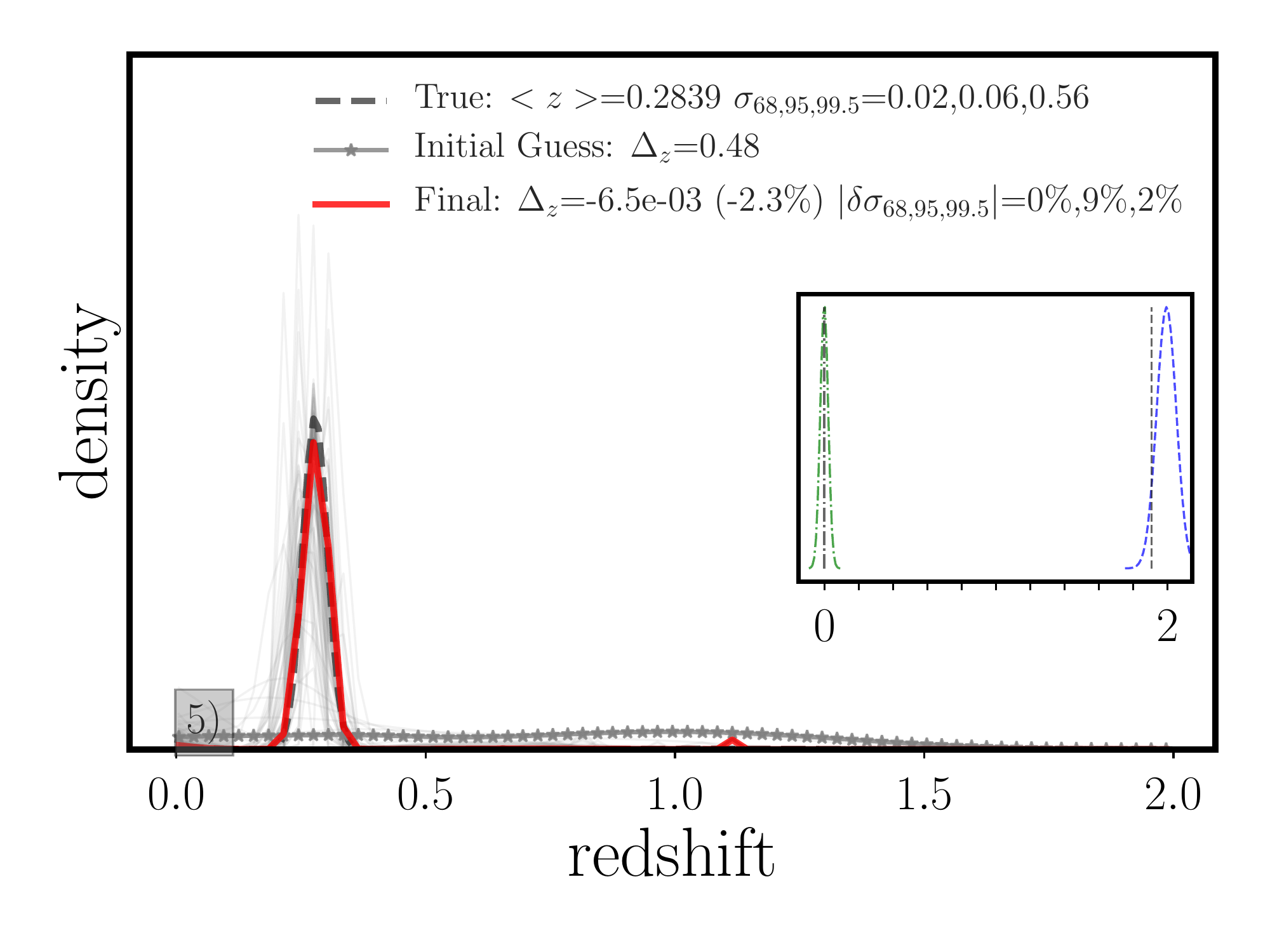}\\
\includegraphics[scale=0.3125,clip=true,trim=15 85 25 25]{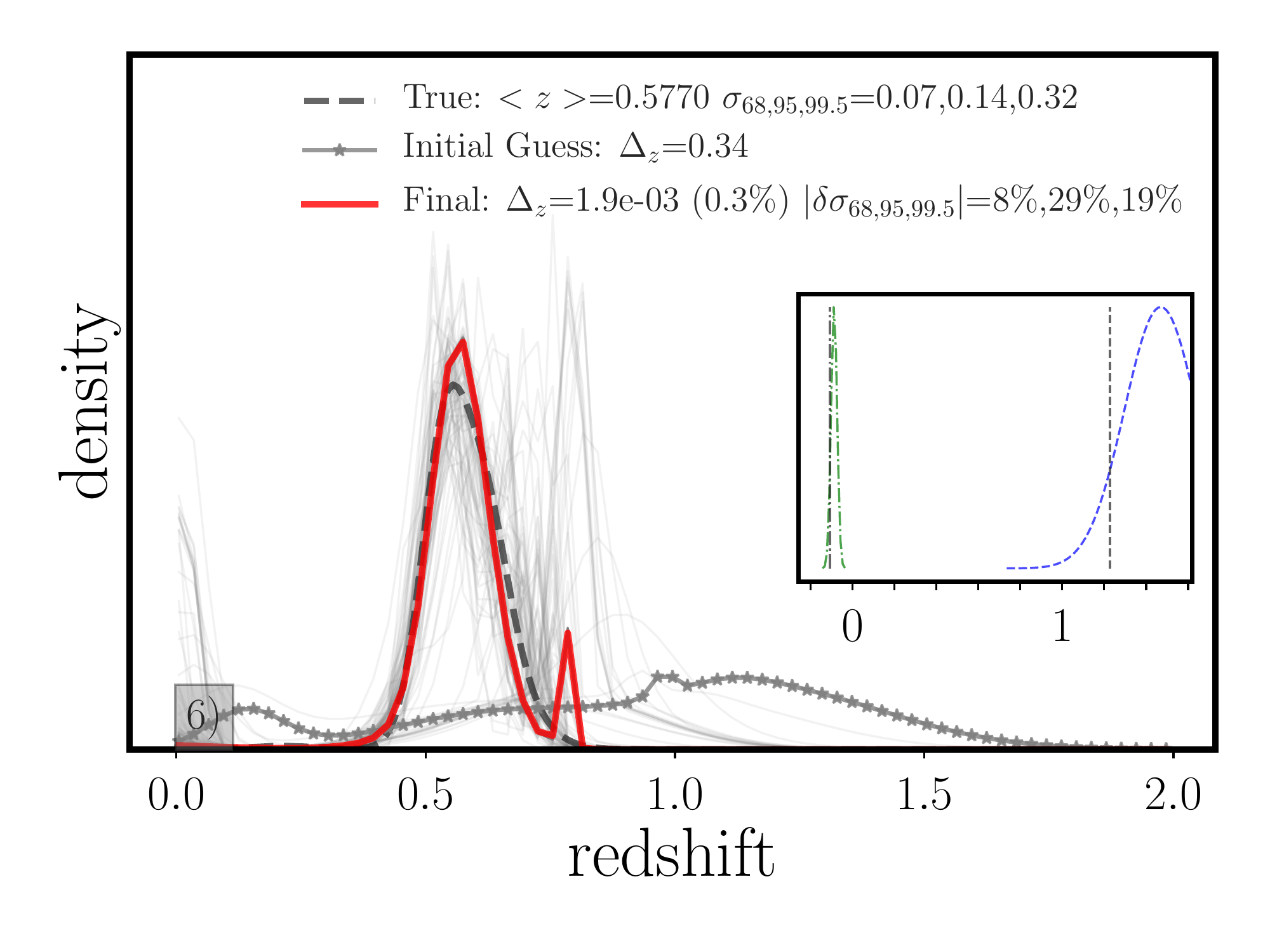}
\includegraphics[scale=0.3125,clip=true,trim=55 85 25 25]{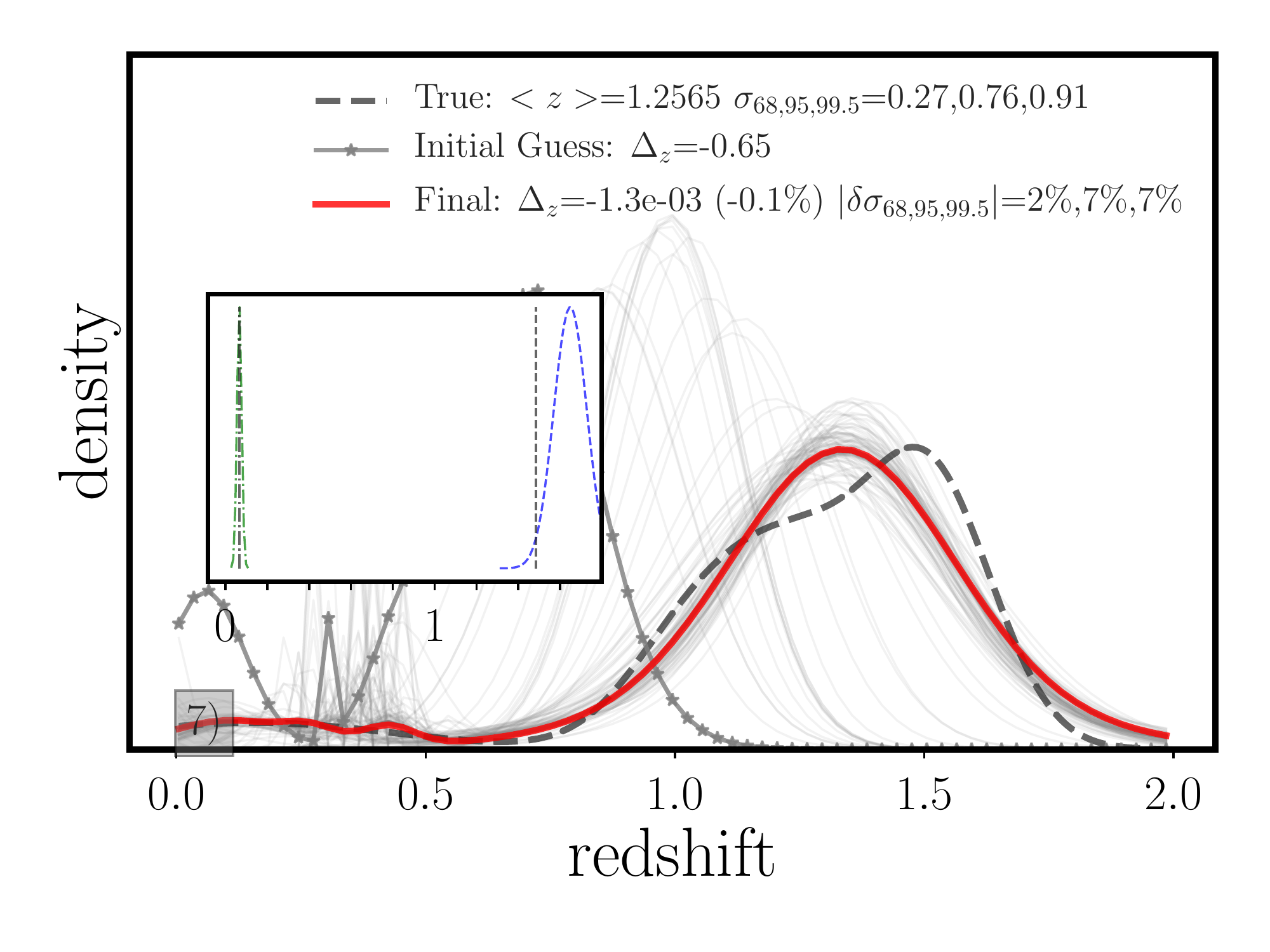}
\includegraphics[scale=0.3125,clip=true,trim=55 85 25 25]{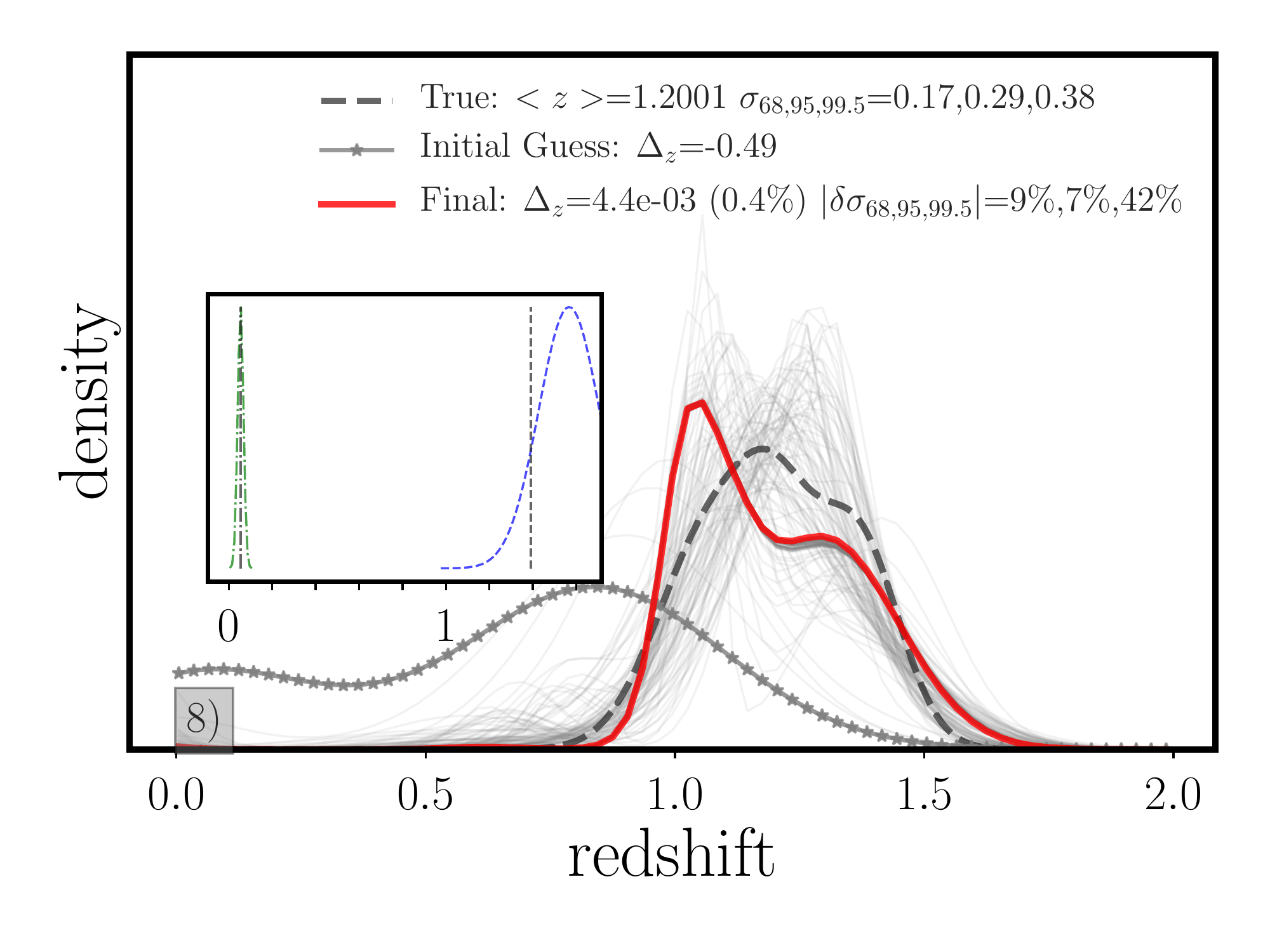}\\
\includegraphics[scale=0.3125,clip=true,trim=15  0 25 25]{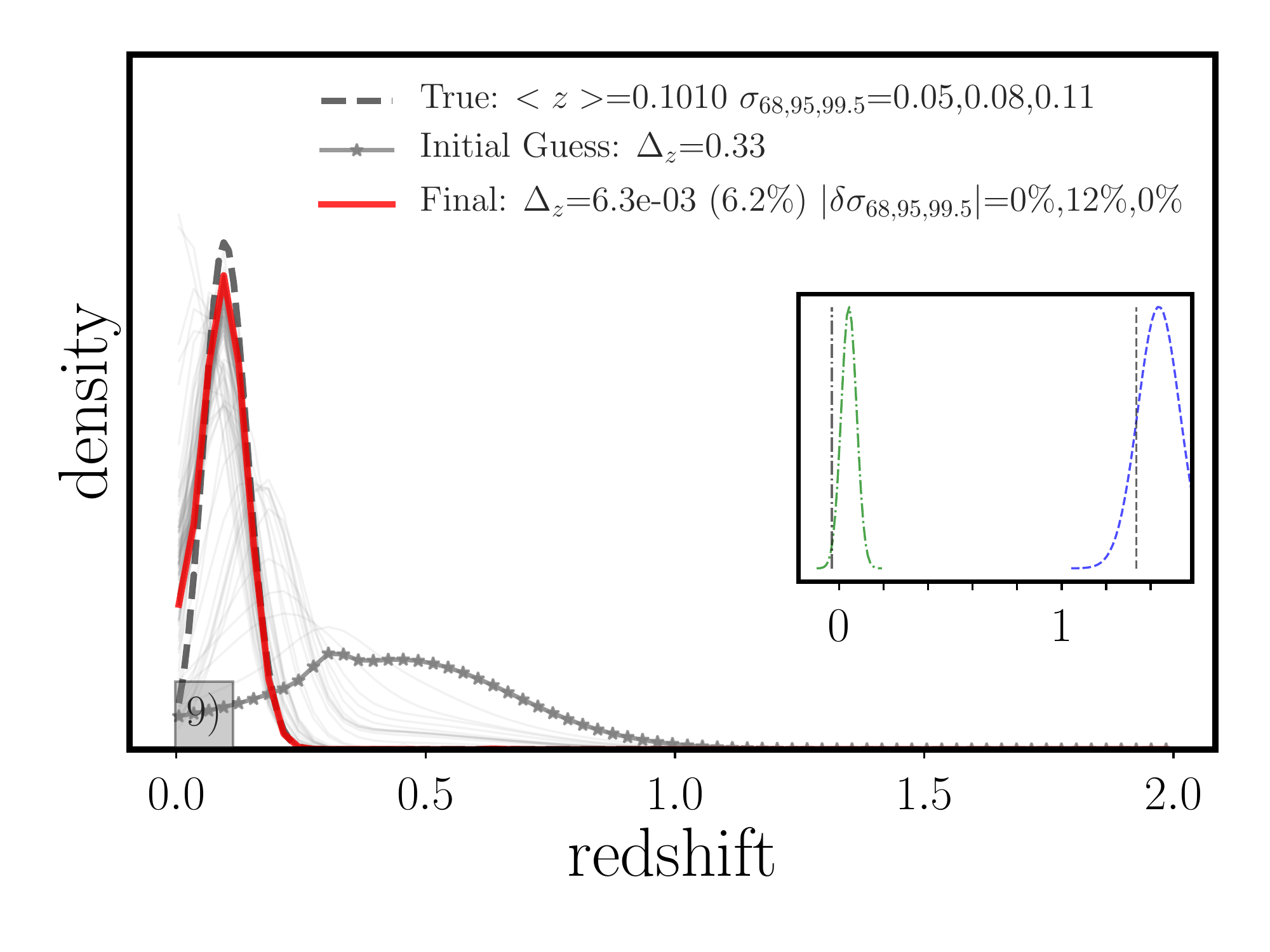}
\includegraphics[scale=0.3125,clip=true,trim=55  0 25 25]{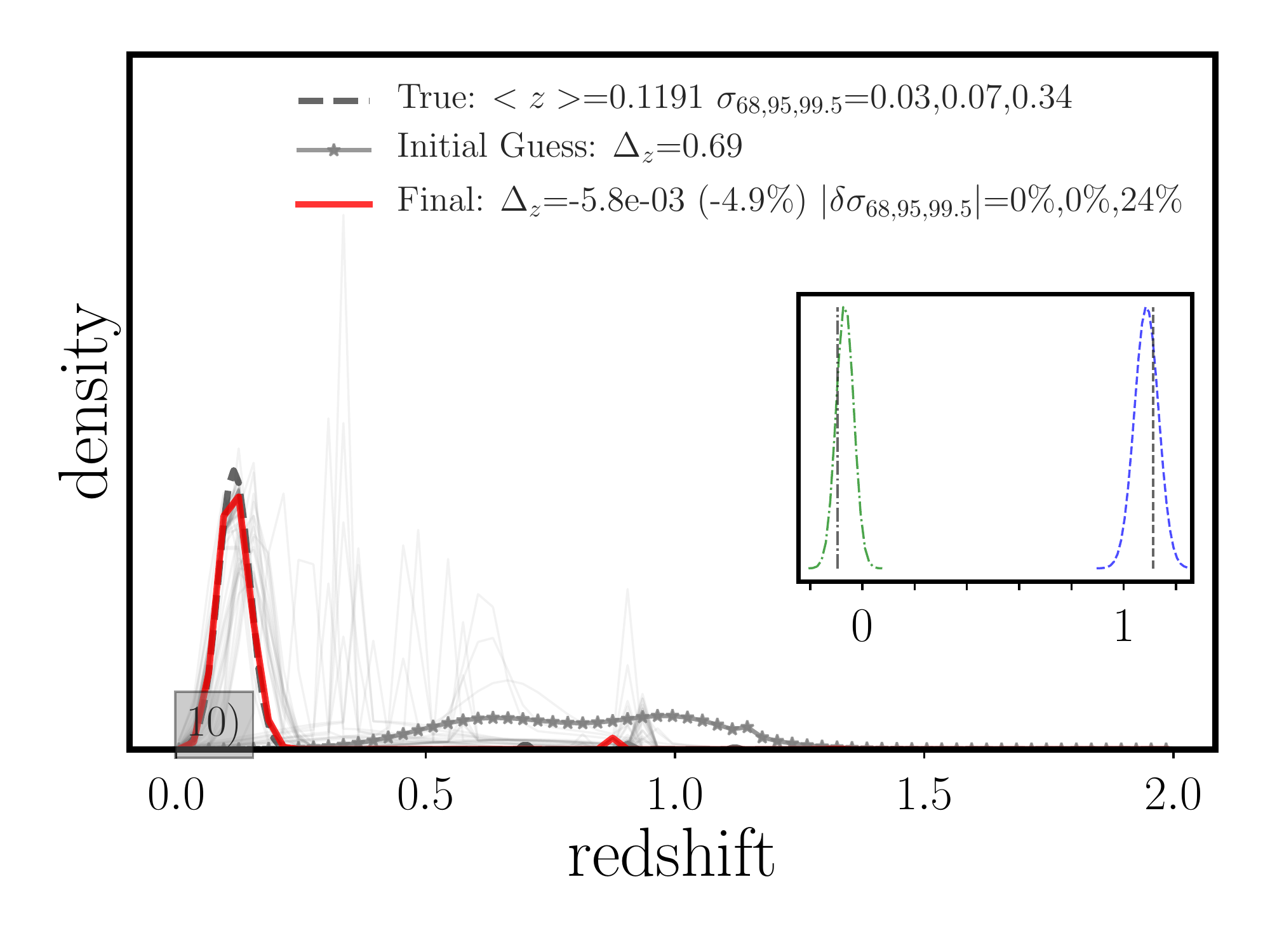}
\includegraphics[scale=0.3125,clip=true,trim=55  0 25 25]{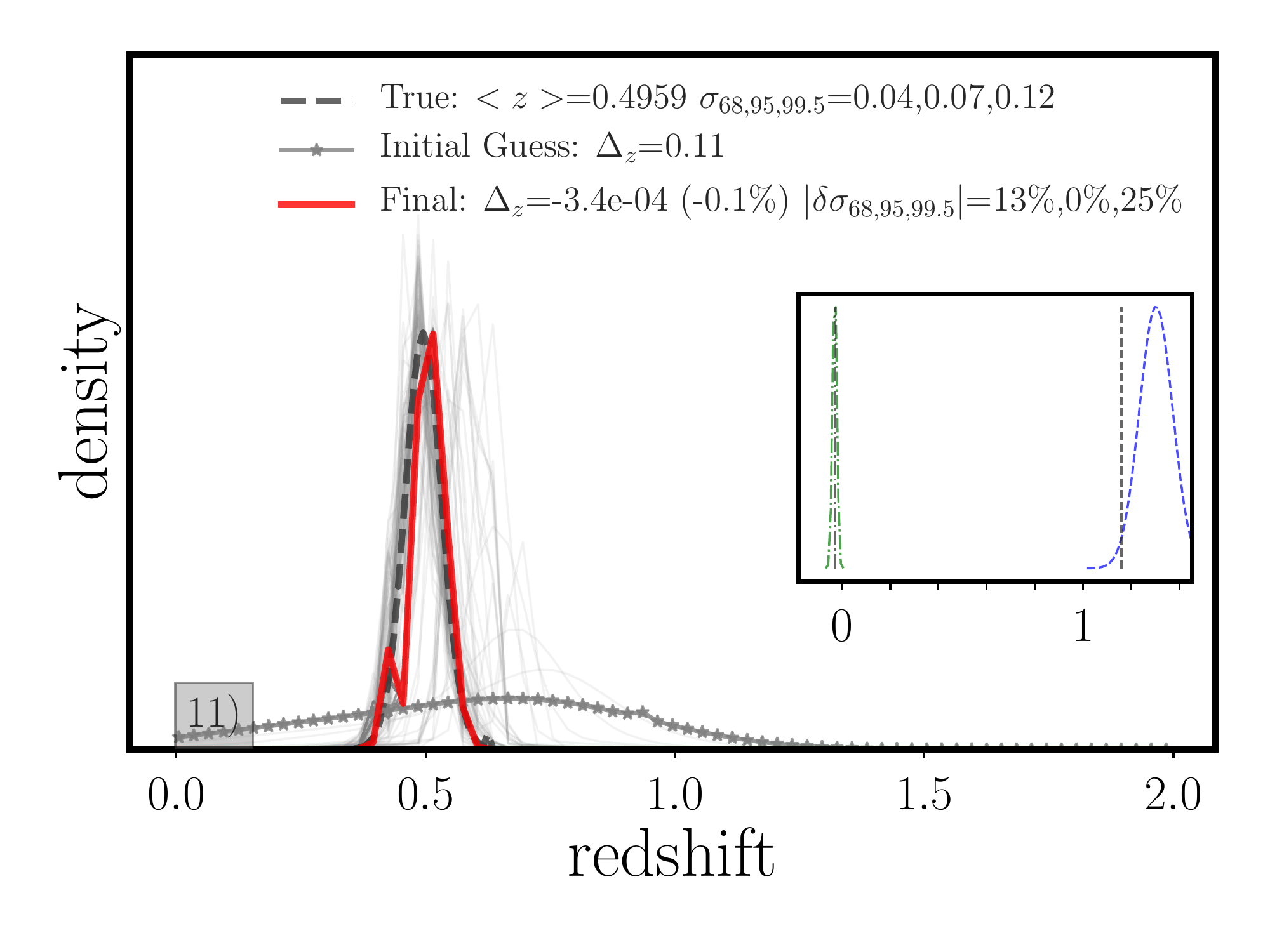}
\caption{The redshift distributions, and galaxy-dark matter bias parameter values estimated using {\rrr simulated data vectors that would correspond to measurements of} the angular positions of galaxies in each colour cell, the galaxy-galaxy lensing between colour cells, and the correlation of galaxy positions and CMB lensing. This figure shows the results of using SDSS redshift distributions as the truth redshift distributions. This figure can be understood as Fig. \ref{dndnz_ncllclscl}. {\bhP The analysis in these panels corresponds to that described in \S\ref{gg-corrl-sdss}}.}
\label{dndnz_sdss_ncl}
\end{figure*}

In this section we replace the {\rrr simulated data vectors using} idealised redshift distributions with {\rrr simulated data vectors} using {\bhP real spectroscopic redshift distributions as measured in the colour cells of the SDSS data.} Fig. \ref{dndnz_sdss_ncl} is the same as Fig. \ref{dndnz_ncllclscl} but with this change in redshift distributions, and a different set of chosen galaxy-dark matter bias model parameter values for each colour cell. We again proceed assuming that all of the galaxy and lensing correlation functions have been measured, as per \S\ref{gg-lcorr}. To again show the unimportance of the initial starting conditions, we use the same initial guess distribution and galaxy-dark matter parameter values as in the previous two sections for each of the colour cells. 

The redshift distributions in Fig. \ref{dndnz_sdss_ncl} are drawn from real galaxy redshifts, and are often more peaked than the redshift distributions from the previous sections. This may be because of the relatively few numbers of galaxies used to approximate these distributions. However, we note that a final draw from the posterior distribution almost sits on top of the true distribution. Exceptions to this are panels 7 (8), which still provide good (reasonable) matches to the full shape of the redshift distribution. We note that the explored range of  distributions for panel 8 is quite large, and indeed covers the true redshift distribution. The summary metric values for all of these distributions are reasonable, and we again expect the agreement to improve as the code continues to explore the high dimensional parameter space. Of particular interest is panel 4, which shows a double peaked redshift distribution in which one very dominant peak is at $z=1.0$ and a much smaller peak is at $z\sim0.5$. This data driven  correlation method appears to be sensitive to the nuances of this distribution and a draw from the posterior is very consistent with the locations and amplitudes of both peaks.

The sub-panels of Fig. \ref{dndnz_sdss_ncl} show that the scale and composite galaxy-dark matter bias parameters are recovered with an offset and scatter which is statistically acceptable $(0.7\pm1.1)\sigma$ and $(1.9\pm0.8)\sigma$.

\subsection{Gal-gal \& lensing correlation functions III}
\label{galgallenserr}
\begin{figure*}
\includegraphics[scale=0.3125,clip=true,trim=15 85 15 25]{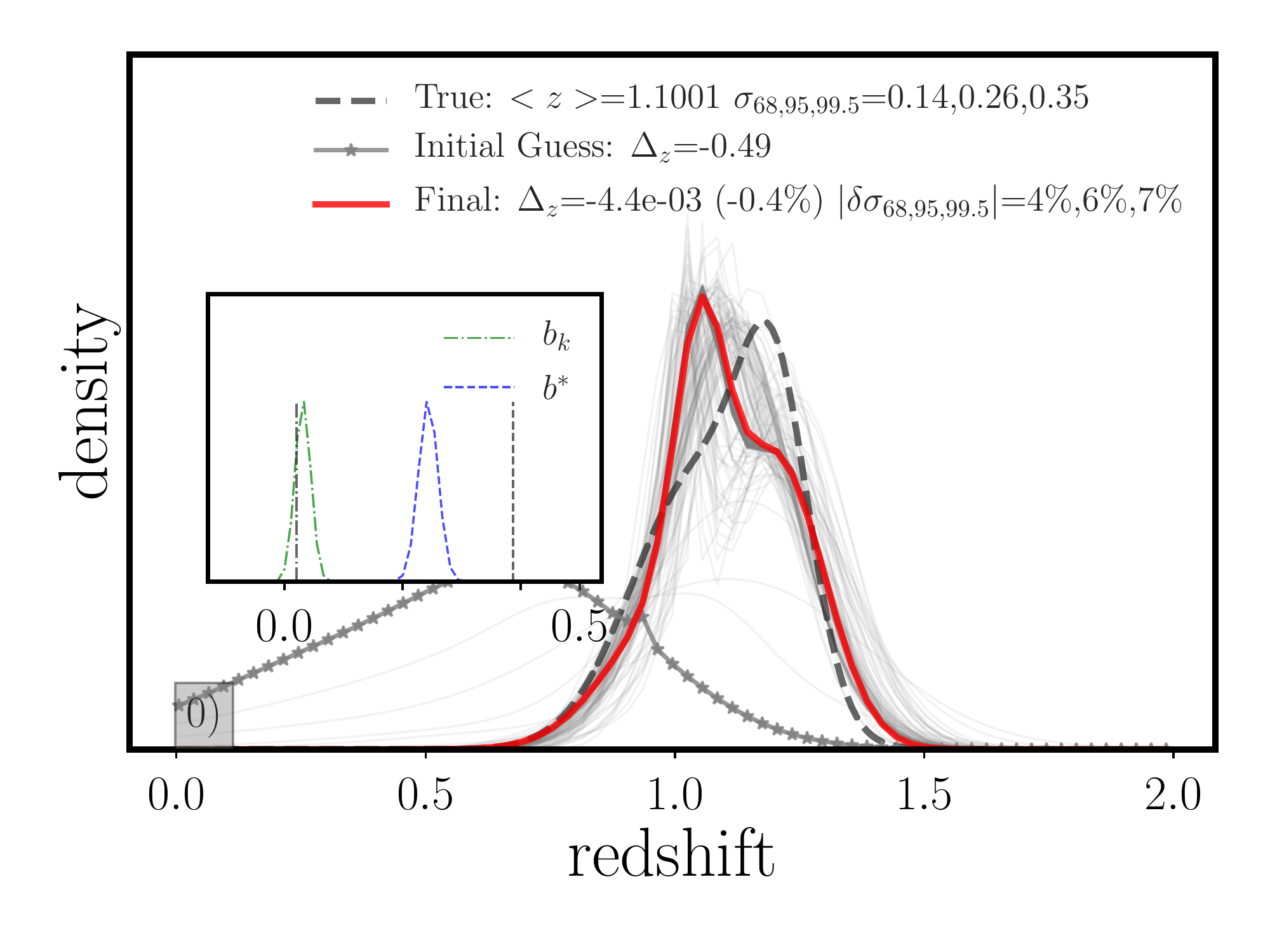}
\includegraphics[scale=0.3125,clip=true,trim=55 85 15 25]{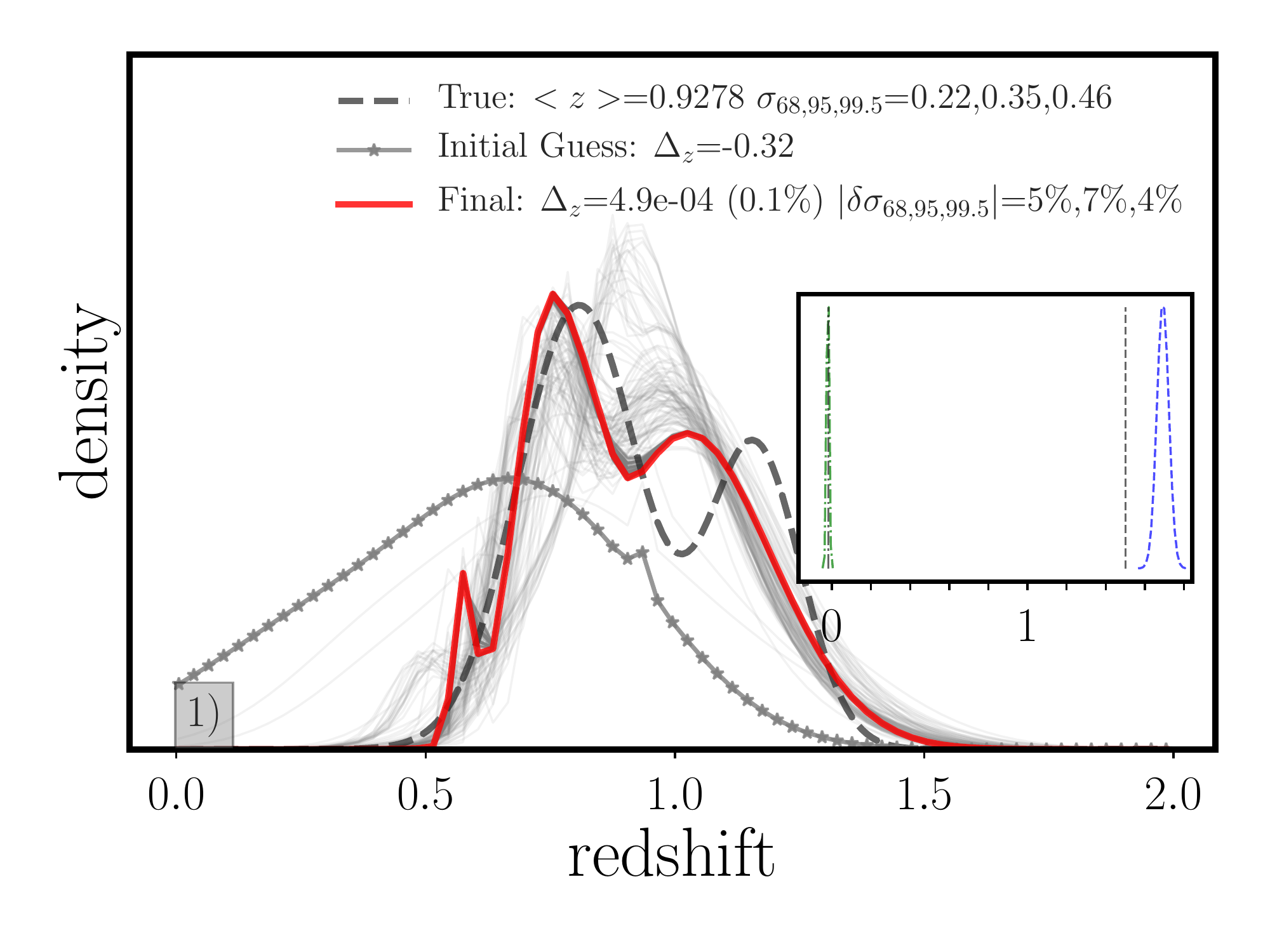}
\includegraphics[scale=0.3125,clip=true,trim=55 85 15 25]{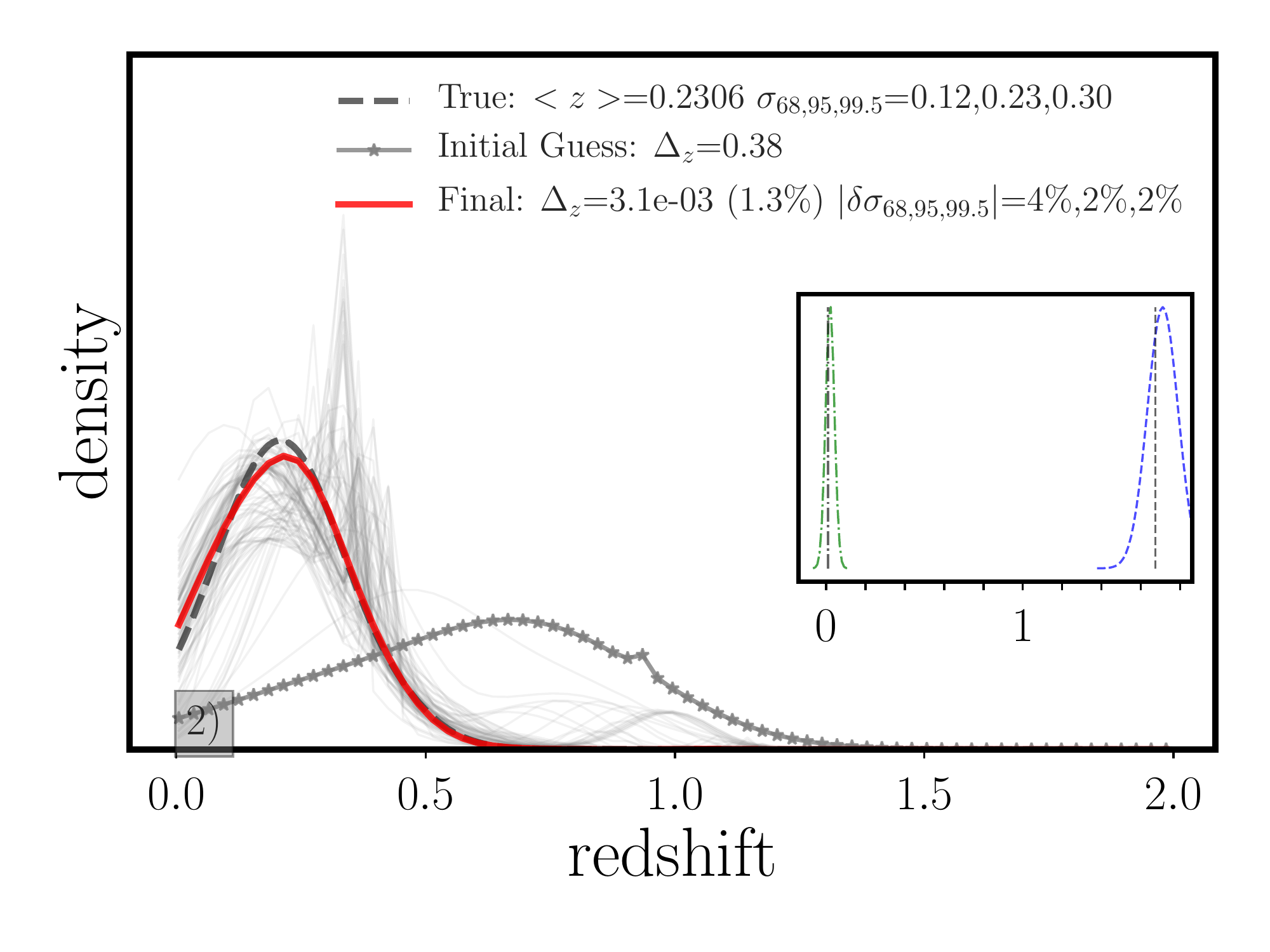}\\
\includegraphics[scale=0.3125,clip=true,trim=15 85 15 25]{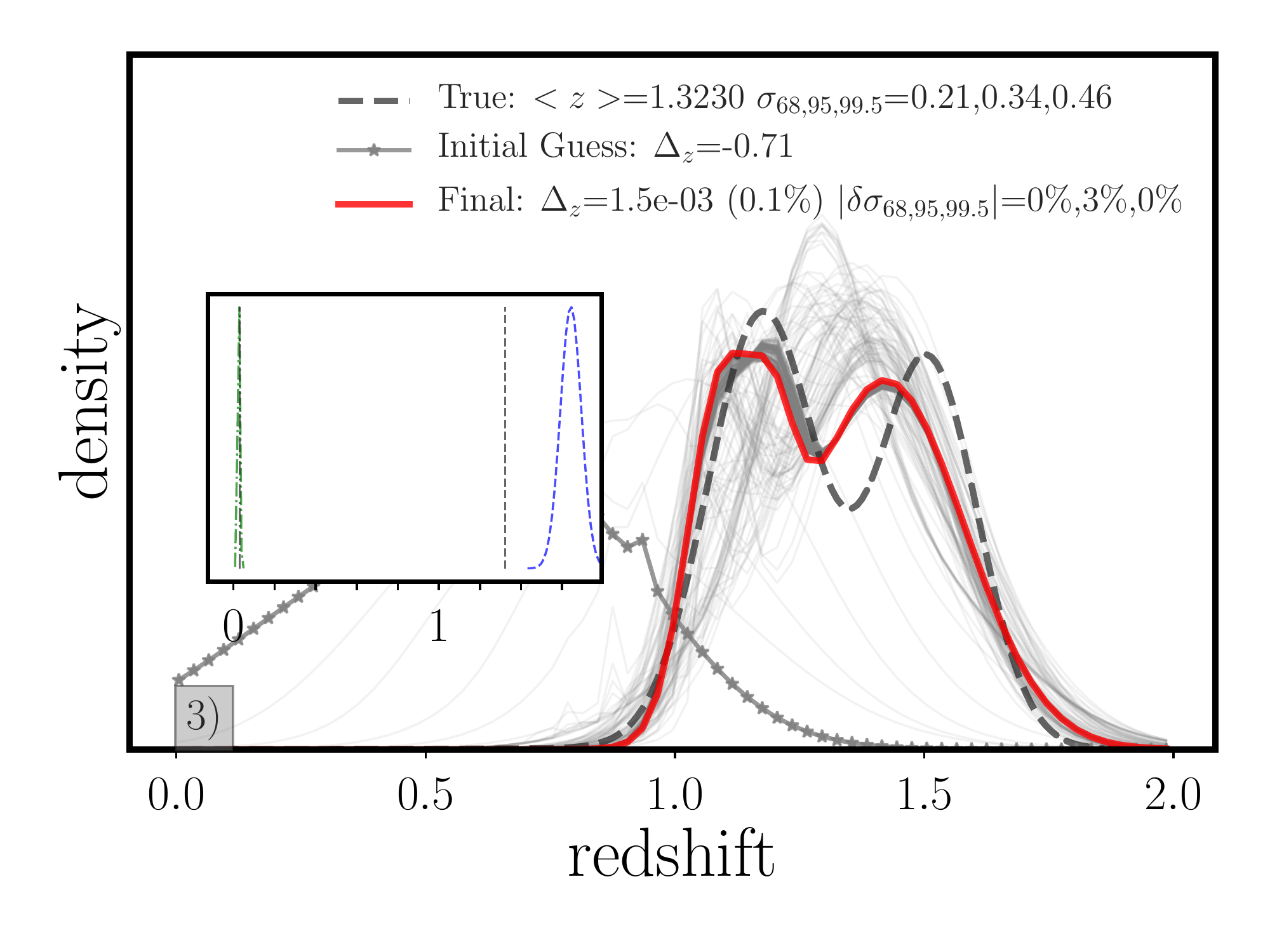}
\includegraphics[scale=0.3125,clip=true,trim=55 85 15 25]{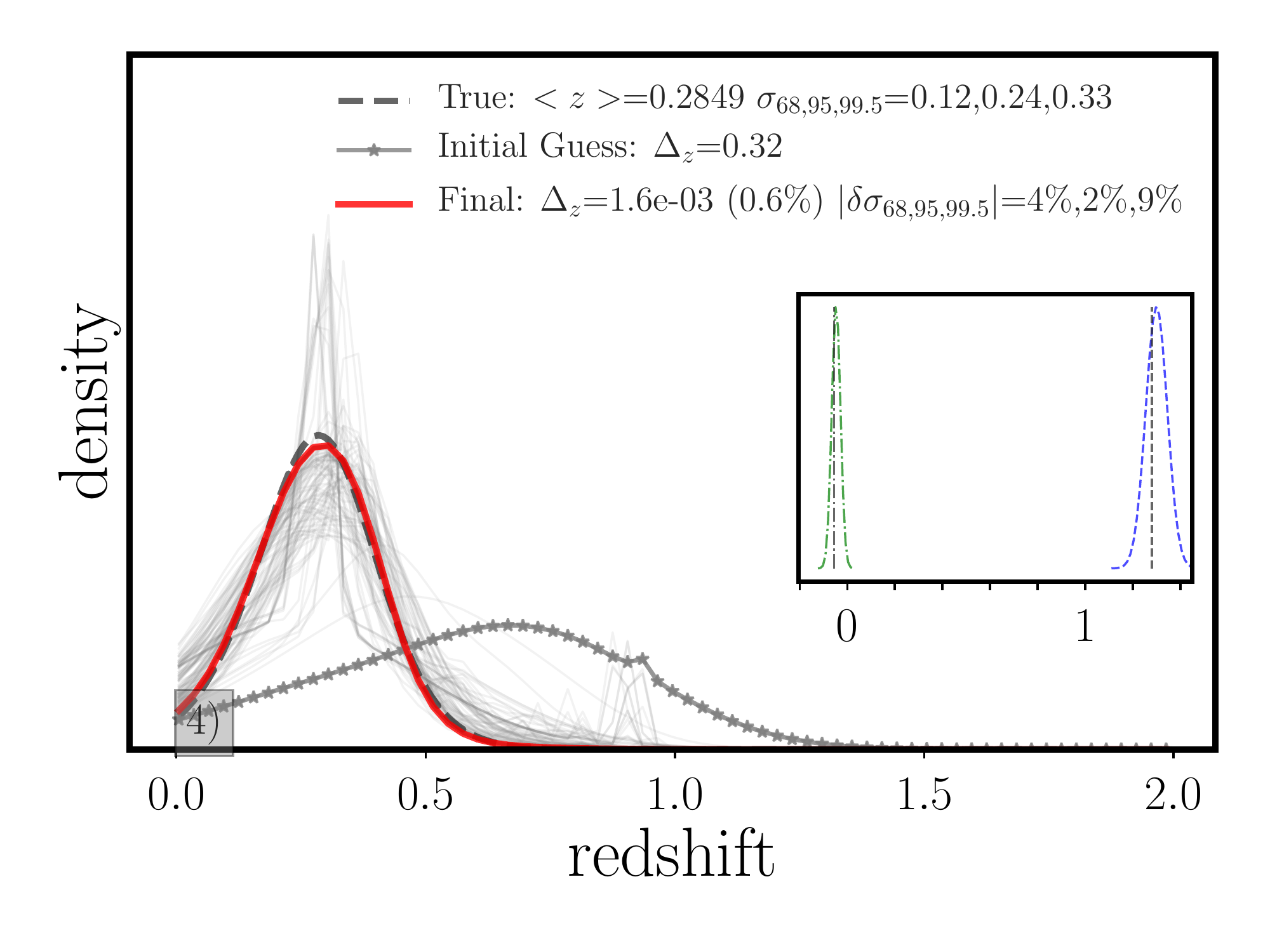}
\includegraphics[scale=0.3125,clip=true,trim=55 85 15 25]{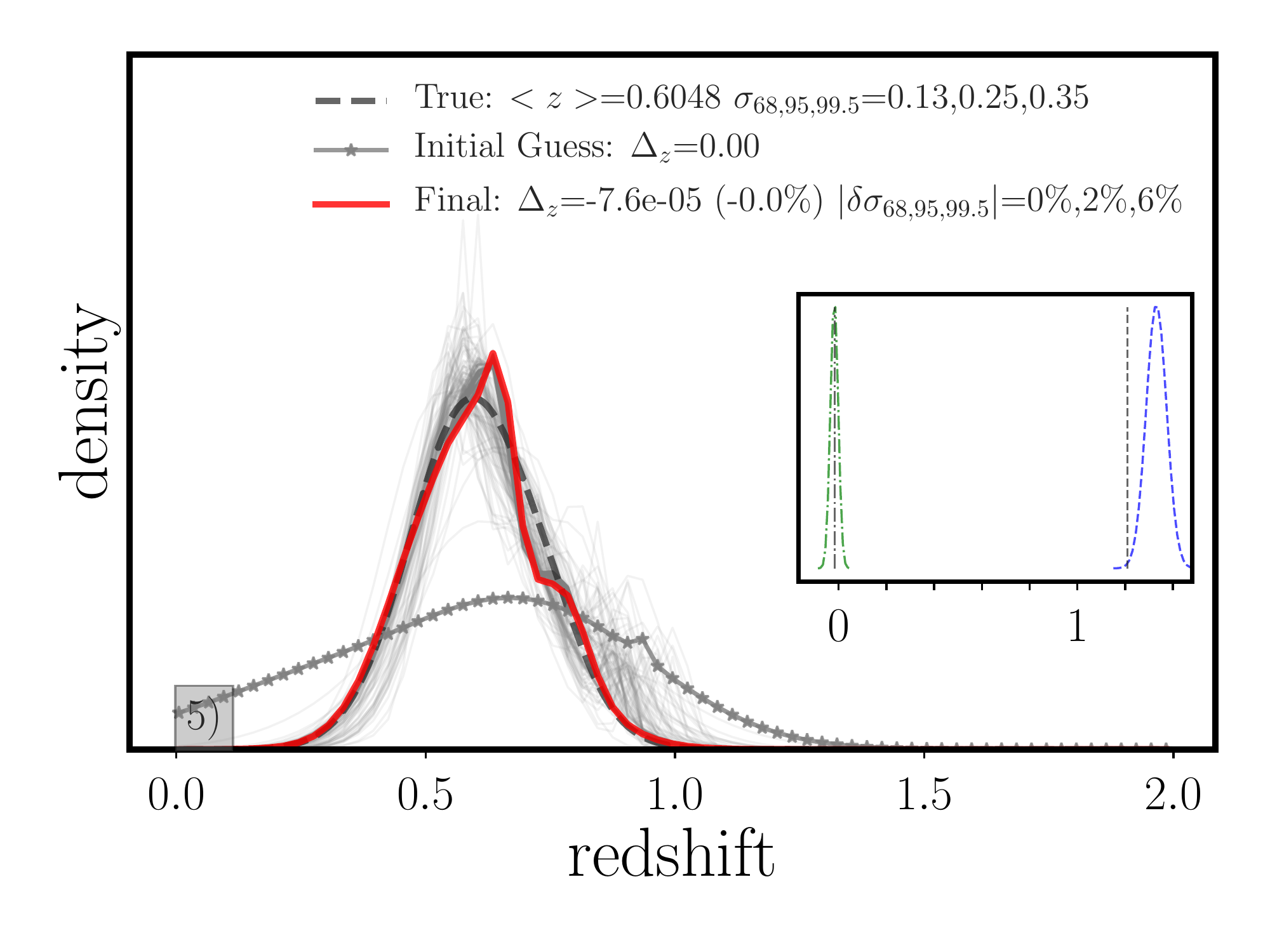}\\
\includegraphics[scale=0.3125,clip=true,trim=15 85 15 25]{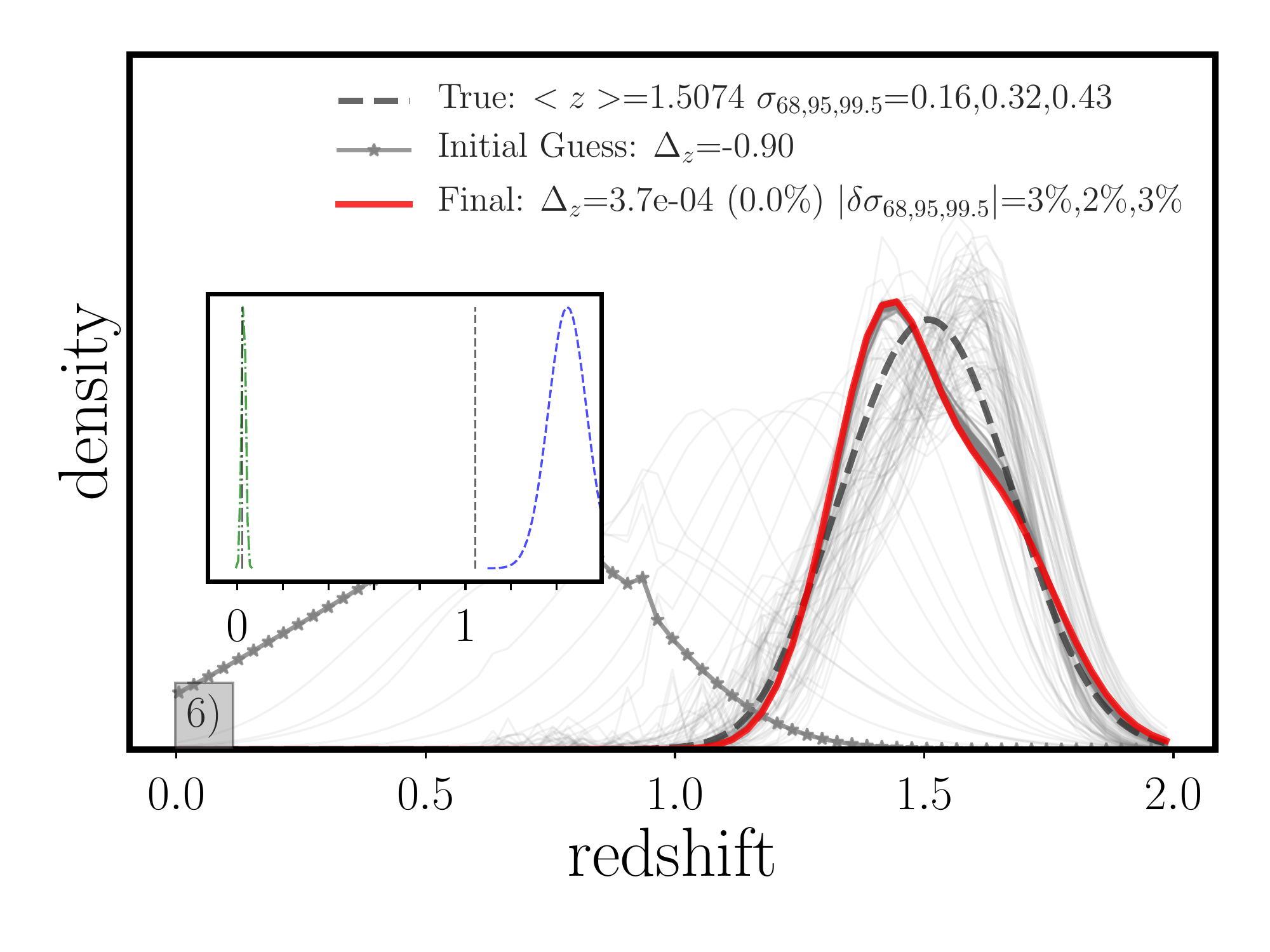}
\includegraphics[scale=0.3125,clip=true,trim=55 85 15 25]{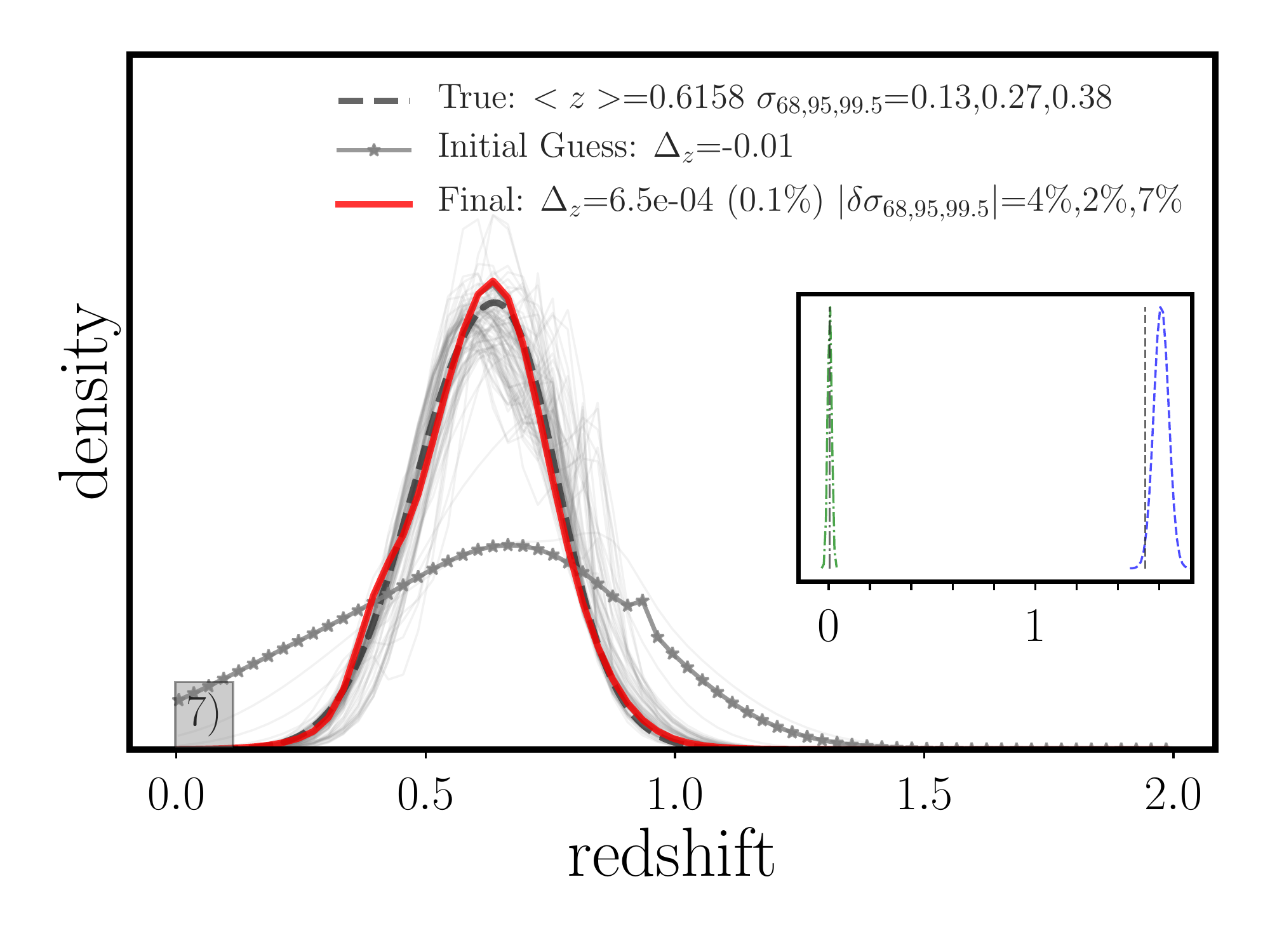}
\includegraphics[scale=0.3125,clip=true,trim=55 85 15 25]{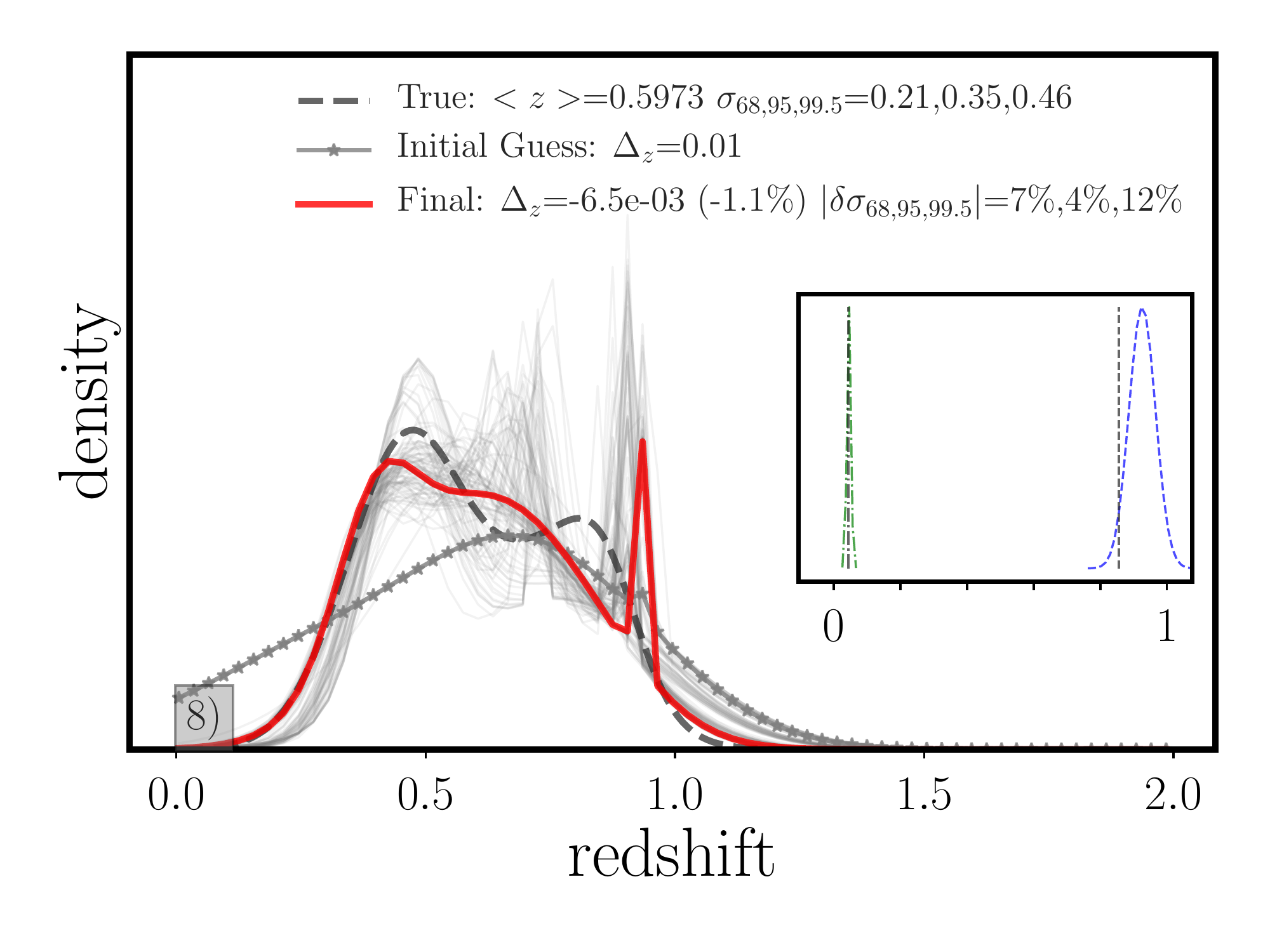}\\
\includegraphics[scale=0.3125,clip=true,trim=15 0 15 25]{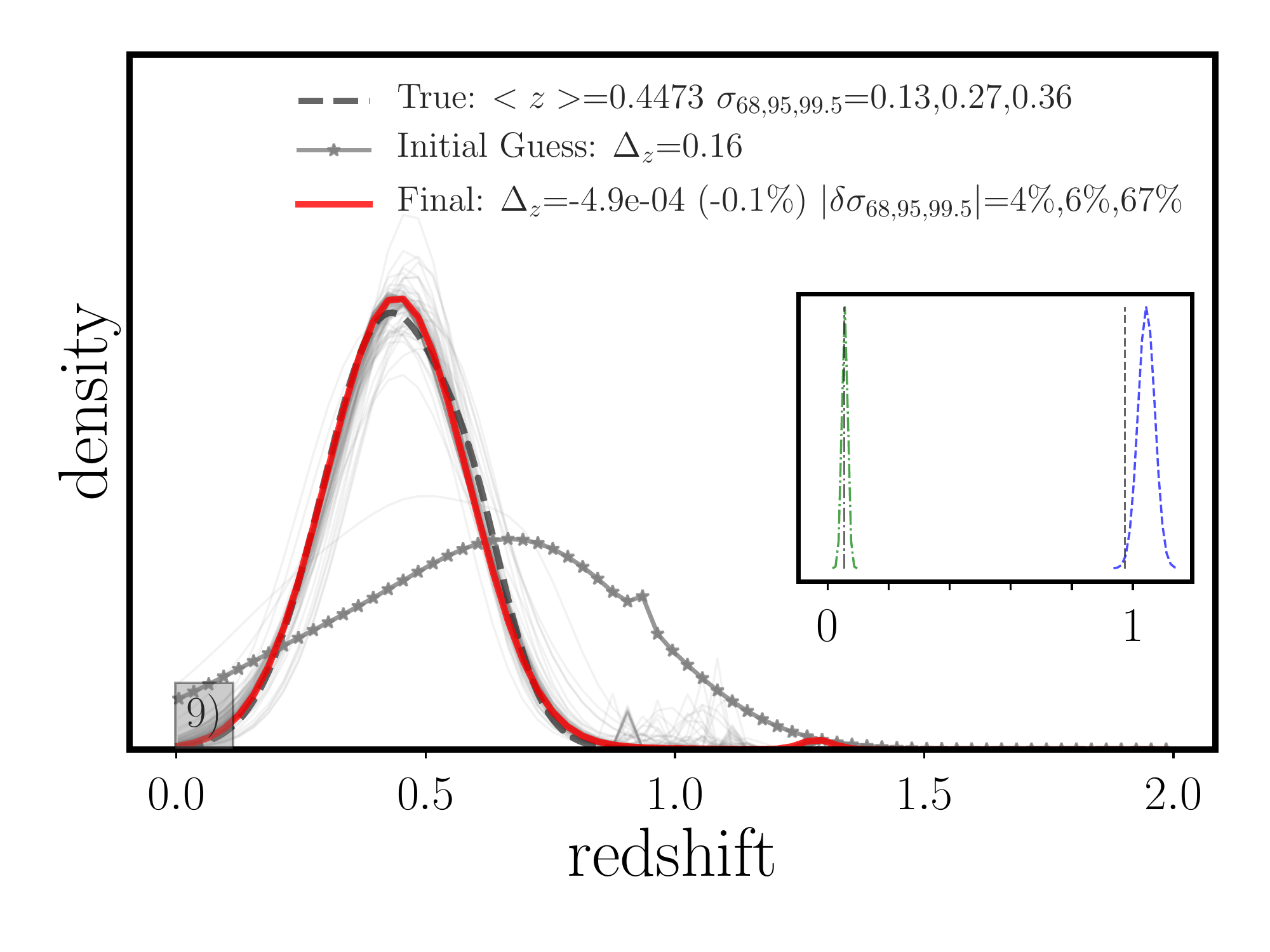}
\includegraphics[scale=0.3125,clip=true,trim=55 0 15 25]{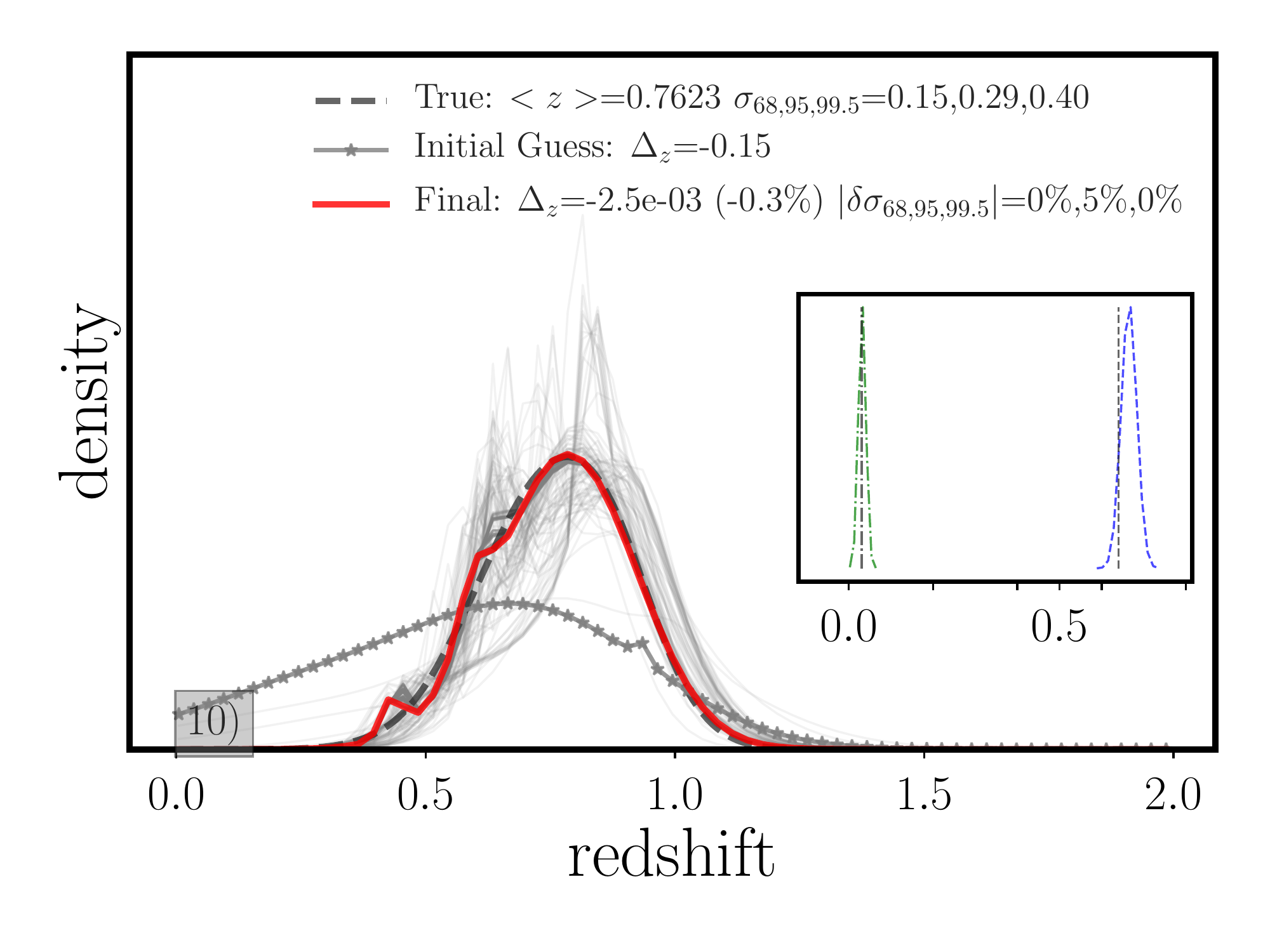}
\includegraphics[scale=0.3125,clip=true,trim=55 0 15 25]{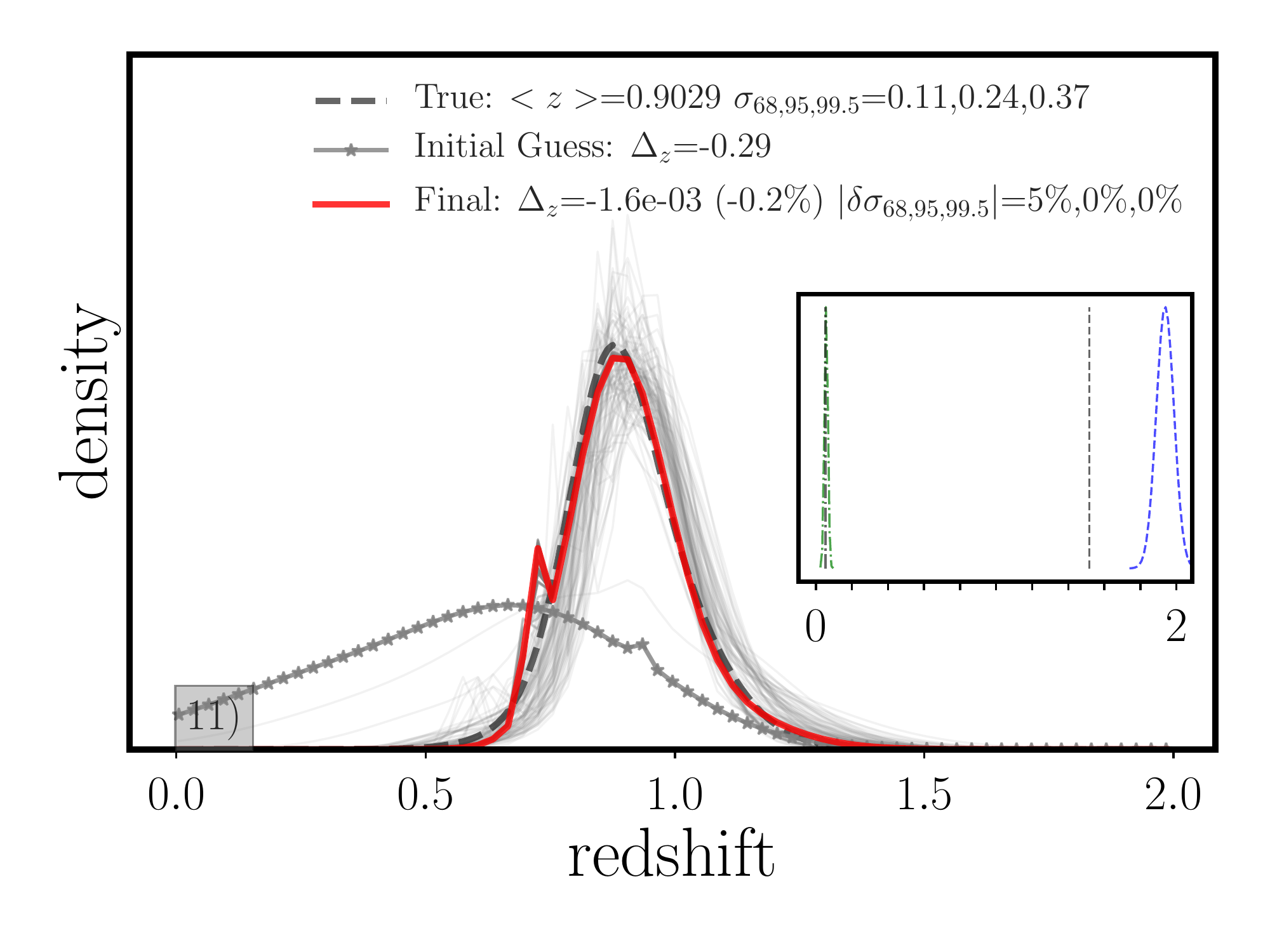}
\caption{The redshift distributions, and galaxy-dark matter bias parameter values as estimated using the {\rrr simulated data vectors which would correspind to the} angular positions of galaxies in each colour cell, the galaxy-galaxy lensing between colour cells, and the correlation of galaxy positions and CMB lensing. This figure shows the results of using {\rrr these idealised data vectors} but with the inclusion of realistic noise properties on the $C_l$'s. This figure can be understood as Fig. \ref{dndnz_ncllclscl}. {\bhP The analysis in these panels corresponds to that described in \S\ref{galgallenserr}}.}
\label{dndnz_sim_ncl_noise}
\end{figure*}

In this section we use the idealised redshift distributions as in \S\ref{gg-lcorr},  but include realistic noise on the measured values of $C_l$. We use the theoretical estimate of $\sigma(C_l)$ given by Equ. \ref{eq:cl_err}, and re-sample the mean value of $C_l$ assuming a Gaussian distribution. We again use {\rrr simulated data vectors corresponding to} all of the galaxy and lensing correlation functions, as per \S\ref{gg-lcorr}. Fig. \ref{dndnz_sim_ncl_noise} shows the estimated redshift distributions and should be understood as Fig. \ref{dndnz_ncllclscl}. In this section we further show how the results are insensitive to the initial guess of the redshift distribution, by using the same initial guess for all colour cells.

We find that this data driven redshift method still performs well when some realistic noise {\rrr is added to the simulated} data vectors. Many of the panels in Fig. \ref{dndnz_sim_ncl_noise} have true redshift distributions, and draws from the posteriors which are almost indistinguishable by eye. The metrics measuring the shape of the distributions are correctly estimated to within a few percent as per \S\ref{gg-lcorr}. In future works we will continue to make the data vectors more closely resemble those obtainable from observational surveys. {\bhM The galaxy-dark matter bias parameters $b_k$ and the composite parameter are also quiet consistent with the truth values within the measured uncertainties.}

\subsection{Summary of results from the correlation function method}
\label{summ_res}
In Table \ref{final_results} we show a summary of performance metrics comparing the true redshift distribution parameters, and the galaxy-dark matter bias model parameters, with those estimated from the posterior distribution using the correlation function method. We present the mean and the 68\% spread of the metric values, as measured between each of the 12 different colour cells. The metrics correspond to the differences between the averages of the redshift distributions $\Delta_z$, and the \% differences between the true and estimated widths, measured at the 68\%, 95\% and 99.5\% quantiles of the distributions. We also show the accuracy at recovering the three galaxy-dark matter bias model parameters, in terms of the standard deviation $\sigma$ as measured from the posterior distributions in each colour cell. Each row corresponds to a different set of analysis using, from top to bottom, {\rrr simulated data vectors corresponding to;} the auto- and cross- correlations (GG) of the galaxy positions in the 12 colour cells, GG with galaxy-galaxy lensing and galaxy-CMB lensing correlation functions. The top three rows show the results for these idealised {\rrr data vectors} without any scatter applied to the $C_l$ correlation function. The third row shows the ability to recover redshift distributions drawn from real SDSS data {\rrr but still using simulated data vectors}. The fourth rows presents the results  {\rrr obtained when errors are placed on the idealised $C_l$ data vectors}. 

\begin{table*}
\centering
\renewcommand{\footnoterule}{}
\begin{tabular}{| c | c | c | c | c | c | c | c |}
\hline
$\Delta_z$ & $\sigma_{68}[\%]$  & $\sigma_{95}[\%]$  & $\sigma_{99.5}[\%]$  & $b_0$ [$\sigma$]  & $b_1$ [$\sigma$]  & $b_k$  [$\sigma$]  & $b^*$ [$\sigma$] \\ \hline
\multicolumn{8}{c}{Idealised sims: using only galaxy-galaxy (GG) $C_l$'s \S\ref{gg-corr}} \\
$0.0405\pm0.0158$&$3.5\pm3.4$&$6.8\pm7.4$&$31.5\pm49.9$& $-5.6\pm8.1$ & $7.2\pm11.9$ & $-1.2\pm0.6$ & $0.1\pm4.8$ \\ \hline
\multicolumn{8}{c}{Idealised sims: using GG and lensing $C_l$'s \S\ref{gg-lcorr}} \\
$-0.0007\pm0.0007$&$1.1\pm1.8$&$2.5\pm2.1$&$6.4\pm3.6$& $-5.9\pm6.7$ & $6.5\pm6.2$ & $0.0\pm0.4$ & $2.2\pm2.9$ \\ \hline
\multicolumn{8}{c}{SDSS like dndz: using GG and lensing $C_l$'s \S\ref{gg-corrl-sdss}} \\
$0.0011\pm0.0055$&$5.1\pm5.0$&$12.7\pm12.3$&$18.9\pm21.9$& $-3.0\pm6.5$ & $4.1\pm7.2$ & $0.7\pm1.1$ & $1.9\pm0.8$ \\ \hline
\multicolumn{8}{c}{Idealised sims with $C_l$ errors: using GG and lensing $C_l$'s \S\ref{galgallenserr}} \\
$-0.0007\pm0.0022$&$3.0\pm2.3$&$3.5\pm1.9$&$10.2\pm5.4$& $-4.2\pm6.5$ & $5.9\pm8.0$ & $0.2\pm0.5$ & $2.5\pm3.1$ \\ \hline
\end{tabular}
\caption{A summary of performance metrics comparing the true redshift distribution parameters, and the galaxy-dark matter bias model parameters, with those estimated from the posterior distribution. We present the mean and the 68\% spread of the metric values, as measured between each of the 12 different colour cells. The metrics correspond to the differences between the averages of the redshift distributions $\Delta_z$, and the \% differences between the true and estimated widths, measured at the 68\%, 95\% and 99.5\% quantiles of the distributions. We also show the accuracy at recovering the three galaxy-dark matter bias model parameters, in terms of the standard deviation $\sigma$ as measured in each colour cell. {\bhN The final row corresponds to the composite parameter $b^*$ c.f. Equ. \ref{eq:bias1}, calculated at the mean of the reshift distribution $\bar{z}$}. Each row corresponds to a different set of analysis using, from top to bottom, {\rrr simulated data vectors that would correspond to;} auto- and cross- correlations (GG) of the galaxy positions in 12 colour cells, GG with galaxy-galaxy lensing and galaxy-CMB lensing correlation functions. The top two, and last rows show the results for idealised {\rrr data vectors} without any scatter applied to the $C_l$ correlation functions. The third rows presents the results using idealised {\rrr data vectors}  with expected scatter on the $C_l$ measurements. The fourth row shows the ability to recover redshift distributions drawn from real SDSS data.}
\label{final_results}
\end{table*}

Concentrating on the metric values as measured on the redshift distributions in the top two rows of Table \ref{final_results}, we find that the uncertainty on the $\Delta_z$ metric is improved by almost two orders of magnitude when considering all the galaxy-galaxy and galaxy-galaxy lensing and galaxy-CMB lensing correlation functions, compared to only the galaxy-galaxy correlation functions. The uncertainties on the metrics describing the widths of the distributions are likewise improved from a few factors for the innermost quantiles $\sigma_{68}$, to an order of magnitude for the outermost quantiles $\sigma_{99.5}$. This shows that the additional correlation functions provide levers which produce more sensitivity to the outer shapes of the redshift distributions. 

We find that in each of the analyses, the initial starting values of each of the parameters do not play a role in the ability of the method to estimate reasonable redshift distributions.

\subsection{Evolution of the $\chi^2$ function}
\label{evol_chi2}
\begin{figure}
\includegraphics[scale=0.4,clip=true,trim=5 10 15 15]{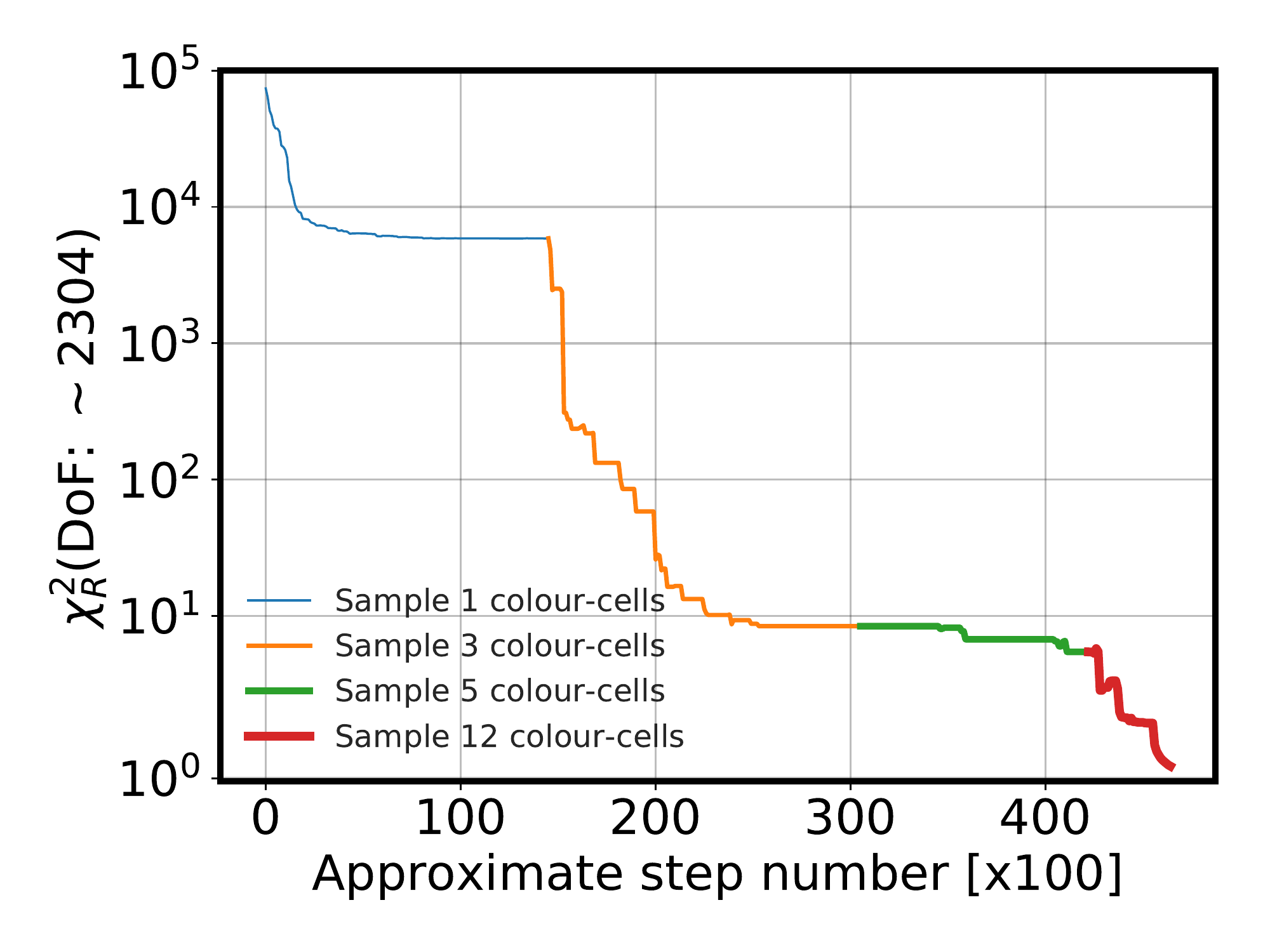}
\caption{Exploring the multi-dimensional parameter space describing the redshift parametrisation and galaxy-dark matter bias model using Gibbs sampling. The value of the reduced $\chi^2$ with approximately 2304 degrees of freedom is shown as a function of the approximate number of Gibbs steps. {\bhP The different line thickness and colours highlight which parts of the distributions present results during the exploration of parameter space using a different number of colour cells.}}
\label{chi2_chains}
\end{figure}

To speed up the exploration of the high dimensional parameter space, we initially walk through parameter space using the predicted and {\rrr idealised} correlation functions associated to only one colour cell. This process is parallelisable for each of the N colour cells. We use the evolution of the $\chi^2$ value, as evaluated after approximately 100 calls to the likelihood function using all of the N colour cells and including all of the available auto- and cross- correlations, as a guide to determine when is an appropriate time to add more colour cells into the analysis. In Fig. \ref{chi2_chains} we show the evolution of the reduced value of $\chi^2$, denoted by $\chi^2_R$ as a function of the number of steps though the high dimensional parameter space {\bhN for the analysis presented in \S\ref{gg-lcorr}.} For the full analysis there are 12*18 parameters and 2520 datapoints (using the chosen $\Delta l$ bin width). This results in the number of degrees of freedom equal to approximately 2304 \citep[see however][for a discussion about the numbers of degrees of freedom in a non-linear model]{2010arXiv1012.3754A}. Recall that the initial random guess will be a very poor fit to each redshift distribution, and for all of the galaxy-dark matter parameter values in each colour cell, therefore we expect the inital $\chi^2$ value to be very large. 

After the $\chi^2_R$ stabilises during the exploration of only one colour cell, we then randomly select three out of N colour cells and continue the search for good parameter values. This increase in the number of colour cells increases the computational time not only because of the increase in the total number of calls to the likelihood function that is required to perform one complete Gibbs sampling, but also in the total time for each individual call to the likelihood function. Again this process is heavily parallelisable, however ocassionaly the same colour cell may be chosen and updated simultaneously. This can produce large wiggles in the $\chi^2_R$ correlation function as seen in Fig. \ref{chi2_chains}. Furthermore, we adjust the thickness and colour of each section of the line in  Fig. \ref{chi2_chains} {\bhP to approximately highlight the stages of the analysis when the number of colour-cells increased. }

We then randomly choose five of the N colour cells and repeat the procedure. Finally we use full set of 12 colour cells in the analysis. Each complete (partial block) Gibbs sample takes of order 2.5 hours (20 minutes) using a machine with 16 CPU cores. We note that all of the other analyses presented here have very similar evolving $\chi^2_R$ functions.

\section{Discussion}
\label{discussion}

\label{discuss}

The size of available and upcoming photometric datasets are ripe for such an analysis described in this work. For example the SDSS has observed the photometric properties of order of one hundred million galaxies, and DES \citep[][]{2005astro.ph.10346T} has order 500 million galaxies \citep[][]{DESDR1}, and Euclid \citep[][]{2011arXiv1110.3193L} will have order 3 thousand million galaxies. In this section we discuss difficulties that may occur when applying this method, and some future prospects.

\subsection{Clean galaxy samples}
In this paper we have not addressed problems that arise when galaxy samples are not pure and complete. We have neglected the problems of star-galaxy separation, and of variations in galaxy number densities across a survey due, for example, to seeing. In practice some amount of stellar contamination of a galaxy data set will result in additional power in the projected correlation function at the largest scales. We could rely on methods to perform star-galaxy separation, such as template or machine learning codes \citep[][]{2013arXiv1306.5236S,2017arXiv171203970M}. Alternatively one could add nuisance parameters to the measurement of the correlations functions describing the amount of stellar contamination \citep[see, e.g.][]{2017arXiv171206211C}. This would further increase the number of parameters in the analyses, but only by a few per colour cell.

\subsection{How many colour cells should be chosen?}
\label{ncolourcells}
In this work we have concentrated on using twelve colour cells, however this number could be pushed arbitrarily large, at the expense of computing resources. We chose twelve colour cells using the following, very soft argument.

Given that we would like to constrain 18 parameters per colour cell, consisting of 15 redshift distribution parameters, and 3 galaxy-dark matter bias model parameters, we would like to have at least as many different measurements as parameters. In fact, of the 15 redshift distribution parameters, only 14 are independent due to the enforcement that the integral of the redshift distribution is unity. Furthermore, the estimation of the number of degrees of freedom of a non-linear model is non-trivial \citep[see][]{2010arXiv1012.3754A}, but we ignore these details in what follows.

Therefore there are $N\times 18$ parameters to estimate for the $N$ colour cells. Concentrating on the available number of galaxy-galaxy auto- and cross- correlation functions, we have $N$ auto-correlations, and ($N-1$) different cross-correlations for the first colour cell, ($N-2$) for the second colour cell etc. The number of different correlation functions scales as $N*(N+1)/2$. This means there will be approximately equality between the number of unknowns and possible measurements with $N$=36 colour cells. 

In this work we also include galaxy-galaxy lensing correlations, which adds another factor $N*(N+1)/2$, and CMB lensing which provides another $N$ correlation functions, this reduces the required number of colour cells to $N$=16. Until now we have neglected the number of data points within each correlation function, which in this work, we have chosen to use bins of {\rr width} $\Delta l=15$ between the range $20<l<1000$, resulting in 65 data points. We find that using 12 colour cells there are an order of magnitude more data than parameters to be estimated. 

We have not explored a larger number of colour cells because the computation time of each call to the likelihood function, and each full Gibbs sample, scales very poorly with $N$.  However we have also explored using less than 12 colour cells. We find that using 2 or 4 colour cells results in reasonable estimate for the difference in the mean of the distributions ($\Delta_z=[0.1\pm0.06, 0.06\pm0.03]$ respectively), but that measurements of the width of the distributions are less constraining, and that the estimation of the galaxy-dark matter bias model parameters are poorly constrained. The complete results of these analyses are shown in Table. \ref{final_results24}, but we caution that the quoted estimate of the 1-$\sigma$ error bars is large, given the small number (2 and 4) of samples. 

We can compare the estimate of the error on the offset of the mean of the distribution $\sigma(\Delta_z)=0.03$, coming from this idealised analysis using 4 colour-cells in which no spectroscopic or photometric redshift data is used, and which marginalises over galaxy-dark matter bias parameters, with that of the results from the Dark Energy Survey \citep{2005astro.ph.10346T} who selected data to fall within 4 tomographic redshift bins and produce an estimate of $\sigma(\Delta_z) \sim 0.02$ \citep[][]{2017arXiv170801532H} by carefully matching the science sample to galaxies with 30-band COSMOS data, and accounting for cosmic variance using simulations. If we increase the number of colour-cells to 12, the measurement of $\sigma(\Delta_z)$ in the idealised work presented here, which is $\sim 7\times10^{-4}$, appears very competitive and suggests this method is worth further investigation.

\begin{table*}
\centering
\renewcommand{\footnoterule}{}
\begin{tabular}{| c | c | c | c | c | c | c | c |}
\hline
$\Delta_z$ & $\sigma_{68}[\%]$  & $\sigma_{95}[\%]$  & $\sigma_{99.5}[\%]$  & $b_0$ [$\sigma$]  & $b_1$ [$\sigma$]  & $b_k$  [$\sigma$] & $b^*$ [$\sigma$]  \\ \hline
\multicolumn{8}{c}{Idealised sims: using GG and lensing $C_l$'s with only 2 colour cells} \\
$-0.099\pm0.060$&$59.6\pm10.8$&$85.7\pm30.0$&$44.4\pm24.1$& $-287.0\pm104.1$ & $1515.8\pm12.2$ & $47.3\pm2.8$ & $332.5\pm94.0$ \\ \hline
\multicolumn{8}{c}{Idealised sims: using GG and lensing $C_l$'s with only 4 colour cells} \\
$-0.063\pm0.032$&$12.8\pm13.0$&$19.5\pm18.9$&$16.4\pm9.1$& $10.3\pm108.6$ & $21.4\pm48.0$ & $48.7\pm15.0$ & $31.0\pm9.7$ \\ \hline \hline
\multicolumn{8}{c}{Idealised sims: using GG and lensing $C_l$'s with 12 colour cells around the global minimum} \\
$0.0000\pm0.0001$&$0.9\pm1.8$&$0.3\pm0.9$&$0.5\pm0.7$& $-0.1\pm1.3$ & $0.0\pm2.0$ & $-0.1\pm1.5$ & $0.1\pm1.1$ \\ \hline
\end{tabular}
\caption{As per Table \ref{final_results}, but with the results of this correlation method applied to only {\bh 2 (4) colours cells first (second) row from random initial parameter values. The final row presents the obtainable constraints found by starting the Gibbs sampling around the global minimum and using 12 colour cells. All galaxy-galaxy and galaxy-galaxy-lensing and galaxy-CMB lensing correlation functions are used in the analyses producing these results. {\bhN The uncertainties presented in the top two rows are estimated from the low number (2 and 4) of colour cells.}}}
\label{final_results24}
\end{table*}

\subsection{Parameter space convergence}
\label{space_conv}
Even with the approaches to speed up this method as described in \S\ref{highd_space} and \S\ref{evol_chi2}, it is still computationally expensive. Indeed the chains are still slowly rolling down the $\chi^2$ surface and converging. Given that many of the experiments did not include noise in the correlation functions, we can exactly examine how well these codes are performing by exploring how close the value of $\chi^2$ is from 0. For example, if we concentrate on the analysis presented in \S\ref{gg-lcorr}, we find that we are close to, but still approaching, this global minimum. However we expect the results presented in this work will only improve with further iterations through the high dimensional parameter space {\rr as the chains converge around the best fit parameter values}. 

\begin{figure}
\includegraphics[scale=0.35,clip=true,trim=0 20 25 25]{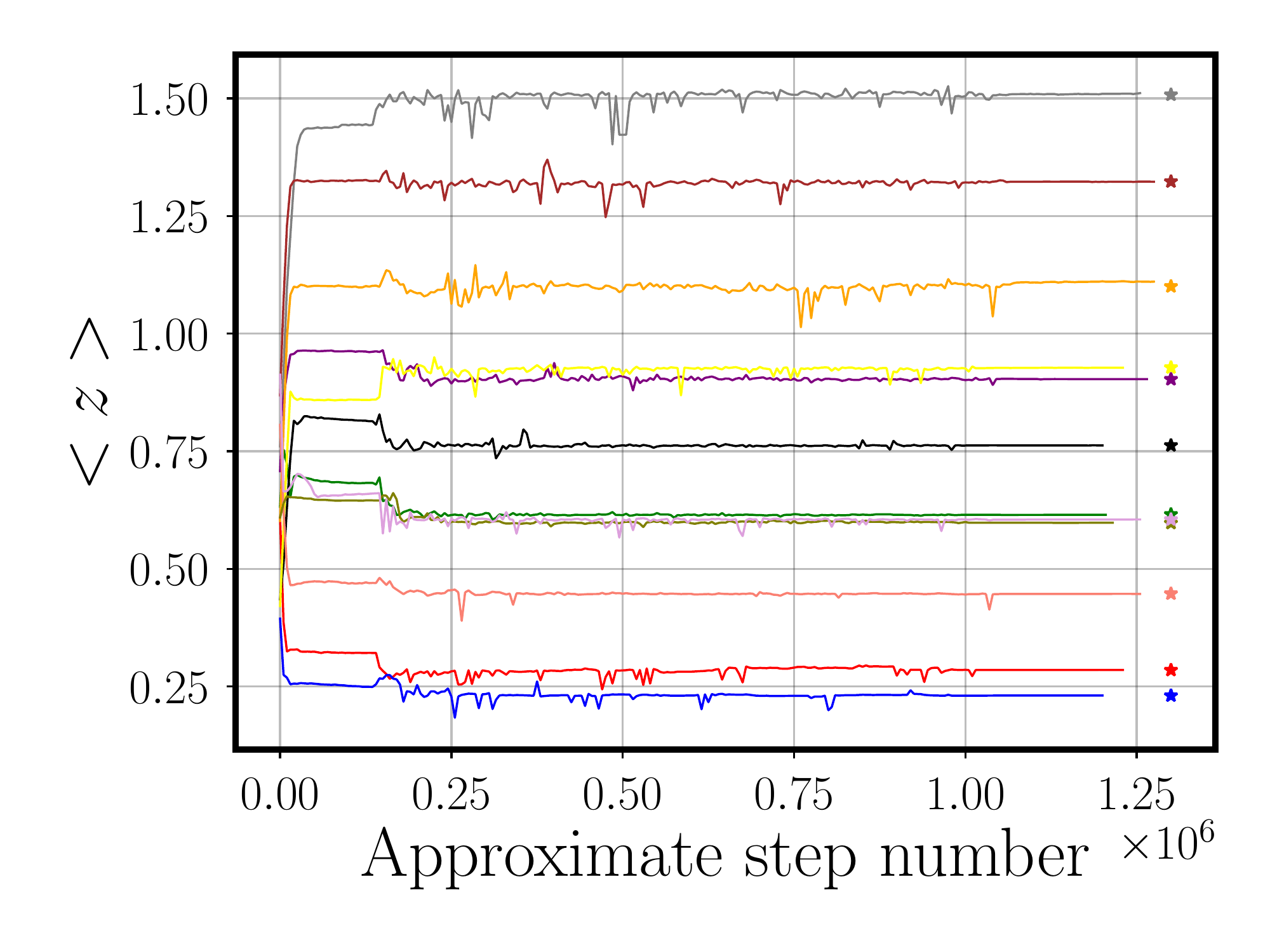}
\includegraphics[scale=0.375,clip=true,trim=20 20 25 25]{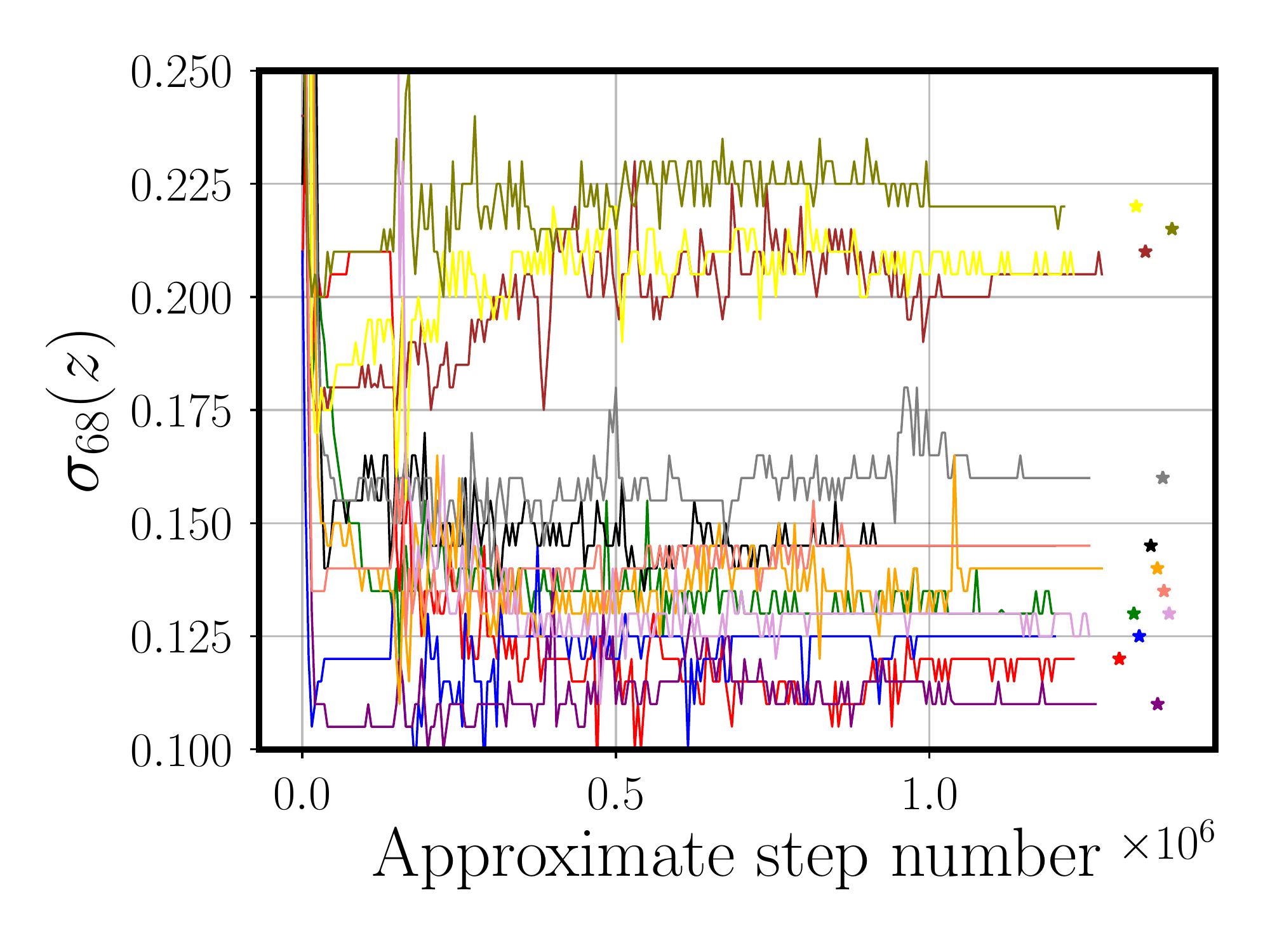}
\includegraphics[scale=0.375,clip=true,trim=20 20 25 25]{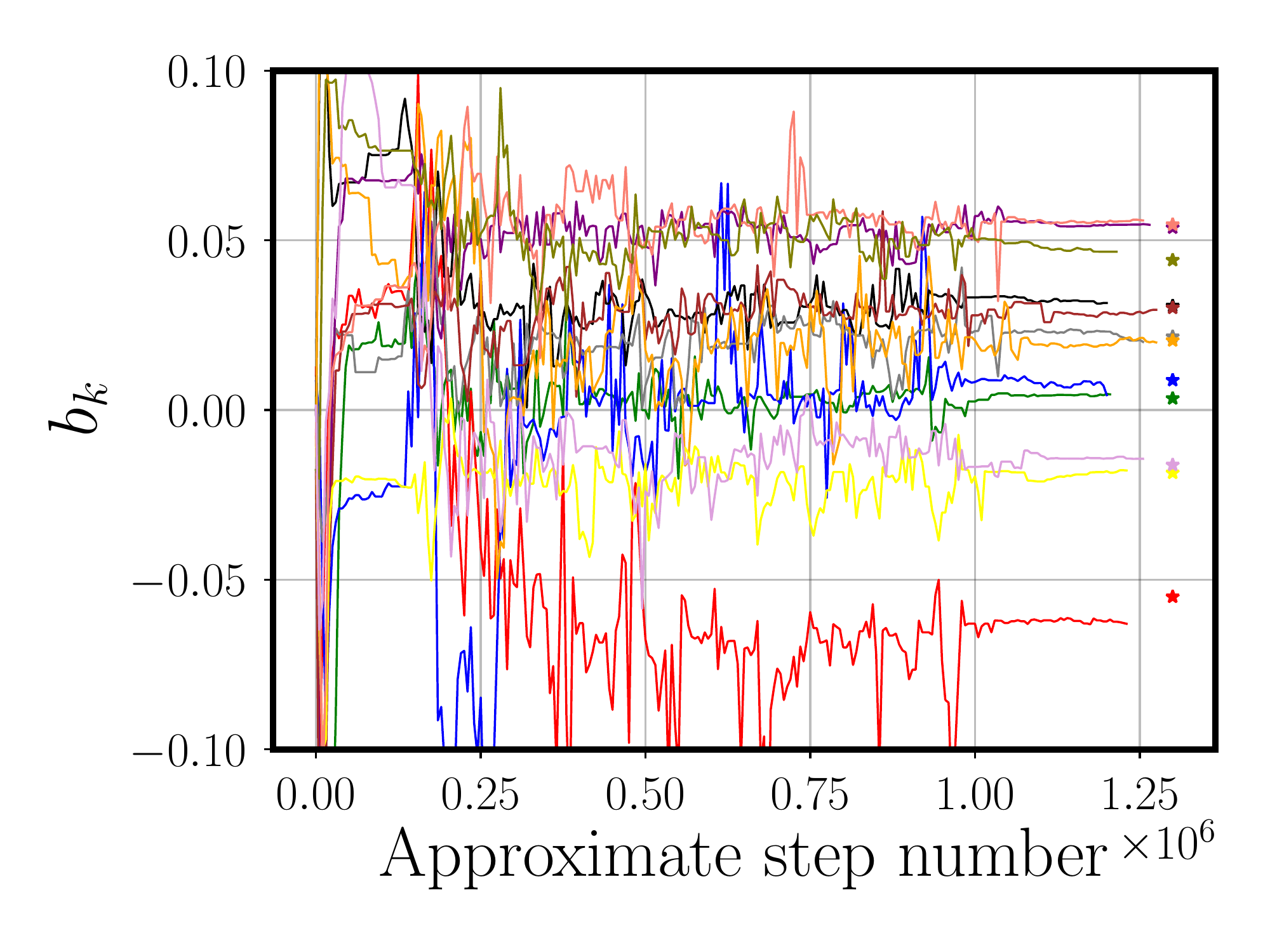}
\includegraphics[scale=0.35,clip=true,trim=-15 20 25 25]{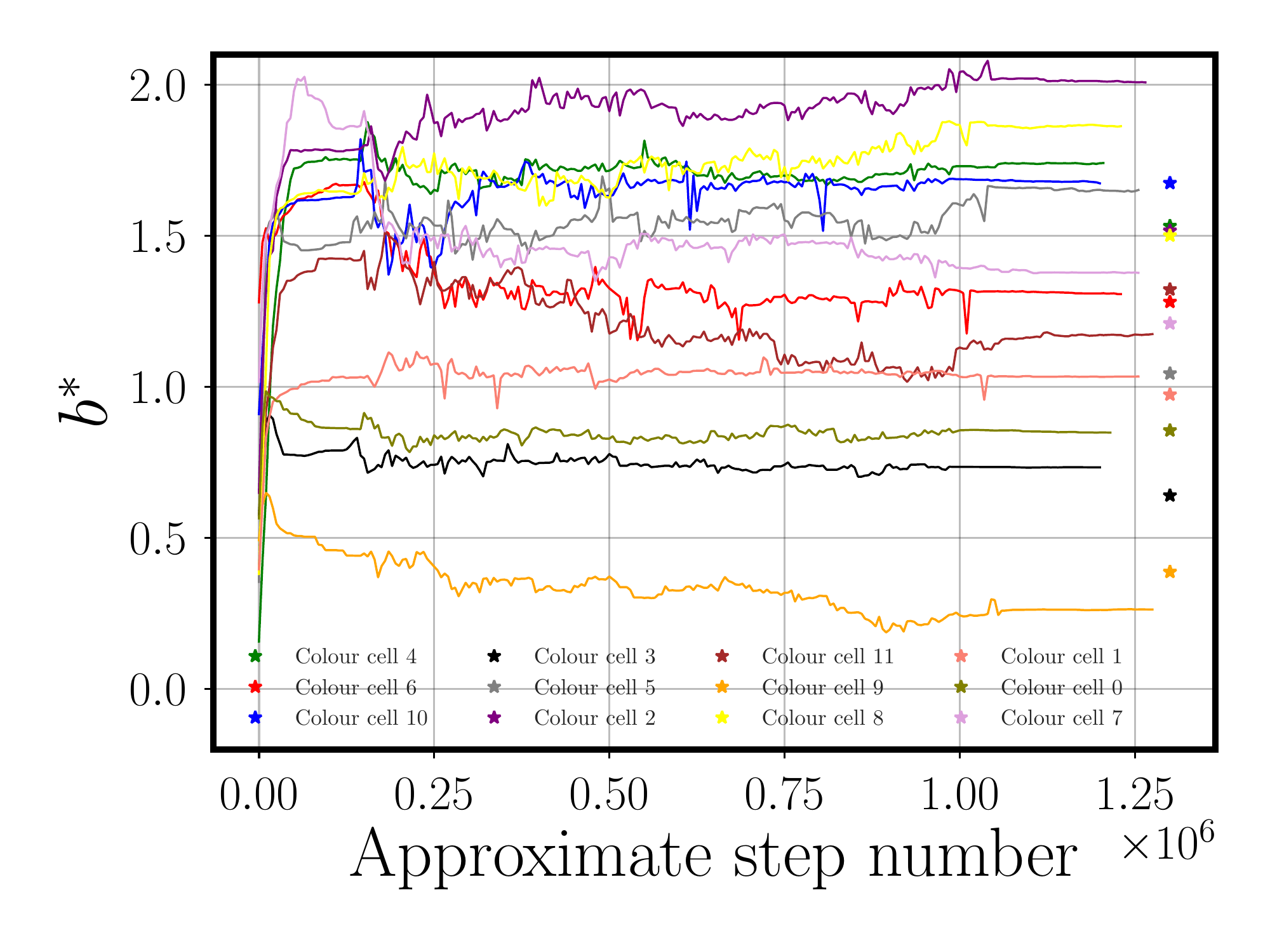}
\caption{Trace plots showing the evolution of the mean and 68\% width of the redshift distributions, and the galaxy-dark matter bias model parameters $b_k$ and the composite parameter $b^*$ (see Equ.\ref{eq:bias1}) for each colour cell, as the code explores the high dimensional parameter space. The truth value is shown by an offset star of the same colour. The analysis in these plots refers to that described in \S\ref{gg-lcorr}.}
\label{evol_chi2_dndz}
\end{figure}

We motivate this statement by examining the evolution of various metrics which are measured on the redshift distribution, and on the galaxy-dark matter bias model parameters in each colour cell, as a function of the Gibbs sampling step number. In Fig. \ref{evol_chi2_dndz} we show the evolution from their initial starting positions of the mean $<z>$, and the 68\% width of the redshift distribution $\sigma_{68}$, the scale dependent galaxy-dark matter bias model parameters $b_k$, and the composite parameter $b^*$. The truth value for these parameters are either measured from the true redshift distributions, or are parameter values assigned to each colour cell. We mark the corresponding truth values on Fig. \ref{evol_chi2_dndz} with the starred data points, which are offset from the end of the curves for viewing purposes.
 
Examining Fig. \ref{evol_chi2_dndz}, we see that many metrics are converging towards the truth values, for example, the redshift metrics, and the galaxy-dark matter bias parameter $b_k$. The most individually discrepant parameter is the redshift dependent galaxy-dark matter bias parameter $b_1$, but as discussed in \S\ref{summ_res} this is to be expected for our choice of galaxy-dark matter model and true $b_1$ values. We do however find that the composite parameter $b^*$ is well constrained and approaches the truth values. The lines in each of the trace plots have varying length due to the random selection of colour cells in the early analysis, before moving to using all 12 colour cells, and therefore not all cells have been selected the same number of times.

\subsection{Exploring around the global minimum}
\label{global_min}
Assuming that the algorithm is able to correctly identify the true global minimum in the high dimensional parameter space, then a Markov walk around this position will constrain each parameter and the covariances between parameters. To explore how this final result may look for our chosen data sample {\bhh we run a Gibbs sampler starting at the true global minimum} in parameter space, and used the data sets and correlation functions as described in \S\ref{gg-lcorr}. We present the results of this analysis in the last row of Table. \ref{final_results24}. 

We find that exploring around the minimum of the $\chi^2$ function provides parameter results consistent within $\approx 1\sigma$ of the truth values for all parameters, including the parameter $b_1$. The uncertainties on the parameters are a few factors smaller than those presented in \S\ref{gg-lcorr}, which further highlight that the chains are still exploring the high dimensional parameter space {\rr and have yet to fully converge}.

We construct similar trace plots as Fig. \ref{evol_chi2_dndz}, showing the evolution of the mean and widths of the redshift distributions, and the scale dependent and composite galaxy-dark matter bias parameters, but with the Gibbs sampler initiated at the global minimum.
\begin{figure}
\includegraphics[scale=0.35,clip=true,trim=0 20 25 25]{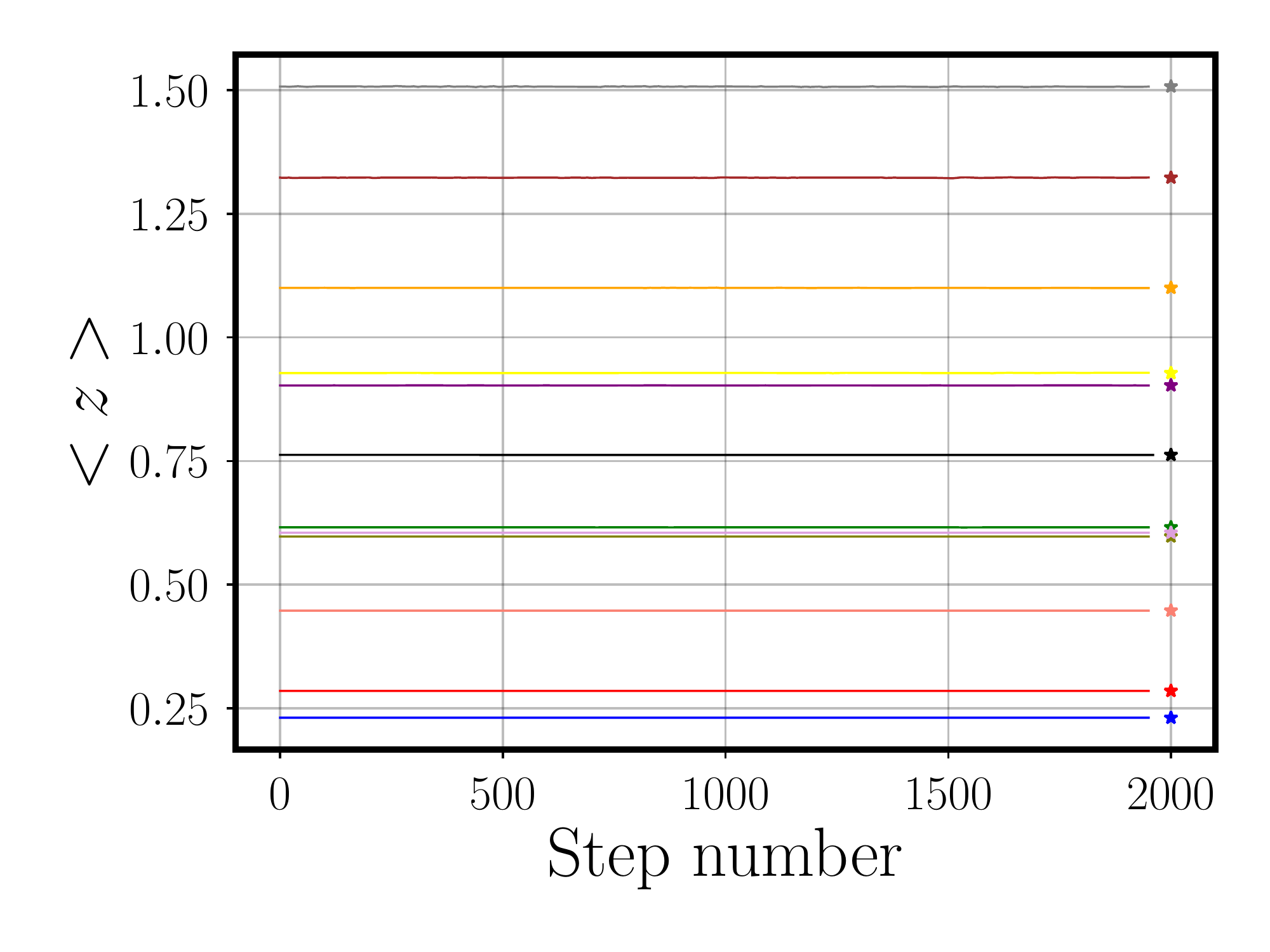}
\includegraphics[scale=0.36,clip=true,trim=20 20 25 25]{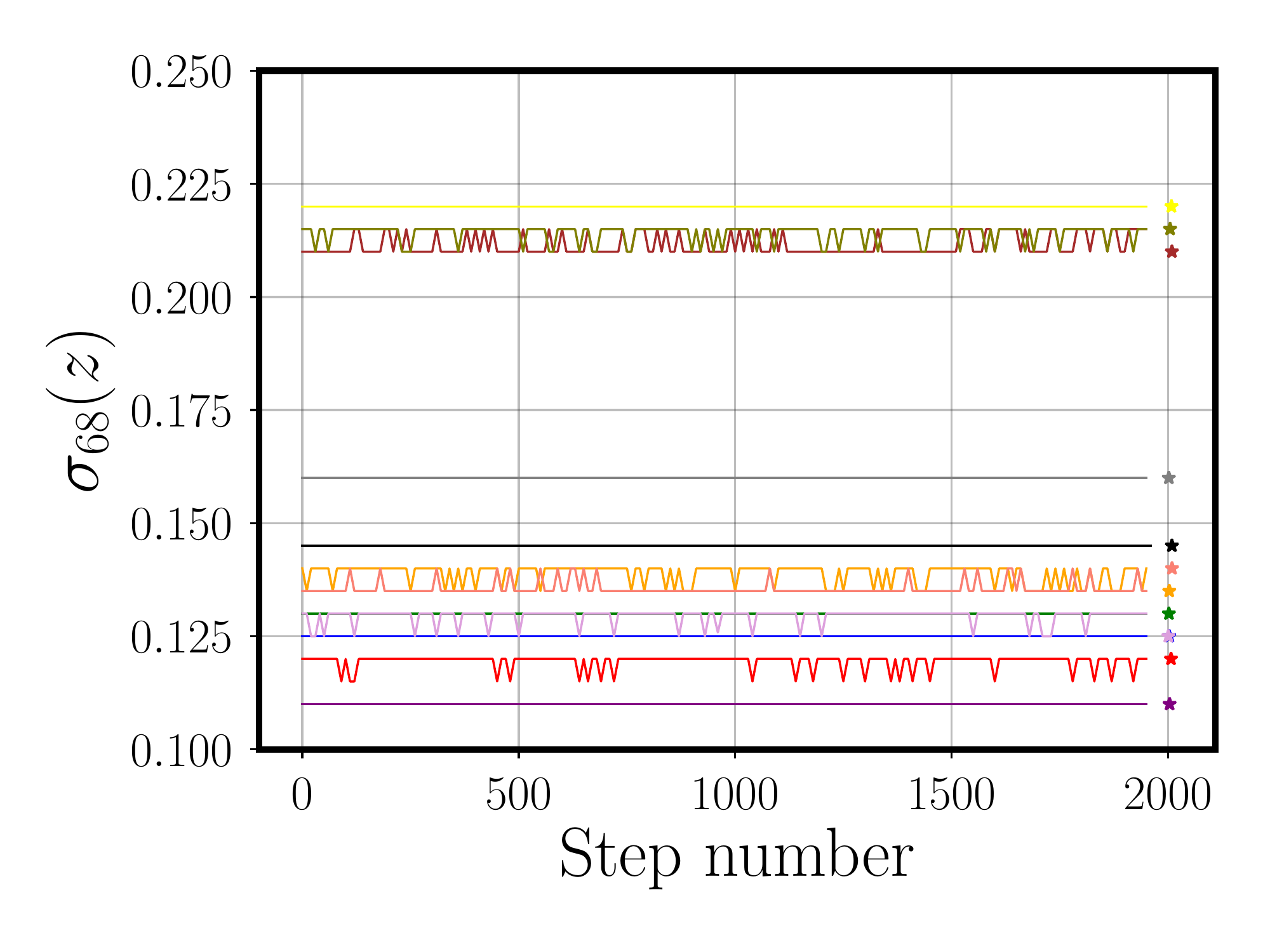}
\includegraphics[scale=0.36,clip=true,trim=20 20 25 25]{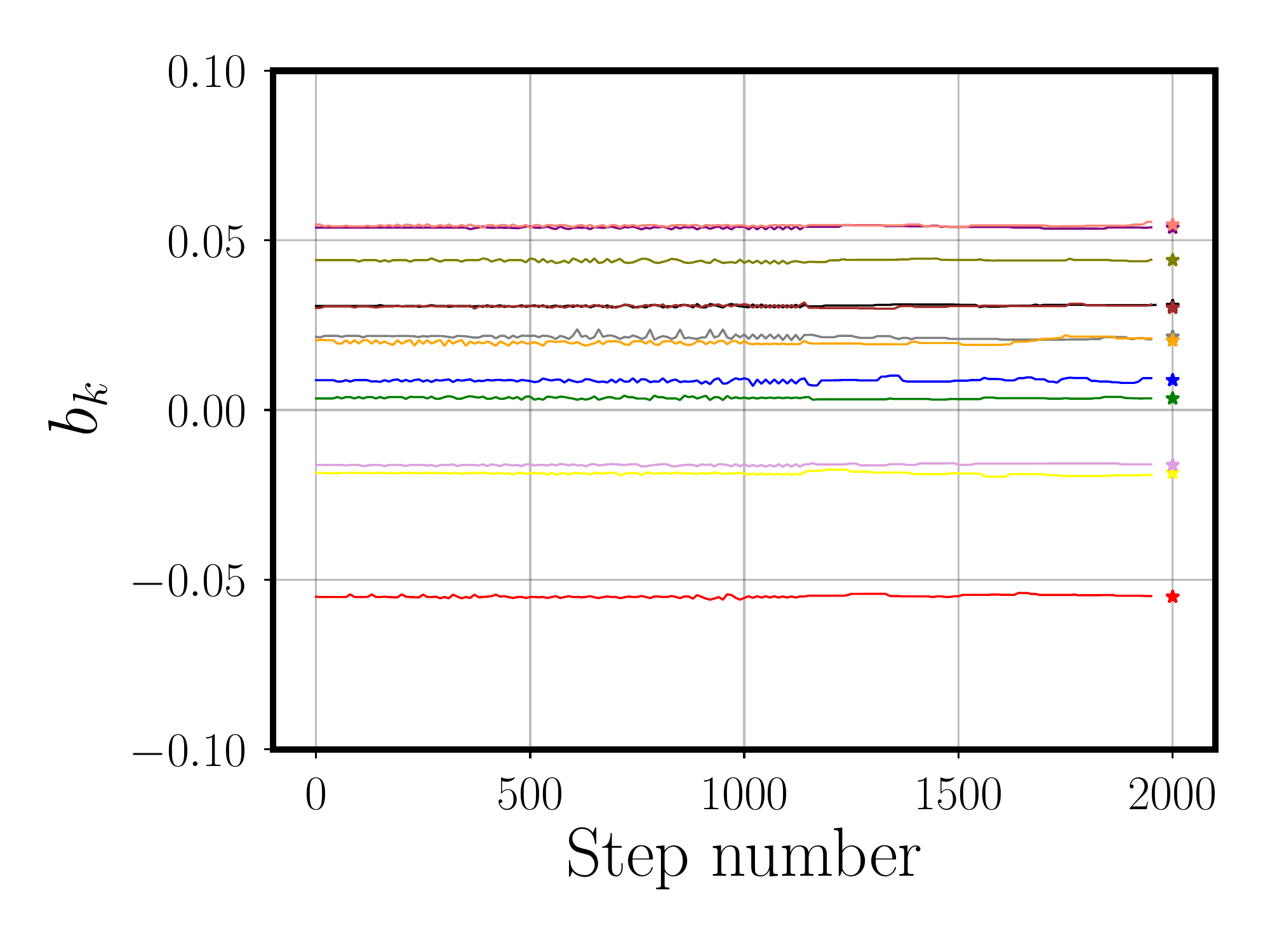}
\includegraphics[scale=0.35,clip=true,trim=-15 20 25 25]{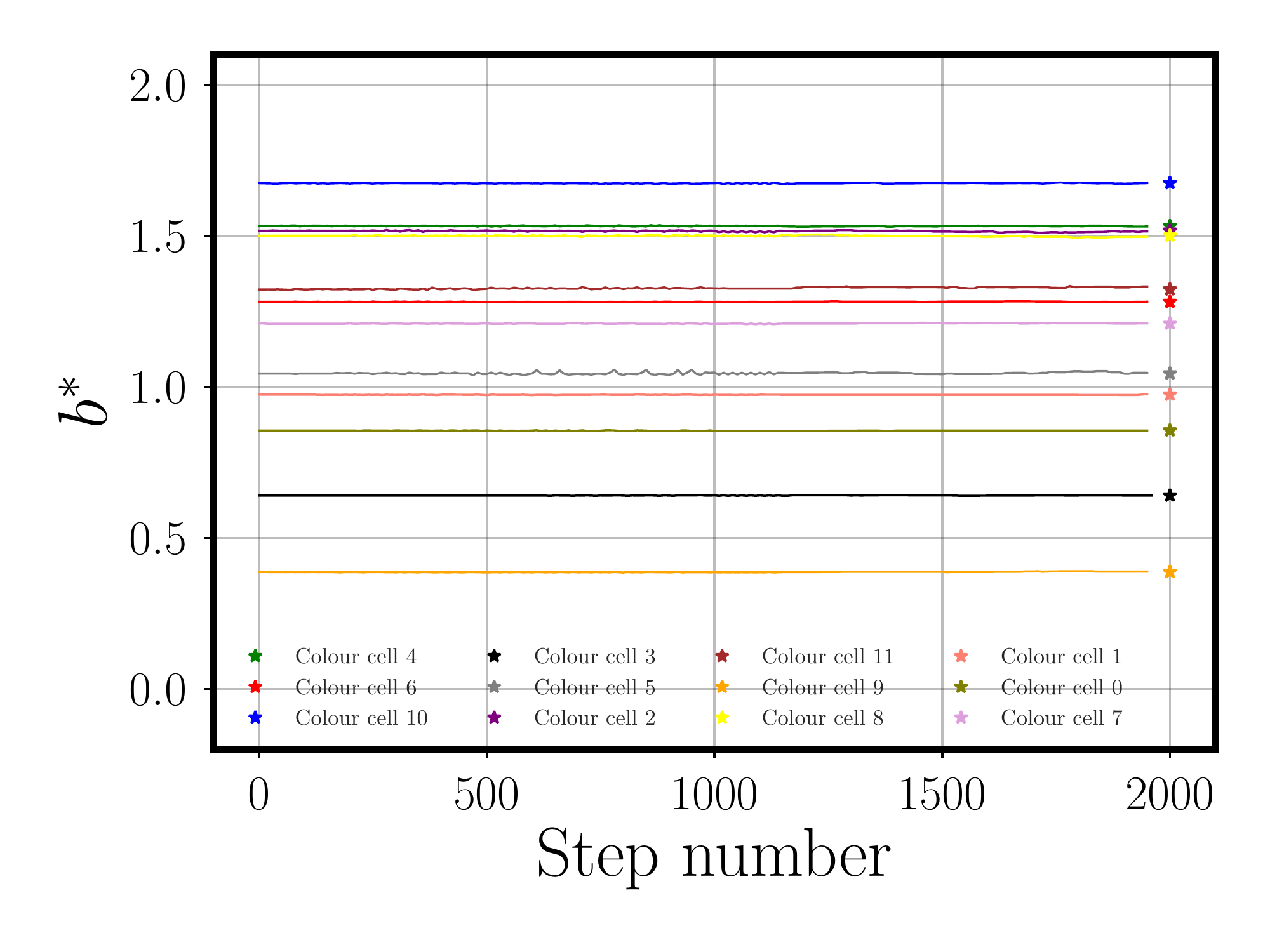}
\caption{Similar trace plots as Fig. \ref{evol_chi2_dndz}, but starting from the global minimum in parameter space. The analysis in these plots refers to that described in the discussion section \S\ref{global_min}.}
\label{evol_chi2_dndz_min}
\end{figure}

Examining Fig. \ref{evol_chi2_dndz_min}, we see that all metrics are stable around the truth values, as also seen by the summary statistics in the last row of Table \ref{final_results24}. These are the results we would expect to find after hitting the global minimum in the parameter space, even if we start from a random point in the parameter space, given a robust sampler. We note that the mean of the redshift distribution seems to be constant. It does however vary but on a scale much to small to be seen at this resolution. This further highlights how well this parameter can be constrained by the redshift estimation technique presented here.

\subsection{Redshift pdfs for individual galaxies}
This method does not provide redshift probability distribution function (pdf) estimates for individual galaxies. One could imagine making the colour cells very small such that each cell contains only a few galaxies. The resulting distribution would then approach an individual galaxy redshift pdf. However the noise in the correlation function would become very {\bh dominant}, and the length of time required to make calls to the likelihood code would make this approach infeasible in practice.

One could however partition the data space many times using many different decision trees, and perform the analysis described in this work many times. This approach would mimic a random forest from machine learning, in so far as each colour cell, that a galaxy resided in, could contribute to the final redshift pdf of that galaxy. The testing and implementation of this method is left to future work. We note that this approach would be trivially parallelizable.

\subsection{A more generalised sample selection}
The paper has concentrated on the selection of galaxy samples using simple bins in photometric colour space. However any such binning in any observable space is possible, for example the selecting of objects which sit in photometric redshift bins \citep[see, e.g.][]{2017MNRAS.466.3558M}. One may also select objects selected in bins of quantities observed in the radio or H1 continuum \citep[see, e.g.][]{NVSS,SKA}. The most important choice of observable is that the quantity must be uniformly measured (or calibrated) across an observed footprint, otherwise the correlation signal would become corrupted. We note that a smaller footprint would lead to larger uncertainties on the correlation functions due to cosmic variance, however these uncertainties will be naturally propagated though to the uncertainties on redshift distributions, and galaxy-dark matter bias model parameters.

\label{conclusions}

\section{Summary and Conclusions} 
\label{conclusions}

In this paper we {\rrr showcase} a new method to estimate redshift distributions and galaxy-dark matter bias model parameter values for sub-samples of galaxies without requiring any knowledge of spectroscopic or photometric redshifts anywhere in the analysis. This method compares the observed projected correlation functions with predicted correlation functions generated using initial guesses for the redshift distribution and galaxy-dark matter bias parameter values. The redshift distribution and galaxy-dark matter bias parameter values are adjusted until the predicted and observed correlation functions agree statistically. This method is equivalent to marginalising over either or both, of the redshift distributions and the galaxy-dark matter bias model parameters for each sub-sample of galaxies. Although this method is computationally very expensive, it does provide joint constraints on redshift distributions and on galaxy-dark matter bias parameter values for each sub-sample of galaxies. The method also naturally incorporates any apriori information one may have about either the redshift distributions, or galaxy-dark matter bias parameter values in each sub-sample in the form of good initial guesses. Most importantly this method does not require any spectroscopic or photometric redshift information apriori.

We propose to move away from the current methodology of redshift estimation and validation, which rely upon samples of galaxies with either spectroscopic redshifts or high quality multi-band photometric redshifts, because of the difficulty of obtaining numerous, high quality samples which are representative of the interesting science samples coming from current and upcoming large area photometric surveys (e.g., KIDS: \cite{2016arXiv160605338H}, DES: \cite{2005astro.ph.10346T}, Euclid: \cite{2011arXiv1110.3193L}, LSST: \cite{2009arXiv0912.0201L}, and current and future large area radio surveys, e.g., LOFAR: \citep{LOFAR} and the SKA: \cite{SKA}). This lack of representative high quality redshifts is also the reason we suggest using any previous knowledge about parameters for any sub-sample of galaxies as initial best guesses, {\bhh or potentially as priors, if the priors have been constructed with unbiased samples.}

This leads us to choose to rely on the projected auto- and cross- correlation functions which are measurable without any knowledge of either redshifts or of galaxy-dark matter bias parameter values. We compare these measured ({\rrr actually in this work, {\it idealised})} correlation functions with those predicted from cosmological codes, such as obtained from the Cosmic Linear Anisotropy Solving System \citep[{\tt CLASS},][]{2011arXiv1104.2932L} or {\tt CAMB} \citep[][]{CAMB}. 

We model the redshift distribution of each sub-sample of galaxies using a five component Gaussian Mixture Model, and furthermore adopt a flexible three parameter galaxy-dark matter bias model for each sub-sample when subdivided into cells of color. We show that the five component Gaussian Mixture Model is flexible enough to recover the redshift distributions of {\rr different sub-samples of SDSS} galaxies.  We implement a galaxy-dark matter bias model in {\tt CLASS} similar to that of \cite{1538-4357-461-2-L65}, but with an additional scale dependent pre-factor, and allow the parameter values of this model to vary independently for each color cell. {\bhN We find that this is actually a poor model for the chosen true parameter values and redshift distributions, and we suggest that any other well motivated and flexible galaxy-dark matter bias model could be adopted, for example that of \cite{2015MNRAS.448.1389C}. However we find a composite parameter being a combination of the scale independent and redshift dependent galaxy-dark matter bias model parameters, is well constrained by our analyses.}

We use the standard $\chi^2$ metric to explore the range of statistically consistent model correlation functions, and by identifying the aforementioned parameters that generate those consistent correlation functions, we are able to infer the posterior distributions of redshifts and galaxy-dark matter bias model parameters for each sub-sample.  By including auto- and cross- correlation functions from both galaxy-galaxy lensing and galaxy-CMB lensing, we find that we are able to improve the uncertainties on the allowable ranges of the redshift posterior distributions by an order of magnitude, compared to the analysis performed using only galaxy-galaxy positional auto- and cross- correlations.

In this paper we use idealised {\rrr data vectors} in which knowledge of all parameters is known. This is not directly possible even using N-body simulations populated by galaxies, because is it unclear from first principles how the intrinsic galaxy-dark matter bias will be interpreted by the adopted model, although this could be measured using correlation functions and adopting the known redshift distributions. Furthermore in real data we may not be completely confident of the accuracy of the redshift distributions.

We highlighte a method to select galaxy sub-samples in SDSS data, using bins in color-magnitude space, which we generically refer to as color cells. We note that any method to select galaxy sub-samples would be reasonable, as long as the number density of objects does not vary systematically across the survey footprint in a manner that cannot be modelled. For example, one could make galaxy sub-samples by selecting galaxies which reside in tomographic bins of photometric redshift, as output by some photometric redshift code, or by selecting in some other observed property, such as radio flux or angular size.

We further highlight procedures to speed up this computationally expensive method, by initially just examining one, and then a few, color cells at a time rather than exploring the parameter space of all of the color-cells simultaneously. This reduces the computational resources by many orders of magnitude. We are still exploring other techniques to decrease the computation cost of this approach but have yet to identify any extremely promising avenues. Ideally one could move away from the very expensive Gibbs sampling, which for 12 color cells requires around 3 hours per step on 16 CPUs, but still retain high acceptance rates.

In this work we have not allowed the cosmological parameters to vary, which do effect the modelled correlation functions. We see the number of additional cosmological parameters ($\sim 10$) as a low dimensional addition to the already very high dimensional space ($\sim 216$). Indeed motivated by the results of \cite{2017MNRAS.466.3558M} we would not expect this to produce drastically different posteriors, but in the case that they were to be very different, they would be self consistent and  data driven.

Finally we note that this entire procedure of jointly estimating redshift distributions and galaxy-dark matter bias parameter values from the data using correlation functions, would have to be repeated should there be a change to assumed model, for example if one adopts a more complex model of gravity, or a galaxy-dark matter bias model with more degrees of freedom. However performing this analysis once, would already suggest reasonable starting positions for the chains in the new parameter space.

\section*{Acknowledgments} 
\label{ack}
BH would like to thank Stella Seitz, Kerstin Peach, Olivier Freidrich and Matteo Constanzi  for useful discussion, and J. Dietrich for highlighting the importance of being first \citep[][]{2008PASP..120..224D} on the arXiv, {\rr and an anonymous referee who's comments have helped improve the clarity of the manuscript. }MMR is supported by DOE grant DESC0011114.
 
Funding for the SDSS and SDSS-II has been provided by the Alfred
P. Sloan Foundation, the Participating Institutions, the
National Science Foundation, the U.S. Department of
Energy, the National Aeronautics and Space Administration,
the Japanese Monbukagakusho, the Max Planck
Society, and the Higher Education Funding Council for
England. The SDSS Web Site is http://www.sdss.org/.  Some of the results in this paper have been derived using the HEALPix \citep[][]{2005ApJ...622..759G} package.

\appendix

\label{appendix}

\subsection{MYSQL QUERIES}
\label{mysql}
To extract SDSS data we use the following query in the SDSS Data Release 12 schema:
\begin{verbatim}
SELECT s.specObjID, s.objid, s.ra,s.dec, 
s.z as specz, s.zerr as specz_err,
s.type as specType, q.type as photpType,
q.dered_u, q.dered_g,
q.dered_r, q.dered_i,
q.dered_z,
    
INTO mydb.specPhotoDR10v4 FROM SpecPhotoAll AS s 
   
JOIN photoObjAll AS q 
ON s.objid=q.objid  AND q.dered_g>0 
AND q.dered_r>0 AND q.dered_z>0 

LEFT OUTER JOIN Photoz AS p ON s.objid=p.objid
\end{verbatim}

and further post process these data as described in \S\ref{data}.

\bibliographystyle{mn2e}
\bibliography{photoz}

\end{document}